\documentclass[longauth]{aa} 

\bibpunct{(}{)}{;}{a}{}{,} 
\usepackage{graphicx}

\usepackage[table]{xcolor}
\usepackage{txfonts}
\usepackage{enumitem}
\usepackage{subcaption}
\usepackage{verbatim}
\usepackage{multicol}
\usepackage{multirow}
\usepackage{float}
\usepackage{euclid}
\usepackage{tabularx}
\newcolumntype{R}{>{\raggedright\let\newline\\\arraybackslash\hspace{0pt}}X}
\newcolumntype{L}{>{\raggedleft\let\newline\\\arraybackslash\hspace{0pt}}X}
\usepackage{geometry}

\definecolor{darkblue}{RGB}{13, 0, 120}
\definecolor{cyan}{RGB}{13, 255, 255}
\definecolor{yellow}{RGB}{253, 255, 11}
\definecolor{darkred}{RGB}{161, 17, 26}

\newcommand{\YEuc} {$Y_{\scriptscriptstyle\rm E}$} 
\newcommand{\JEuc} {$J_{\scriptscriptstyle\rm E}$} 
\newcommand{\HEuc} {$H_{\scriptscriptstyle\rm E}$}
\newcommand{\IEuc} {$I_{\scriptscriptstyle\rm E}$}

\newcommand{\mYEuc} {Y_{\scriptscriptstyle\rm E}} 
\newcommand{\mJEuc} {J_{\scriptscriptstyle\rm E}} 
\newcommand{\mHEuc} {H_{\scriptscriptstyle\rm E}}
\newcommand{\mIEuc} {I_{\scriptscriptstyle\rm E}}

\usepackage[]{hyperref}
\begin{document} 

\title{\Euclid preparation. XXI. Intermediate-redshift contaminants in the search for z > 6 galaxies within the \Euclid Deep Survey}

\newcommand{\orcid}[1]{} 
\author{Euclid Collaboration: S.E.~van Mierlo\orcid{0000-0001-8289-2863}$^{1}$\thanks{\email{mierlo@astro.rug.nl}}, K. I.~Caputi$^{1,2}$, M.~Ashby\orcid{0000-0002-3993-0745}$^{3}$, H.~Atek\orcid{0000-0002-7570-0824}$^{4}$, M.~Bolzonella\orcid{0000-0003-3278-4607}$^{5}$, R.A.A.~Bowler\orcid{0000-0003-3917-1678}$^{6,7}$, G.~Brammer\orcid{0000-0003-2680-005X}$^{2,8}$, C.J.~Conselice$^{6}$, J.~Cuby$^{9}$, P.~Dayal$^{1}$
% ...
} 

\institute{$^{1}$ Kapteyn Astronomical Institute, University of Groningen, PO Box 800, 9700 AV Groningen, The Netherlands\\
$^{2}$ Cosmic Dawn Center (DAWN)\\
$^{3}$ Center for Astrophysics | Harvard \& Smithsonian, 60 Garden St., Cambridge, MA 02138, USA\\
$^{4}$ Institut d'Astrophysique de Paris, 98bis Boulevard Arago, F-75014, Paris, France\\
$^{5}$ INAF-Osservatorio di Astrofisica e Scienza dello Spazio di Bologna, Via Piero Gobetti 93/3, I-40129 Bologna, Italy\\
$^{6}$ Jodrell Bank Centre for Astrophysics, Department of Physics and Astronomy, University of Manchester, Oxford Road, Manchester M13 9PL, UK\\
$^{7}$ Department of Physics, Oxford University, Keble Road, Oxford OX1 3RH, UK\\
$^{8}$ Niels Bohr Institute, University of Copenhagen, Jagtvej 128, 2200 Copenhagen, Denmark\\
$^{9}$ Aix-Marseille Univ, CNRS, CNES, LAM, Marseille, France\\
$^{10}$ Departamento F\'isica Aplicada, Universidad Polit\'ecnica de Cartagena, Campus Muralla del Mar, 30202 Cartagena, Murcia, Spain\\
$^{11}$ The University of Texas at Austin, Austin, TX, 78712, USA\\
$^{12}$ Leiden Observatory, Leiden University, Niels Bohrweg 2, 2333 CA Leiden, The Netherlands\\
$^{13}$ Instituto de Astrof\'isica e Ci\^encias do Espa\c{c}o, Universidade do Porto, CAUP, Rua das Estrelas, PT4150-762 Porto, Portugal\\
$^{14}$ Institut d'Astrophysique de Paris, UMR 7095, CNRS, and Sorbonne Universit\'e, 98 bis boulevard Arago, 75014 Paris, France\\
$^{15}$ Dark Cosmology Centre, Niels Bohr Institute, University of Copenhagen, Juliane Maries Vej 30, DK-2100 Copenhagen, Denmark\\
$^{16}$ Department of Astronomy, University of Geneva, ch. d\'Ecogia 16, CH-1290 Versoix, Switzerland\\
$^{17}$ Dipartimento di Fisica e Astronomia "G.Galilei", Universit\'a di Padova, Via Marzolo 8, I-35131 Padova, Italy\\
$^{18}$ Max-Planck-Institut f\"ur Astronomie, K\"onigstuhl 17, D-69117 Heidelberg, Germany\\
$^{19}$ Department of Physics \& Astronomy, University of Sussex, Brighton BN1 9QH, UK\\
$^{20}$ NRC Herzberg, 5071 West Saanich Rd, Victoria, BC V9E 2E7, Canada\\
$^{21}$ Institute of Cosmology and Gravitation, University of Portsmouth, Portsmouth PO1 3FX, UK\\
$^{22}$ Dipartimento di Fisica e Astronomia, Universit\'a di Bologna, Via Gobetti 93/2, I-40129 Bologna, Italy\\
$^{23}$ INFN-Sezione di Bologna, Viale Berti Pichat 6/2, I-40127 Bologna, Italy\\
$^{24}$ Max Planck Institute for Extraterrestrial Physics, Giessenbachstr. 1, D-85748 Garching, Germany\\
$^{25}$ Universit\"ats-Sternwarte M\"unchen, Fakult\"at f\"ur Physik, Ludwig-Maximilians-Universit\"at M\"unchen, Scheinerstrasse 1, 81679 M\"unchen, Germany\\
$^{26}$ INAF-Osservatorio Astrofisico di Torino, Via Osservatorio 20, I-10025 Pino Torinese (TO), Italy\\
$^{27}$ Department of Mathematics and Physics, Roma Tre University, Via della Vasca Navale 84, I-00146 Rome, Italy\\
$^{28}$ INFN-Sezione di Roma Tre, Via della Vasca Navale 84, I-00146, Roma, Italy\\
$^{29}$ INAF-Osservatorio Astronomico di Capodimonte, Via Moiariello 16, I-80131 Napoli, Italy\\
$^{30}$ Dipartimento di Fisica, Universit\'a degli Studi di Torino, Via P. Giuria 1, I-10125 Torino, Italy\\
$^{31}$ INFN-Sezione di Torino, Via P. Giuria 1, I-10125 Torino, Italy\\
$^{32}$ INAF-IASF Milano, Via Alfonso Corti 12, I-20133 Milano, Italy\\
$^{33}$ Port d'Informaci\'{o} Cient\'{i}fica, Campus UAB, C. Albareda s/n, 08193 Bellaterra (Barcelona), Spain\\
$^{34}$ Institut de F\'{i}sica d'Altes Energies (IFAE), The Barcelona Institute of Science and Technology, Campus UAB, 08193 Bellaterra (Barcelona), Spain\\
$^{35}$ INAF-Osservatorio Astronomico di Roma, Via Frascati 33, I-00078 Monteporzio Catone, Italy\\
$^{36}$ INFN section of Naples, Via Cinthia 6, I-80126, Napoli, Italy\\
$^{37}$ Department of Physics "E. Pancini", University Federico II, Via Cinthia 6, I-80126, Napoli, Italy\\
$^{38}$ Dipartimento di Fisica e Astronomia "Augusto Righi" - Alma Mater Studiorum Universit\'a di Bologna, Viale Berti Pichat 6/2, I-40127 Bologna, Italy\\
$^{39}$ INAF-Osservatorio Astrofisico di Arcetri, Largo E. Fermi 5, I-50125, Firenze, Italy\\
$^{40}$ Centre National d'Etudes Spatiales, Toulouse, France\\
$^{41}$ Institut national de physique nucl\'eaire et de physique des particules, 3 rue Michel-Ange, 75794 Paris C\'edex 16, France\\
$^{42}$ Institute for Astronomy, University of Edinburgh, Royal Observatory, Blackford Hill, Edinburgh EH9 3HJ, UK\\
$^{43}$ European Space Agency/ESRIN, Largo Galileo Galilei 1, 00044 Frascati, Roma, Italy\\
$^{44}$ ESAC/ESA, Camino Bajo del Castillo, s/n., Urb. Villafranca del Castillo, 28692 Villanueva de la Ca\~nada, Madrid, Spain\\
$^{45}$ Univ Lyon, Univ Claude Bernard Lyon 1, CNRS/IN2P3, IP2I Lyon, UMR 5822, F-69622, Villeurbanne, France\\
$^{46}$ Institute of Physics, Laboratory of Astrophysics, Ecole Polytechnique F\'{e}d\'{e}rale de Lausanne (EPFL), Observatoire de Sauverny, 1290 Versoix, Switzerland\\
$^{47}$ Departamento de F\'isica, Faculdade de Ci\^encias, Universidade de Lisboa, Edif\'icio C8, Campo Grande, PT1749-016 Lisboa, Portugal\\
$^{48}$ Instituto de Astrof\'isica e Ci\^encias do Espa\c{c}o, Faculdade de Ci\^encias, Universidade de Lisboa, Campo Grande, PT-1749-016 Lisboa, Portugal\\
$^{49}$ Universit\'e Paris-Saclay, CNRS, Institut d'astrophysique spatiale, 91405, Orsay, France\\
$^{50}$ INFN-Padova, Via Marzolo 8, I-35131 Padova, Italy\\
$^{51}$ AIM, CEA, CNRS, Universit\'{e} Paris-Saclay, Universit\'{e} de Paris, F-91191 Gif-sur-Yvette, France\\
$^{52}$ INAF-Osservatorio Astronomico di Trieste, Via G. B. Tiepolo 11, I-34143 Trieste, Italy\\
$^{53}$ Aix-Marseille Univ, CNRS/IN2P3, CPPM, Marseille, France\\
$^{54}$ Istituto Nazionale di Astrofisica (INAF) - Osservatorio di Astrofisica e Scienza dello Spazio (OAS), Via Gobetti 93/3, I-40127 Bologna, Italy\\
$^{55}$ Istituto Nazionale di Fisica Nucleare, Sezione di Bologna, Via Irnerio 46, I-40126 Bologna, Italy\\
$^{56}$ INAF-Osservatorio Astronomico di Padova, Via dell'Osservatorio 5, I-35122 Padova, Italy\\
$^{57}$ Institute of Theoretical Astrophysics, University of Oslo, P.O. Box 1029 Blindern, N-0315 Oslo, Norway\\
$^{58}$ Jet Propulsion Laboratory, California Institute of Technology, 4800 Oak Grove Drive, Pasadena, CA, 91109, USA\\
$^{59}$ von Hoerner \& Sulger GmbH, Schlo{\ss}Platz 8, D-68723 Schwetzingen, Germany\\
$^{60}$ Technical University of Denmark, Elektrovej 327, 2800 Kgs. Lyngby, Denmark\\
$^{61}$ Universit\'e Paris-Saclay, Universit\'e Paris Cit\'e, CEA, CNRS, Astrophysique, Instrumentation et Mod\'elisation Paris-Saclay, 91191 Gif-sur-Yvette, France\\
$^{62}$ Mullard Space Science Laboratory, University College London, Holmbury St Mary, Dorking, Surrey RH5 6NT, UK\\
$^{63}$ Universit\'e de Gen\`eve, D\'epartement de Physique Th\'eorique and Centre for Astroparticle Physics, 24 quai Ernest-Ansermet, CH-1211 Gen\`eve 4, Switzerland\\
$^{64}$ Department of Physics and Helsinki Institute of Physics, Gustaf H\"allstr\"omin katu 2, 00014 University of Helsinki, Finland\\
$^{65}$ European Space Agency/ESTEC, Keplerlaan 1, 2201 AZ Noordwijk, The Netherlands\\
$^{66}$ NOVA optical infrared instrumentation group at ASTRON, Oude Hoogeveensedijk 4, 7991PD, Dwingeloo, The Netherlands\\
$^{67}$ Argelander-Institut f\"ur Astronomie, Universit\"at Bonn, Auf dem H\"ugel 71, 53121 Bonn, Germany\\
$^{68}$ Dipartimento di Fisica e Astronomia "Augusto Righi" - Alma Mater Studiorum Universit\`{a} di Bologna, via Piero Gobetti 93/2, I-40129 Bologna, Italy\\
$^{69}$ Department of Physics, Institute for Computational Cosmology, Durham University, South Road, DH1 3LE, UK\\
$^{70}$ Universit\'e C\^{o}te d'Azur, Observatoire de la C\^{o}te d'Azur, CNRS, Laboratoire Lagrange, Bd de l'Observatoire, CS 34229, 06304 Nice cedex 4, France\\
$^{71}$ Department of Physics and Astronomy, University of Aarhus, Ny Munkegade 120, DK-8000 Aarhus C, Denmark\\
$^{72}$ Institute of Space Science, Bucharest, Ro-077125, Romania\\
$^{73}$ Departamento de F\'isica, FCFM, Universidad de Chile, Blanco Encalada 2008, Santiago, Chile\\
$^{74}$ IFPU, Institute for Fundamental Physics of the Universe, via Beirut 2, 34151 Trieste, Italy\\
$^{75}$ Centro de Investigaciones Energ\'eticas, Medioambientales y Tecnol\'ogicas (CIEMAT), Avenida Complutense 40, 28040 Madrid, Spain\\
$^{76}$ Instituto de Astrof\'isica e Ci\^encias do Espa\c{c}o, Faculdade de Ci\^encias, Universidade de Lisboa, Tapada da Ajuda, PT-1349-018 Lisboa, Portugal\\
$^{77}$ Universidad Polit\'ecnica de Cartagena, Departamento de Electr\'onica y Tecnolog\'ia de Computadoras, 30202 Cartagena, Spain\\
$^{78}$ Infrared Processing and Analysis Center, California Institute of Technology, Pasadena, CA 91125, USA\\
$^{79}$ INAF-Osservatorio Astronomico di Brera, Via Brera 28, I-20122 Milano, Italy\\
$^{80}$  Universit\'e Paris Cit\'e, CNRS, Astroparticule et Cosmologie, F-75013 Paris, France\\
$^{81}$ Dipartimento di Fisica "Aldo Pontremoli", Universit\'a degli Studi di Milano, Via Celoria 16, I-20133 Milano, Italy\\
$^{82}$ INFN-Sezione di Milano, Via Celoria 16, I-20133 Milano, Italy\\
$^{83}$ INFN, Sezione di Trieste, Via Valerio 2, I-34127 Trieste TS, Italy\\
$^{84}$ SISSA, International School for Advanced Studies, Via Bonomea 265, I-34136 Trieste TS, Italy\\
$^{85}$ Instituto de Astrof\'isica de Canarias, Calle V\'ia L\'actea s/n, E-38204, San Crist\'obal de La Laguna, Tenerife, Spain\\
$^{86}$ Departamento de Astrof\'{i}sica, Universidad de La Laguna, E-38206, La Laguna, Tenerife, Spain\\
$^{87}$ Institut de Recherche en Astrophysique et Plan\'etologie (IRAP), Universit\'e de Toulouse, CNRS, UPS, CNES, 14 Av. Edouard Belin, F-31400 Toulouse, France\\
$^{88}$ Dipartimento di Fisica - Sezione di Astronomia, Universit\'a di Trieste, Via Tiepolo 11, I-34131 Trieste, Italy\\
$^{89}$ Dipartimento di Fisica e Scienze della Terra, Universit\'a degli Studi di Ferrara, Via Giuseppe Saragat 1, I-44122 Ferrara, Italy\\
$^{90}$ INAF, Istituto di Radioastronomia, Via Piero Gobetti 101, I-40129 Bologna, Italy\\
$^{91}$ INFN-Bologna, Via Irnerio 46, I-40126 Bologna, Italy\\
$^{92}$ Institute for Theoretical Particle Physics and Cosmology (TTK), RWTH Aachen University, D-52056 Aachen, Germany\\
$^{93}$ Department of Physics \& Astronomy, University of California Irvine, Irvine CA 92697, USA\\
$^{94}$ University of Lyon, UCB Lyon 1, CNRS/IN2P3, IUF, IP2I Lyon, France\\
$^{95}$ Institut de Ciencies de l'Espai (IEEC-CSIC), Campus UAB, Carrer de Can Magrans, s/n Cerdanyola del Vall\'es, 08193 Barcelona, Spain\\
$^{96}$ Institute of Space Sciences (ICE, CSIC), Campus UAB, Carrer de Can Magrans, s/n, 08193 Barcelona, Spain\\
$^{97}$ INFN-Sezione di Genova, Via Dodecaneso 33, I-16146, Genova, Italy\\
$^{98}$ INAF-Istituto di Astrofisica e Planetologia Spaziali, via del Fosso del Cavaliere, 100, I-00100 Roma, Italy\\
$^{99}$ Instituto de F\'isica Te\'orica UAM-CSIC, Campus de Cantoblanco, E-28049 Madrid, Spain\\
$^{100}$ Department of Physics, P.O. Box 64, 00014 University of Helsinki, Finland\\
$^{101}$ Department of Physics, Lancaster University, Lancaster, LA1 4YB, UK\\
$^{102}$ Universit\'e de Paris, F-75013, Paris, France, LERMA, Observatoire de Paris, PSL Research University, CNRS, Sorbonne Universit\'e, F-75014 Paris, France\\
$^{103}$ Instituto de Astrof\'isica de Canarias (IAC); Departamento de Astrof\'isica, Universidad de La Laguna (ULL), E-38200, La Laguna, Tenerife, Spain\\
$^{104}$ Code 665, NASA Goddard Space Flight Center, Greenbelt, MD 20771 and SSAI, Lanham, MD 20770, USA\\
$^{105}$ Dipartimento di Fisica, Sapienza Universit\`a di Roma, Piazzale Aldo Moro 2, I-00185 Roma, Italy\\
$^{106}$ Zentrum f\"ur Astronomie, Universit\"at Heidelberg, Philosophenweg 12, D- 69120 Heidelberg, Germany\\
$^{107}$ Institut f\"ur Theoretische Physik, University of Heidelberg, Philosophenweg 16, 69120 Heidelberg, Germany\\
$^{108}$ INAF-Sezione di Lecce, c/o Dipartimento Matematica e Fisica, Via per Arnesano, I-73100, Lecce, Italy\\
$^{109}$ INFN, Sezione di Lecce, Via per Arnesano, CP-193, I-73100, Lecce, Italy\\
$^{110}$ Department of Mathematics and Physics E. De Giorgi, University of Salento, Via per Arnesano, CP-I93, I-73100, Lecce, Italy\\
$^{111}$ Institute for Computational Science, University of Zurich, Winterthurerstrasse 190, 8057 Zurich, Switzerland\\
$^{112}$ Department of Astrophysical Sciences, Peyton Hall, Princeton University, Princeton, NJ 08544, USA\\
$^{113}$ Department of Physics, P.O.Box 35 (YFL), 40014 University of Jyv\"askyl\"a, Finland\\
$^{114}$ Helsinki Institute of Physics, Gustaf H{\"a}llstr{\"o}min katu 2, University of Helsinki, Helsinki, Finland}
   \date{Received 4 May 2022; accepted 18 July 2022}

  \abstract
   {The \Euclid\ mission is expected to discover thousands of $z>6$ galaxies in three deep fields, which together will cover a $\sim$50 deg$^2$ area. However, the limited number of \Euclid\ bands (four) and the low availability of ancillary data could make the identification of $z>6$ galaxies challenging. }
   {In this work we assess the degree of contamination by intermediate-redshift galaxies ($z=1$--5.8) expected for $z>6$ galaxies within the Euclid Deep Survey.}
   {This study is based on $\sim$176\,000 real galaxies at $z=1$--8 in a $\sim$0.7 deg$^2$ area selected from the UltraVISTA ultra-deep survey and $\sim$96\,000 mock galaxies with $25.3\leq H<27.0$, which altogether cover the range of magnitudes to be probed in the Euclid Deep Survey. We simulate \Euclid\ and ancillary photometry from fiducial 28-band photometry and fit spectral energy distributions to various combinations of these simulated data.}
   {We demonstrate that identifying $z>6$ galaxies with \Euclid\ data alone will be very effective, with a $z>6$ recovery of 91\% (88\%)\, for bright (faint) galaxies. For the UltraVISTA-like bright sample, the percentage of $z=1$--5.8 contaminants amongst apparent $z>6$ galaxies as observed with \Euclid\ alone is 18\,\%, which is reduced to 4\% (13\%)\, by including ultra-deep Rubin (\textit{Spitzer}) photometry. Conversely, for the faint mock sample, the contamination fraction with \Euclid\ alone is considerably higher at 39\,\%, and minimised to 7\,\% when including ultra-deep Rubin data. For UltraVISTA-like bright galaxies, we find that \Euclid\ $(\mIEuc-\mYEuc)>2.8$ and $(\mYEuc-\mJEuc)<1.4$ colour criteria can separate contaminants from true $z>6$ galaxies, although these are applicable to only 54\,\% of the contaminants as many have unconstrained $(\mIEuc-\mYEuc)$ colours. In the best scenario, these cuts reduce the contamination fraction to 1\,\% whilst preserving 81\,\% of the fiducial $z>6$ sample. For the faint mock sample, colour cuts are infeasible; we find instead that a $5\sigma$ detection threshold requirement in at least one of the \Euclid\ near-infrared bands reduces the contamination fraction to 25\,\%.}
  {}
   \titlerunning{Intermediate-redshift contaminants of \Euclid\ $z>6$ galaxies}
   \authorrunning{S.E. van Mierlo et al.}
   \maketitle
   

\section{Introduction}\label{sec:introduction}

Over more than a decade now, numerous works have investigated the presence of galaxies around the epoch of re-ionisation. In particular,  photometric studies of various fields have identified many galaxies at $z>6$, mostly through deep \HST (HST) imaging (e.g. \citealt{bouwens+2010};  \citealt{ellis2013}; \citealt{oesch2014}; \citealt{atek2015}; \citealt{finkelstein2015}; \citealt{bowler2020}; \citealt{salmon2020}; \citealt{roberts2022}). These studies provide us with clues regarding the physical nature of the objects present in the early Universe, which is of key importance for constraining the early phases of galaxy evolution.

The number densities of low-luminosity $z>6$ galaxies are relatively high, enabling a search for these sources in deep, small-area surveys, such as the Cosmic Assembly Near-infrared Deep Extragalactic Legacy Survey (CANDELS; \citealt{grogin2011,koekemoer2011}). Conversely, bright $z>6$ galaxies ($M_{\mathrm{UV}} \lesssim -20.5$) are much rarer and, thus, are only likely to be found in wide-area surveys with reasonable depths at optical and near-infrared (NIR) wavelengths. For example, recent work from \citet{bouwens2021} determined that the number density of $z\sim6$ galaxies with rest-frame magnitudes $M_{1600} \sim -21$ is $1.4 \times 10^{-5}\, \rm{ Mpc^{-3}\, mag^{-1}}$, and increases by a factor of $\sim 10^3$ for sources with $M_{1600} \sim -17$. Consequently, only a minor fraction of $z>6$ galaxy studies have been devoted to exploring the bright end of the galaxy luminosity function at these high redshifts (e.g. \citealt{willott2013}; \citealt{duncan2014}; \citealt{bowler2015}; \citealt{song2015}; \citealt{stefanon2019}). It is around these brightest galaxies where re-ionisation was presumably completed first \citep{pentericci2014,castellano2016}.

To date, the necessary combination of area and depth to search for bright $z>6$ sources is only available in a few fields \citep{mccracken2012,jarvis2013}. However, the forthcoming \Euclid\ mission \citep{laureijs2011} will open up a new era in the search of such objects by mapping a large area of the sky at NIR wavelengths. In addition to its main wide survey, \Euclid\ will perform deep observations of three so-called Euclid Deep Fields, which will encompass a total area of $\sim$50 deg$^2$ \citep{scaramella2021}. \Euclid\ carries four photometric bands: a {\color{black}visible imager (VIS; \citealt{cropper2016})} that has an optical band \IEuc\ \footnote{Originally referred to as the VIS band within the Euclid Consortium but recently renamed the \IEuc\ band.} at $5500$--$9000\,\AA$ and the {\color{black} Near-Infrared Spectrometer and Photometer (NISP; \citealt{maciaszek2016})} that carries three NIR bands, that is, \YEuc, \JEuc, and \HEuc, which together cover the wavelength range $9000$--$20\,000\,\AA$ \citep{schirmer2022}. The expected 5$\sigma$ depths (assuming point-like sources) for the Euclid Deep Fields are $\mIEuc=28.2$ and $\mYEuc \mJEuc \mHEuc \approx 26.4$ (AB magnitude). With these characteristics, the Euclid Deep Fields are expected to reveal thousands of $z>6$ galaxies and therefore enable studies of early galaxy formation and evolution with unprecedented statistical significance.

Since \Euclid\ has a limited number of photometric bands, enormous efforts are being made to provide additional, external coverage of the Euclid Deep Fields, both with ground-based facilities and the \Spitzer\ (e.g. \citealt{moneti2021}; McPartland et al., in prep). In the best-case scenario, Euclid Deep sources will have photometric coverage in at most 10--12 filters, and therefore deriving accurate photometric redshifts and galaxy physical parameters will be challenging. As such, a pre-launch critical assessment of contamination in \Euclid\ galaxy selections at different redshifts is of utmost importance.

Identifying $z>6$ galaxies in particular is challenging for a number of reasons. Extreme emission line galaxies at intermediate redshifts can mimic Lyman-break galaxies due to a combination of large equivalent width emission lines and a faint continuum, therefore contaminating the selection of high-redshift objects \citep{atek2011,huang14}. A second type of degeneracy arises from the blackbody spectra of cool, brown dwarf stars that have similar NIR colours to $z>6$ galaxies \citep{stern2007, bowler2015, salmon2020}. Finally, another main source of contamination in the selection of $z>6$ sources are intermediate-redshift ($z\sim1$--6) galaxies, for which the $4000\,\AA$ break can be misidentified as the Lyman-$\alpha$ break at $\lambda = 1216\,\AA$ of a high-redshift object \citep{vulcani2017}. This latter sort of contamination is the focus of this work. We note that the study of high-redshift contaminants to intermediate-$z$ sources, especially dusty galaxies at $z=4$--6, is an interesting complementary problem, but outside the scope of this paper.  
 
Here we make use of galaxies selected from the third data release (DR3) of the UltraVISTA ultra-deep survey \citep{mccracken2012} and the \textit{Spitzer} Matching Survey of the UltraVISTA ultra-deep Stripes (SMUVS; \citealt{ashby2018}) to assess the degree of contamination produced by intermediate-redshift galaxies in the selection of $z>6$ galaxies in the Euclid Deep Survey. UltraVISTA and SMUVS are uniquely suited for this simulation because of their considerable common area ($\sim$0.66 deg$^2$) and depths ($\sim$25.5 AB mag). However, we emphasise that this analysis is only valid for galaxies at $z=6$--8 due to the limitations of the fiducial sample, and as such we cannot study the photometric redshift recovery of \Euclid\ $z>8$ galaxies.  
 
This paper is organised as follows. In Sect.\,\ref{sec:data} we briefly describe the datasets used in this work. In Sect.\,\ref{sec:sim_data} we describe our source catalogue construction and how the \Euclid\ and ancillary photometry were simulated. We present our estimates on the contamination fraction of bright $z>6$ galaxies in the Euclid Deep Fields in Sect.\,\ref{sec:results}, together with colour selection criteria to separate intermediate-$z$ interlopers from true $z>6$ galaxies. In addition, we analyse the degree of $z>6$ contamination and the effectiveness of the colour criteria for a sample of faint ($25.3\leq H<27.0$) mock galaxies in Sect.\,\ref{sec:mock}. In Sect.\,\ref{sec:discussion} we comment on the validity of our results and finally present our concluding remarks in Sect.\,\ref{sec:conclusion}. Throughout this paper we adopt a cosmology with $\rm{H_0 = 70\ km\ s^{-1}\ Mpc^{-1}}$, $\Omega_{\rm m}=0.3,$ and $\Omega_\Lambda=0.7$. All magnitudes and fluxes are total, with magnitudes referring to the AB system \citep{okegun1983}. Stellar masses correspond to a \citet{chabrier2003} initial mass function (IMF). 
 

\section{COSMOS as a basis to simulate \Euclid\ galaxies}\label{sec:data}

\subsection{UltraVISTA/SMUVS and non-SMUVS galaxy catalogues}

\noindent As a basis to simulate \Euclid\ (+ancillary) photometry, we use real, NIR galaxy surveys in the field of the Cosmic Evolution Survey (COSMOS; \citealt{scoville2006}). Specifically, the ultra-deep UltraVISTA survey \citep{mccracken2012} has provided \textit{Y, J, H, $K_{\rm s}$} images whose depth is relatively similar ($\sim$25--26 mag) to that expected for the Euclid Deep Fields, and therefore constitutes an excellent starting point to simulate \Euclid\ galaxies. However, given that the Euclid Deep Survey will be 1.2 magnitude deeper in the $H$ band than the UltraVISTA survey, we create a complementary catalogue of \textit{Euclid}-like faint mock galaxies from scaled-down versions of the fiducial UltraVISTA spectral energy distributions (SEDs). This process is described in detail in Sect.\,\ref{sec:mock}; here we discuss the construction of the UltraVISTA galaxy catalogue that forms the basis for both the bright UltraVISTA-like sample and the faint mock sample. 

In this work we only consider the three (out of four) UltraVISTA ultra-deep stripes with ultra-deep \Spitzer\ \citep{werner2004} coverage from SMUVS (\citealt{ashby2018}). This programme used the \textit{Spitzer} Infrared Array Camera (IRAC; \citealt{fazio2004}) to map the three UltraVISTA ultra-deep stripes with deepest ancillary data, reaching matching depths in the 3.6 and 4.5~\micron\ bands, over a total area of $0.66 \, \rm deg^2$. 

\citeauthor{deshmukh18}~(2018; hereafter D18) presented a photometric catalogue of approximately 300\,000 SMUVS sources with multi-wavelength ancillary data in COSMOS, for a total of 28 bands from Canada–France–Hawaii Telescope (CFHT) $u$ through IRAC 4.5~\micron. The SMUVS photometry has been obtained using sources detected in the UltraVISTA DR3 $HK_{\rm s}$ stack mosaic as priors, and by requiring that each source has a detection in at least one of the IRAC bands. Given that the SMUVS images suffer from severe source confusion, the IRAC photometry was measured using a point-spread-function (PSF) fitting technique from the {\color{black}Image Reduction and Analysis Facility (IRAF)}, using empirical images of the PSF as constructed from stars in the field. Using this method, $\sim95$\,\% of all UltraVISTA sources are detected in at least one IRAC band. In addition, the IRAC photometry of sources with bright IRAC neighbours was not utilised in the SED fitting to prevent contamination in these bands from affecting the best-fit SED. 

D18 derived photometric redshifts and stellar masses for all these sources, based on SED analysis, as we explain in Sect.\,\ref{sec:sed}. We refer the reader to D18 for detailed information about the SMUVS catalogue. Here we use it as a basis to obtain our \Euclid\ simulated data.

In addition, we considered a second, complementary catalogue consisting of all the UltraVISTA $HK_{\rm s}$-stack sources that are not \textit{Spitzer}-detected, in the same three UltraVISTA ultra-deep stripes 1, 2, and 3, to which we refer as non-SMUVS sources throughout this work. As in D18, we used the $HK_{\rm s}$ stack source positions to measure $2''$ diameter circular photometry, using the code \texttt{SourceExtractor} \citep{bertin1996}, on 26 bands in the COSMOS field:  CFHT \textit{u}; Subaru SuprimeCam \textit{B, V, r, $i^+$, $z^+$, $z^{++}$, IA427, IA464, IA484, IA505, IA527, IA574, IA624, IA679, IA709, IA738, IA767, IA82, NB711}, and \textit{NB816}; HST \textit{F814W}; and UltraVISTA \textit{Y, J, H}, and \textit{$K_{\rm s}$}. 

The measured fluxes were corrected to total fluxes through point-source aperture corrections, based on the curves of growth of non-saturated stars in the field (as derived by D18). These corrections are consistent with those quoted in \citet{mccracken2012} and \cite{laigle2016}. For the \textit{Spitzer} photometry, typical aperture corrections have been tabulated by \citet{ashby2015}. These authors have demonstrated that treating \textit{Spitzer} sources as point-like is valid in virtually all cases at $z>2$ and in a large fraction of sources at $z=1$--2. Moreover, we note that our \Euclid\ (+ancillary) photometry are simulated directly from our COSMOS photometry and therefore, the recovery fraction and contaminants of high-$z$ sources studied in this paper are not influenced by the use of point-like photometry. This is confirmed by the results shown in Appendix \ref{appendix}, where we repeat our $z>6$ recovery tests on the COSMOS2020 catalogue \citep{weaver2021}, which contains independent photometric measurements of the same field. 

The total fluxes were subsequently corrected for Galactic extinction using the dust maps from \citet{schlafly2011}. We used the masks from D18 to mask regions of contaminated light surrounding the brightest sources. This removes $\sim6$\,\% of the considered UltraVISTA region; as a consequence, our masked catalogue covers a total area of $\sim0.7$~deg$^2$. Following the method outlined in \citet[][Fig.~1]{caputi2006}, we cleaned the non-SMUVS catalogue for Galactic stars using the Subaru SuprimeCam $(i^+-z^+)$ and UltraVISTA $(J-K_{\rm s})$ colour diagram. Sources that have an $HK_{\rm s}$-based stellarity parameter greater than 0.8 and reside in the stellar locus were discarded from our non-SMUVS sample, where the stellar locus refers to sources that have $(J-K_{\rm s}) \leq 0.1$. This approach is slightly different from D18, who used the $(J-[3.6])$ versus $(B-J)$ colour diagram to identify Galactic stars. Given that no IRAC 3.6 \micron\ photometry is available for the non-SMUVS sources, we utilised this alternative colour diagnostic to clean the non-SMUVS catalogue from Galactic stars. 

By including the non-SMUVS sources in our analysis, we gain approximately 19\,700 additional sources. The majority of the additional sources ($\sim70$\,\%) resides in the second ultra-deep stripe, as the northern part of it ($\ang{2.61;;}\le {\rm Dec}\le \ang{2.76;;}$) is not covered by SMUVS.

Finally, for both the UltraVISTA SMUVS and non-SMUVS catalogues, we updated the UltraVISTA photometry using the DR4 mosaics to increase the sensitivity of our photometry, by running \texttt{SourceExtractor} on the DR4 images and matching the resulting source catalogue with our DR3 catalogue. We therefore emphasise that our final UltraVISTA catalogue consists exclusively of DR3-selected sources, of which the UltraVISTA $Y,J,H$, and $K_{\rm s}$ bands have been updated with the DR4 photometry. Between DR3 and DR4, the 5$\sigma$ limiting magnitudes in the ultra-deep stripes increase by 0.1, 0.2, and 0.1~mag in the $Y$, $J$, and $H$ bands, respectively, while the $K_{\rm s}$ band depth is unchanged.

\subsection{Galaxy physical parameters obtained with SED fitting}\label{sec:sed}

\noindent We derived photometric redshifts and main physical parameters for all the galaxies in the general (SMUVS and non-SMUVS) UltraVISTA catalogue with updated DR4 photometry, following a similar SED fitting methodology to that applied by D18, but with a few important changes more suitable for high-redshift sources, as follows. We made use of the $\chi^2$-fitting routine \texttt{LePhare} \citep{arnouts1999, ilbert2006}, adopting a broader set of star formation histories (SFHs) than D18, that is, a single stellar population, an exponentially declining SFH, and a delayed exponentially declining SFH. For both declining models, we used the same range of star formation timescales $\tau$ = 0.01, 0.1, 0.3, 1.0, 3.0, 5.0, 10.0, and 15 Gyr. We considered two metallicities: solar  ($Z=\rm{Z_\odot}$) and sub-solar ($Z=0.2 \rm{Z_\odot}$). In total, we considered 36 templates of different combinations of SFH and metallicity. We also included empirical spectra of L, M, and T stars from the SpeX Prism Library \citep{burgasser2014} to minimise the contamination of the high-$z$ galaxy sample by dwarf stars. The effectiveness of this method in removing brown-dwarf contamination was demonstrated by \citet{bowler2014} and \citet{bowler2015}. Finally, we used the redshift range $z=0$--9 for our SED fitting, whereas D18 used the redshift range $z=0$--7.

As in D18, each SED template was attenuated with the \citet{calzetti2000} reddening law and the colour excess was left as a free parameter between $E(B-V)=0$ and 1, with a step of 0.1.   We ran \texttt{LePhare} with emission lines (the recipe based on simple scaling relations from \citealt{kennicutt1998} between the ultraviolet (UV) luminosity and O{\sc ii} line; see \citealt{ilbert2009}) and multiplied the flux errors by a factor of 1.5 since, as shown by \citet{dahlen2013}, this choice improves photometric redshift estimation. We include a flat prior for the absolute magnitude in the Subaru $r$ band such that $-10 < M_r < -26$.  We adopted the same treatment for non-detections as in D18 and \citet{caputi2015}, that is, we substituted them with $3\sigma$ flux upper limits in the broad bands and simply ignored them in the intermediate- and narrow-band data. We then chose the option in \texttt{LePhare} that rejects any SED template that produces fluxes higher than the 3$\sigma$ upper limits in the bands with non-detections. In order to improve the quality of the fit, we applied photometric zero-point corrections as in D18. These were derived as follows: after we obtained best-fit SEDs with \texttt{LePhare}, we calculated the mean offset between the observed and template fluxes in each band, which were subsequently applied to the photometric catalogue. We repeated this process until the offsets converged to obtain our final photometric redshifts. Averaged over all 28 bands, the offset between the observed and template fluxes is 0.06 mag. 

We cleaned the output redshift catalogue returned by \texttt{LePhare} as follows: first, we removed sources that are best fit by stellar (rather than galaxy) templates. This was achieved by comparing the best-fit galaxy $\chi^2_{\rm{\nu,gal}}$ and stellar $\chi^2_{\rm{\nu,star}}$ values for any source with a $HK_{\rm s}$-based stellarity parameter greater than $0.8$ (as measured with \texttt{SourceExtractor} from the $HK_{\rm s}$ detection image); we removed these sources if $\chi^2_{\rm{\nu,star}} < \chi^2_{\rm{\nu,gal}}$ or if $|\chi^2_{\rm{\nu,gal}}-\chi^2_{\rm{\nu,star}}| < 4$. 

Second, for all galaxies at $z>3.6$, we checked if the high-redshift solution is compatible with their detection at short wavelengths, that is, we ensure galaxies with high-redshift solutions do not have flux bluewards of the Lyman break. Therefore, following \citet{caputi2015}, we discarded sources with $z_{\mathrm{phot}} > 3.6$ and a $>2\sigma\ U$-band detection; or with $z_{\mathrm{phot}} > 4.6$ and a $>2\sigma\ B$-band detection; or with $z_{\mathrm{phot}} > 5.6$ and a $>2\sigma\ V$-band detection; or with $z_{\mathrm{phot}} > 6.6$ and a $>2\sigma\ r$-band detection. To further clarify, these conditions are implemented such that any band bluewards of the Lyman break is checked, for instance, we ensured a $z_{\mathrm{phot}} > 6.6$ source also does not have significant detections in the $U$, $B$, and $V$ bands. In addition, for $z_{\mathrm{phot}}>7$ galaxies we do not expect any detection bluewards of the Lyman-$\alpha$ line, due to Lyman series absorption of neutral hydrogen in the intergalactic medium \citep{inoue+2014}. Therefore, we discarded sources with $z_{\mathrm{phot}}>7.0$ and a $>2\sigma\ z^{++}$-band detection; or with $z_{\mathrm{phot}}>8.0$ and a $>2\sigma\ Y$-band detection. Lastly, for all sources with $z_{\mathrm{phot}} \geq 6$, we performed rigorous visual inspection of their broad-band images and removed all sources that are for example contaminated by bright neighbours or appear artificial (e.g. they are aligned exactly on the diffraction spikes of bright stars).   

\begin{figure}
 \centering
  \includegraphics[width=\hsize]{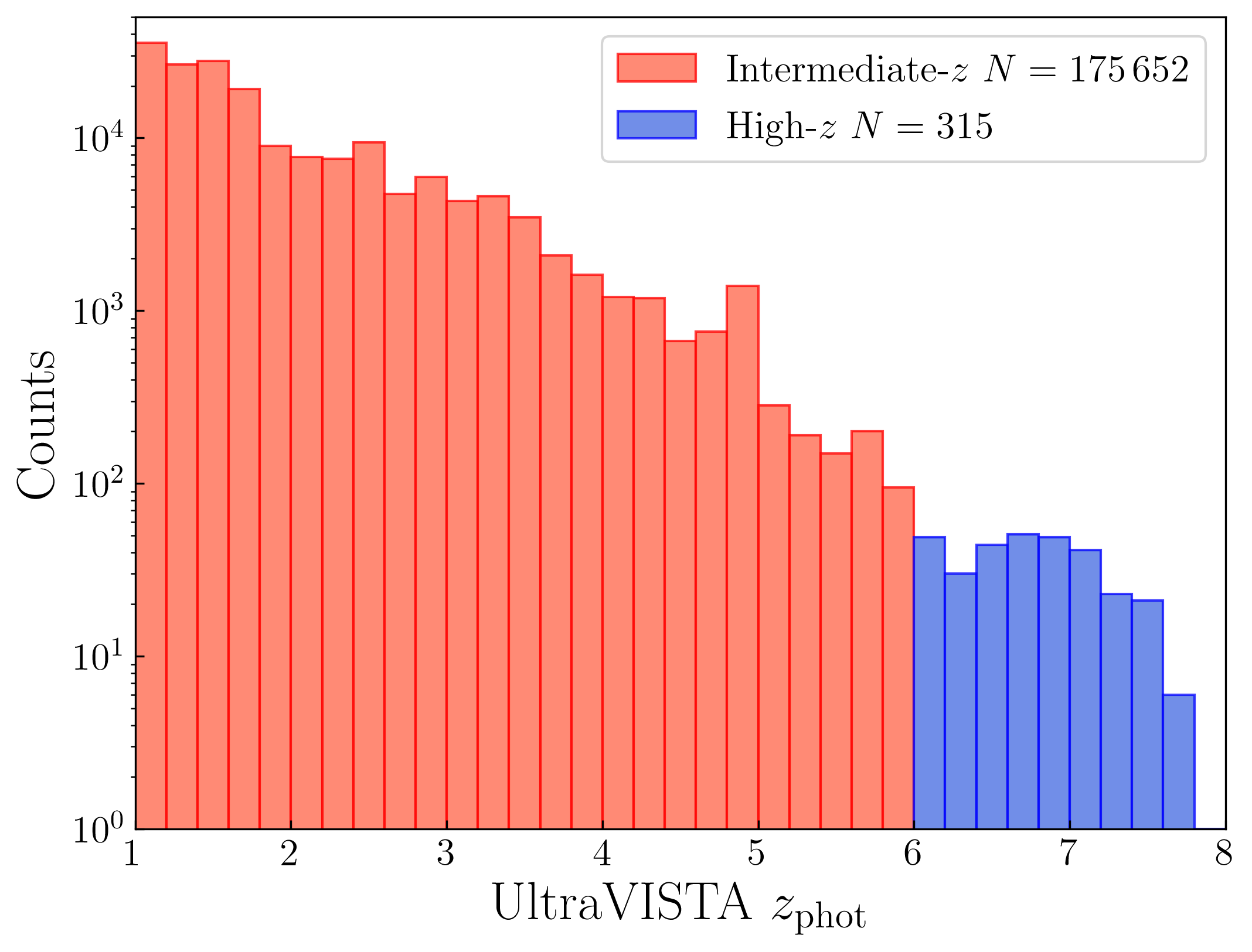}
   \caption{Redshift distribution of the UltraVISTA DR4 $z=1$--8 galaxies in the UltraVISTA ultra-deep stripes 1, 2, and 3. These galaxies constitute the fiducial intermediate ($z=1$--6) and high-$z$ ($z=6$--8) samples in this work. The intermediate-$z$ and high-$z$ samples consist of 175652 and 315 galaxies, respectively.}
 \label{fig:highzsample}
\end{figure}

In total, these measures removed $<0.5$\,\% of the sources. The total (SMUVS and non-SMUVS), clean UltraVISTA catalogue in the three ultra-deep stripes 1, 2, and 3 contains $\sim306\, 000$ galaxies, including $\sim176\,000$ with best photometric redshifts $z=1$--8 (see Fig.\,\ref{fig:highzsample}). These latter objects constitute our fiducial intermediate-redshift ($z=1$--6) and high-redshift ($z=6$--8) galaxy samples.

\section{Simulation of \Euclid, Rubin-LSST, H20 survey, and \textit{Spitzer} photometry of $z=1$--8 galaxies}\label{sec:sim_data}

\noindent The main goal of this analysis is to assess the identification of $z>6$ and $z=1$--6 galaxies based on the data that are (or will be) available in the Euclid Deep Fields. \Euclid\ will observe the sky in four photometric bands: the \IEuc, \YEuc, \JEuc\, and \HEuc\ bands, which together cover the wavelength range $5500$--$20\,000\,\AA$ \citep{schirmer2022}. Given that one of the aims of this research is to investigate how the inclusion of external data improves the photometric redshift of \Euclid\ sources at $z>6$, we considered additional Rubin, CFHT {\color{black}and} Subaru Hyper Suprime Camera (HSC), and \textit{Spitzer} photometry. To simulate photometry in the \Euclid\ (+ancillary) bands, we made use of the above described SMUVS/UltraVISTA galaxy catalogue as a basis. In addition, as described in Appendix \ref{appendix}, we repeated our analysis based on the COSMOS2020 catalogue \citep{weaver2021}, for which all photometric measurements have been obtained in an independent manner. 

In this paper we consider complementary data from the Vera Rubin C. Observatory, which will sample the two southern Euclid Deep Fields in the Legacy Survey of Space and Time (LSST; \citealt{ivezic2019}). The Rubin Observatory will observe in six photometric bands, \textit{ugrizy}, which span the wavelength range $3200$--$11\,000\,\AA$. Given that the Euclid Deep Field North cannot be observed by Rubin, we also {\color{black}consider} the ongoing Hawaii Two-0 Survey (H20), which is currently observing the Euclid Deep Field North and Euclid Deep Field Fornax (McPartland et al., in prep). The H20 survey will consist of deep optical data in the MegaCam $u$ band of the CFHT and the Subaru HSC \textit{g,r,i,z} bands, and will be available long before the Rubin full depth mosaics. Therefore, we consider both simulated Rubin- and H20-like photometry complementary to the \Euclid\ bands. Lastly, \textit{Spitzer}/IRAC observations of the Euclid Deep Fields were presented in \citet{moneti2021}, who combined new observations with all relevant archival IRAC data to produce very deep imaging of these fields in all four IRAC bands. Given that these \textit{Spitzer} mosaics are very similar in depth to SMUVS (5$\sigma$ mag = 24.8), we directly use the observed SMUVS photometry and therefore only consider the IRAC 3.6 and 4.5 \micron\ bands. We note that our choice for including the H20 survey and IRAC bands is based on the Cosmic Dawn Survey (Toft et al., in prep), which is an ongoing effort to obtain multi-wavelength imaging for the Euclid Deep Fields to depths that will match the \Euclid\ data. 

\begin{figure}
 \centering
  \includegraphics[width=\hsize]{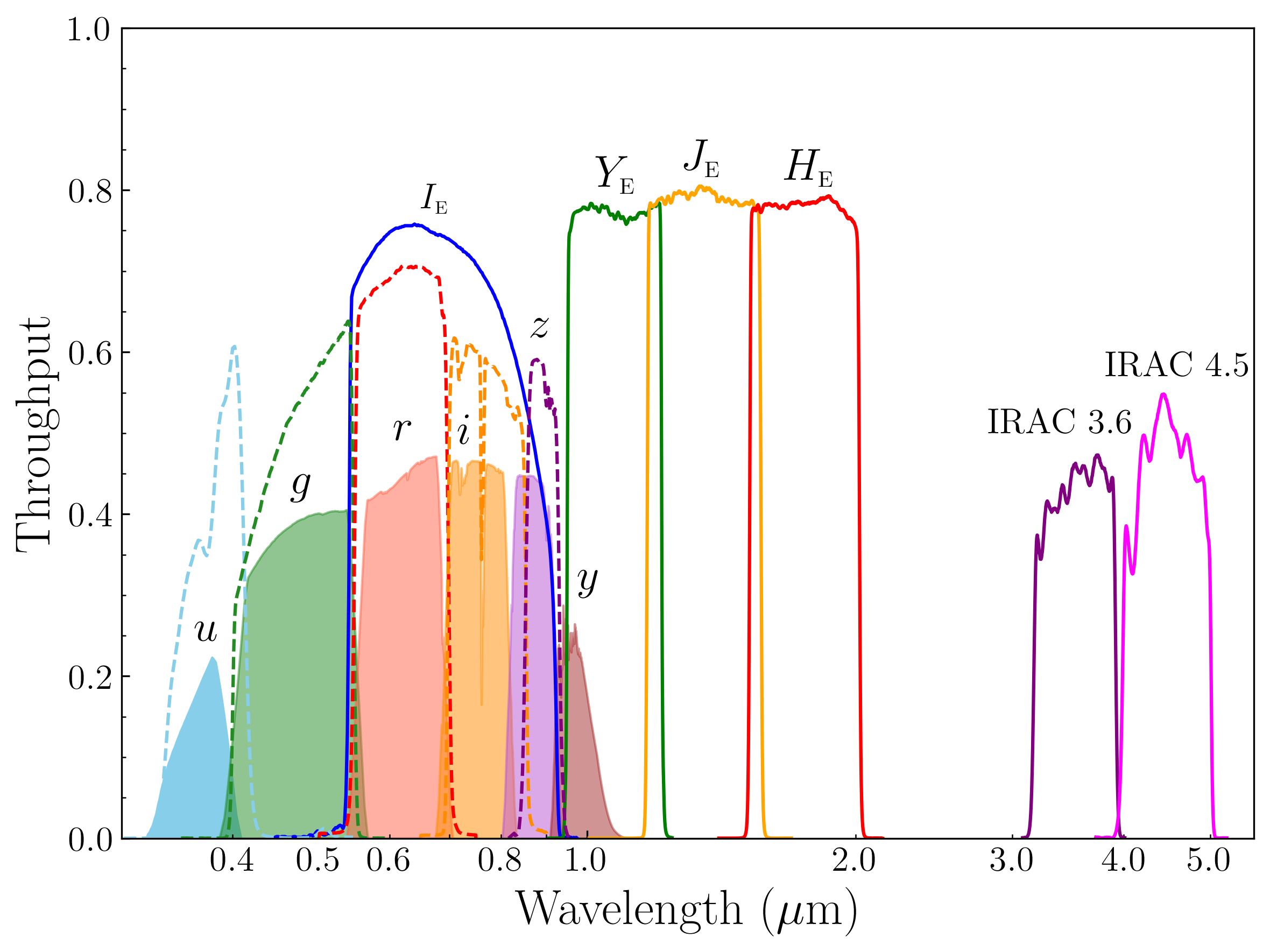}
   \caption{Transmission curves of the \Euclid\ \IEuc, \YEuc, \JEuc, and \HEuc\ filters, the Rubin \textit{u, g, r, i, z}, and \textit{y} filters (filled-in, solid curves), the CFHT $u$ and Subaru HSC \textit{g,r,i}, {\color{black}and $z$ filters} (open, dashed curves), and the \textit{Spitzer} IRAC 3.6 and 4.5 \micron\ filters. }
 \label{fig:throughput}
\end{figure}

In Table\,\ref{table:euclid_rubin_bands} we provide an overview of the expected $5\sigma$ point-like source depths, mean wavelengths and filter widths of \Euclid\ (+ancillary) photometric bands considered in this work. Their corresponding transmission curves are shown in Fig.\,\ref{fig:throughput}. The expected \Euclid\ depths are taken from \citet{scaramella2021}, assuming that the Euclid Deep Survey will be two magnitudes deeper than the Euclid Wide Survey. For our tests we consider two different scenarios for the Rubin 5$\sigma$ point source depth: one that is expected after 10 years of observing and one that is representative for the Rubin Deep Drilling Fields (DDF), which are likely to coincide with the two southern Euclid Deep Fields. We assumed approximate Rubin DDF depths from \citet{foley2018}. It is worth noting that the $5\sigma$ depths presented in Table\,\ref{table:euclid_rubin_bands} are estimates and may vary once all programmes are finalised, with the exception of the already completed \textit{Spitzer} observations.

{\renewcommand{\arraystretch}{1.2}
\begin{table}
\caption{Summary of passbands considered in this work.}             
\label{table:euclid_rubin_bands}      
\centering                          
\begin{tabular}{l c c c}        
\hline\hline                 
Band & 5$\sigma$ depth [AB] & $\lambda_{\mathrm{mean}}$ [\AA] & FWHM [\AA] \\     
\hline                        
   \IEuc & 28.2 & 7140 & 3627 \\      
   \YEuc & 26.3 & 10\,829 & 2667 \\
   \JEuc & 26.5 & 13\,696 & 4053 \\
   \HEuc & 26.4 & 17\,762 & 5032 \\
 \hline
   Rubin/$u$ & 26.1 (26.8)\tablefootmark{a} & 3685 & 516 \\
   Rubin/$g$ & 27.4 (28.4) & 4802 & 1461 \\
   Rubin/$r$ & 27.5 (28.5) & 6231 & 1356 \\
   Rubin/$i$ & 26.8 (28.3) & 7542 & 1248 \\
   Rubin/$z$ & 26.1 (28.0) & 8690 & 1028 \\
   Rubin/$y$ & 24.9 (26.2) & 9736 & 9699 \\
 \hline 
   CFHT/$u$ & 26.2 & 3832 & 3899 \\
   HSC/$g$  & 27.5 & 4816 & 4724 \\
   HSC/$r$  & 27.5 & 6234 & 6136 \\
   HSC/$i$  & 27.0 & 7741 & 7654 \\
   HSC/$z$  & 26.5 & 8911 & 8902 \\
 \hline 
   IRAC/3.6 \micron\ & 24.8 & 35\,634 & 7444\\
   IRAC/4.5 \micron\ & 24.7 & 45\,110 & 10\,119 \\
\hline                                   
\end{tabular}
\tablefoot{ Expected 5$\sigma$ depth (for point-like objects) in the Euclid Deep Fields, mean wavelength, and full width half maximum (FWHM) of the four \Euclid\ filters, the Rubin $ugrizy$ filters, the CFHT $u$ and Subaru/HSC $griz$ filters (McPartland et al., in prep), and the \textit{Spitzer} IRAC $3.6$ and $4.5$ \micron\ filters \citep{moneti2021}. \\
\tablefoottext{a}{Rubin depth after 10 years of observing; in parentheses we show the depths expected for the Rubin DDF \citep{foley2018}.}} 
\end{table}}

We considered our UltraVISTA galaxy catalogue with fiducial $z>1$ redshifts and simulated their \Euclid\ (+ancillary) photometry by convolving their best-fit SEDs based on the 28-band COSMOS photometry with the \Euclid\ (+ancillary) filter curves, which can be easily done with \texttt{LePhare}. We modelled the flux errors following separate methods for each instrument. For the \Euclid\ and H20 photometry, we followed the method presented in \texttt{LePhare} to simulate magnitude errors: 
\begin{equation}
 \sigma_{m} =
    \begin{cases}
  \sigma_{m_5} \times 10^{\ 0.3\ (m-m_5)} & \text{if $m \leq m_5$} \\
  \sigma_{m_5} \times 0.37 \exp\left(10^{\ 0.22\ (m-m_5)}\ \right) & \text{if $m > m_5$}
\end{cases}\ ,
\end{equation}
where $m_5$ is the 5$\sigma$ point-source depth from Table \ref{table:euclid_rubin_bands} and $\sigma_{m_5}$ is the corresponding magnitude error. In addition, we added a 0.03 systemic magnitude error to $\sigma_m$ in quadrature. For the Rubin photometry, we used the formulae provided in \citet{ivezic2019}. The Rubin total photometric error has both a systematic and a random contribution, with the latter being dependent on the expected 5$\sigma$ depth in each band. We refer the reader to \citet{ivezic2019} for a detailed description of the Rubin flux error prescription. For the \textit{Spitzer} photometry we {\color{black}adopted} a strategy where the fluxes are sampled from the fiducial best-fit SED (as was done for the \Euclid, H20, and Rubin photometry) and the flux errors are simply set to the observed \textit{Spitzer} flux errors from SMUVS, as the \textit{Spitzer} observations of the Euclid Deep Fields are similar in depth to SMUVS. Given that not all sources in our UltraVISTA catalogue are \textit{Spitzer}-detected, we only simulated \textit{Spitzer} photometry for galaxies that have a detection in at least one IRAC band. All simulated magnitudes and magnitude errors {\color{black}were} subsequently converted to flux space. 

In total, we address eight scenarios of different combinations of \Euclid, Rubin, H20, and \textit{Spitzer} photometry, as summarised in Table~\ref{table:summary_scenarios}. Throughout this paper, we globally refer to these combinations as \Euclid\ (+ancillary) photometry. The number of final sources in the simulated \Euclid\ photometric catalogues are listed in this table, where the distinction between intermediate-$z$ and high-$z$ galaxies is based on the fiducial redshift. We remind the reader that the number of sources in the catalogues including \textit{Spitzer} photometry is lower as not all galaxies in our UltraVISTA DR4 catalogue are IRAC-detected. For each filter we randomised the simulated photometry of each galaxy by sampling a Gaussian distribution with mean $\mu$ equal to the modelled flux from the fiducial best-fit SED and standard deviation $\sigma$ equal to the flux error, derived as explained above. 

Since our simulated fluxes are directly sampled from the fiducial best-fit galaxy template, they are unaffected by exposure time limits, contrary to real, observed photometry. Therefore, to ensure our simulated photometry is realistically deep, we applied a $2\sigma$ flux limit to each filter, as derived from their expected $5\sigma$ survey depth. For \Euclid, Rubin, and H20 bands with fluxes fainter than their $2\sigma$ detection limits, we adopted $2\sigma$ flux upper limits in the subsequent SED fitting process. For \textit{Spitzer} fluxes fainter than the corresponding $2\sigma$ limits, we did not adopt $2\sigma$ upper limits, but rather excluded the band in the SED fitting process. We implemented this criterion because the $\chi^2$-minimisation technique of the SED fitting naturally has most of its weight at the longest wavelength filters and could be confused rather than helped by the presence of flux upper limits. 

In our analysis, we consider a single realisation of the randomly simulated
photometry. We produced and analysed a few other realisations, but found no significant differences in the results discussed below.  
 {\renewcommand{\arraystretch}{1.5}
\begin{table*}
\caption{Scenarios of simulated data availability in the Euclid Deep Fields considered in this work.}             
\label{table:summary_scenarios}      
\centering     
\small
\begin{tabular}{l l  c c}        
\hline\hline                 
Scenario & Description & $z=1$--6 & $z=6$--8 \\     
\hline                        
   \Euclid\ & \Euclid\ \IEuc,\YEuc,\JEuc,\HEuc & 175\,652 & 315 \\
   \Euclid\ + Rubin & \Euclid\ \IEuc,\YEuc,\JEuc,\HEuc; Rubin $ugrizy$ (depth after 10 years) & 175\,652 & 315 \\
   \Euclid\ + Rubin DDF & \Euclid\ \IEuc,\YEuc,\JEuc,\HEuc; Rubin $ugrizy$ (Deep Drilling Fields depth) & 175\,652 & 315 \\
   \Euclid\ + H20 & \Euclid\ \IEuc,\YEuc,\JEuc,\HEuc; CFHT $u$ and Subaru HSC $griz$ & 175\,652 & 315 \\
   \Euclid\ + \textit{Spitzer} & \Euclid\ \IEuc,\YEuc,\JEuc,\HEuc; \textit{Spitzer} IRAC 3.6 and 4.5 \micron\ & 134\,562 & 203 \\
   \Euclid\ + Rubin + \textit{Spitzer} & \Euclid\ \IEuc,\YEuc,\JEuc,\HEuc; Rubin $ugrizy$ (depth after 10 years); & 134\,562 & 203\\
   & \textit{Spitzer} IRAC 3.6 and 4.5 \micron\  & & \\
   \Euclid\ + Rubin DDF + \textit{Spitzer} & \Euclid\ \IEuc,\YEuc,\JEuc,\HEuc; Rubin $ugrizy$ (Deep Drilling Fields depth); & 134\,562 & 203\\
   & \textit{Spitzer} IRAC 3.6 and 4.5 \micron\ &  &  \\
   \Euclid\ + H20 + \textit{Spitzer} & \Euclid\ \IEuc,\YEuc,\JEuc,\HEuc; CFHT $u$ and Subaru HSC $griz$; \textit{Spitzer} IRAC 3.6 and 4.5 \micron\ & 134\,562 & 203\\
\hline\hline 
Photometry & \multicolumn{3}{l}{2$\sigma$ limiting magnitude [AB]}\\
\hline 
\Euclid\ & \multicolumn{3}{l}{$\mIEuc=29.2$; $\mYEuc=27.3$; $\mJEuc=27.5$; $\mHEuc=27.4$} \\
Rubin (10 yrs depth) & \multicolumn{3}{l}{$u=27.1$; $g=28.4$; $r=28.5$; $i=27.8$; $z=27.1$; $y=25.9$}\\
Rubin (DDF depth) & \multicolumn{3}{l}{$u=27.8$; $g=29.4$; $r=29.5$; $i=29.3$; $z=29.0$; $y=27.2$}\\
H20 & \multicolumn{3}{l}{$u=27.2$; $g=28.5$; $r=28.5$; $i=28.0$; $z=27.5$}\\
\textit{Spitzer} & \multicolumn{3}{l}{$[3.6]=25.8$; $[4.5]=25.7$} \\
\hline                                   
\end{tabular}
\tablefoot{Summary of the eight scenarios of combinations of \Euclid\ and ancillary data considered in this work and the corresponding $2\sigma$ magnitude limits that were applied to the photometry. The number of intermediate-$z$ and high-$z$ galaxies that are inserted in our simulations (based on their fiducial redshift) is indicated in the last two columns.  } 
\end{table*}}

\section{Results}\label{sec:results}

\begin{figure}
\begin{subfigure}{8.5cm}
  \centering
  \includegraphics[trim={0 1.2cm 0 0 },clip,width=\hsize]{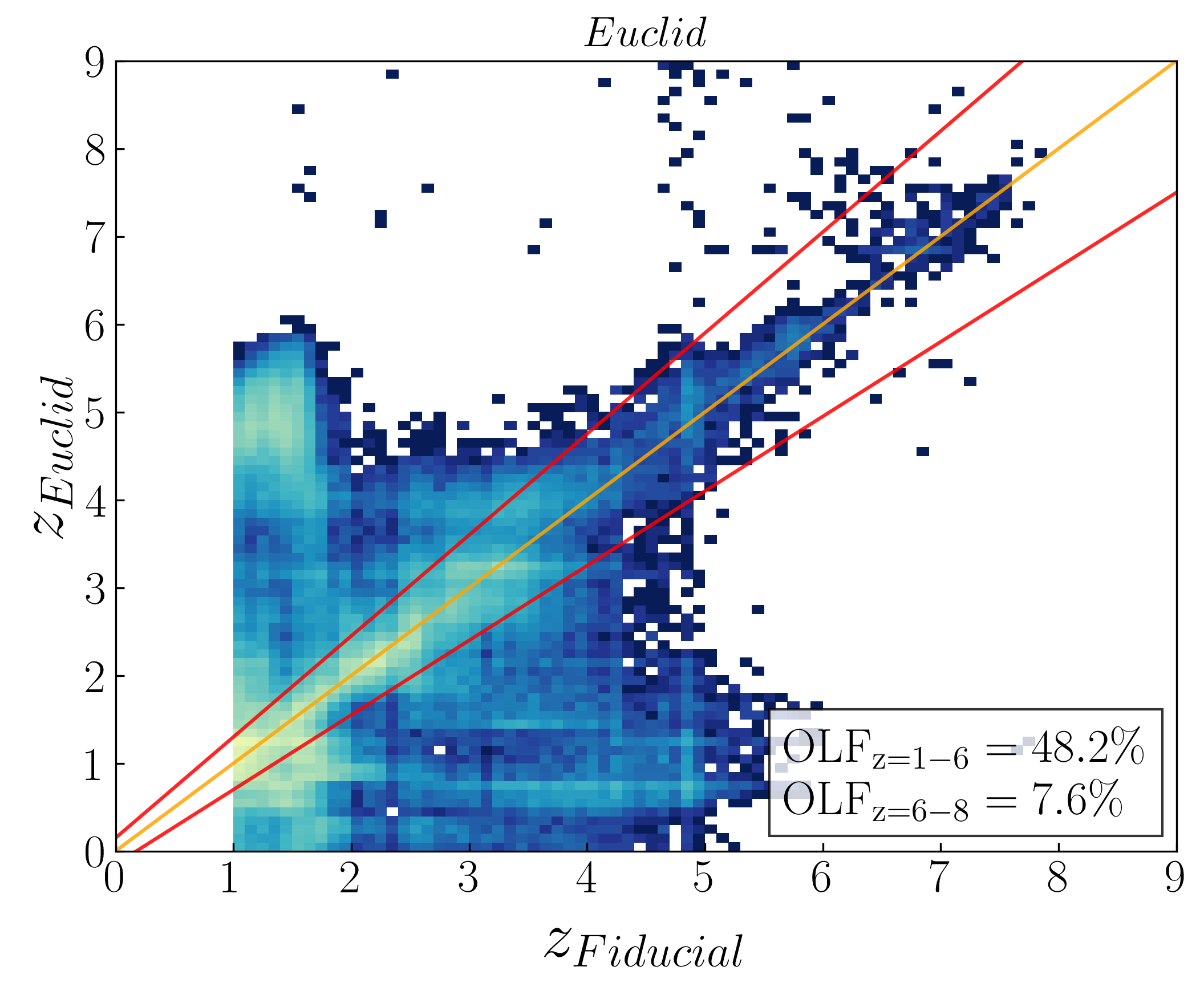}   
\end{subfigure}
\newline
\begin{subfigure}{8.5cm}
  \centering
  \includegraphics[trim={0 1.2cm 0 0 },clip,width=\hsize]{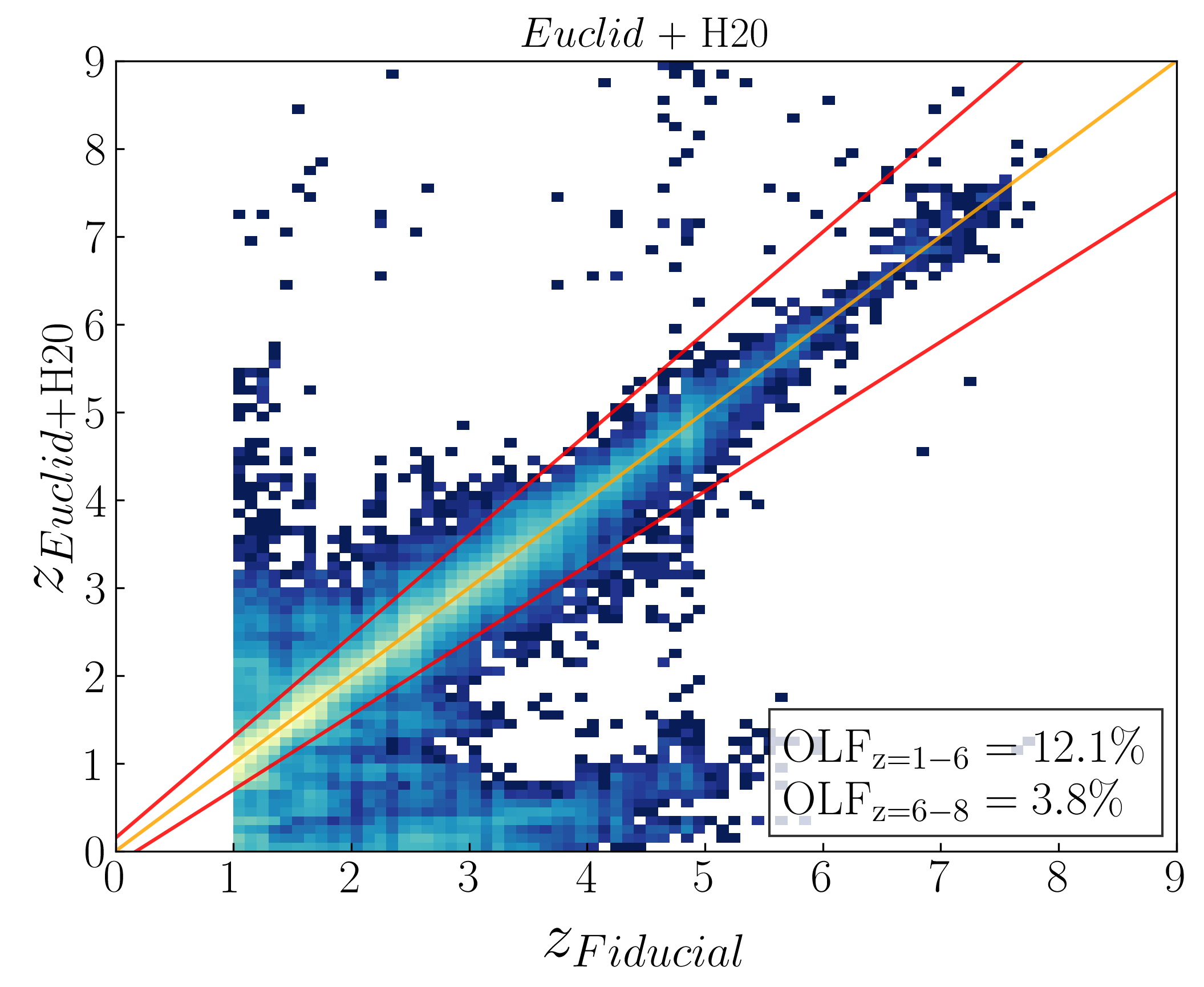} 
\end{subfigure}
\newline
\begin{subfigure}{8.5cm}
  \centering
  \includegraphics[trim={0 0 0 0 },clip,width=1.02\hsize, height=1.02\hsize]{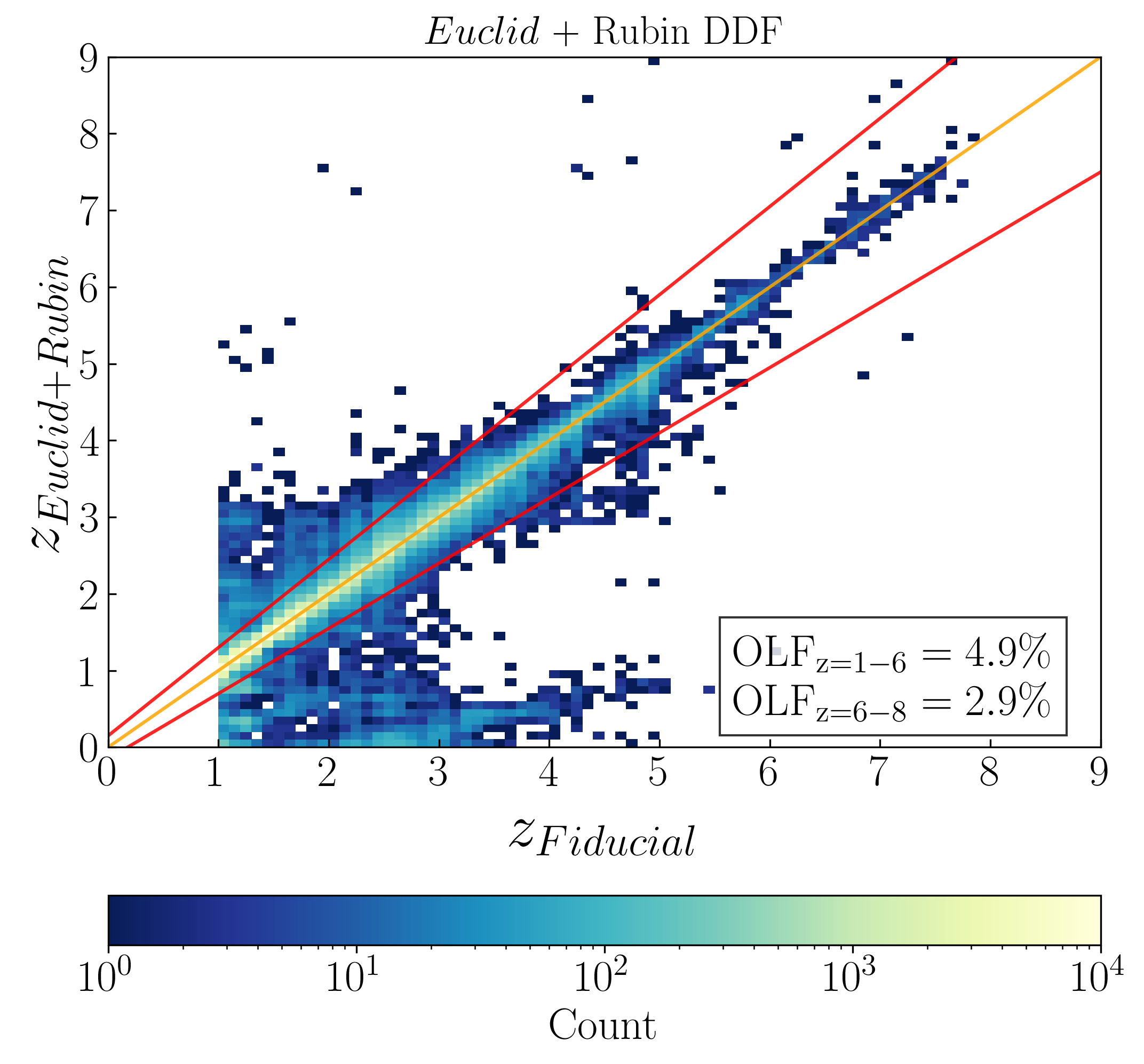} 
\end{subfigure}
\caption{Comparison of the fiducial photometric redshift to the photometric redshift obtained from three combinations of \Euclid\ and ancillary photometry. In each panel, the catastrophic outlier fractions (OLFs) are reported for two fiducial redshift bins, $z=1$--6 and $z=6$--8. The OLF represents the fraction of sources with $|z_{\mathrm{sim}} - z_{\mathrm{fid}}| / (1+z_{\mathrm{fid}}) > 0.15$. The outlier identification boundaries are indicated with solid red  lines.  The data points are presented as 2D histograms with bin size $\Delta z=0.1$. The colour bar corresponds to the number of galaxies in each bin and is the same for all panels. The solid orange line is the identity curve. Data outside the two solid red lines are identified as catastrophic outliers.} 
\label{fig:zphot}
\end{figure}

\subsection{Photometric redshifts based on \Euclid\ and ancillary data} \label{zphot_compare}

We repeated the SED fitting of the sources with fiducial $z>1$ redshifts, using \texttt{LePhare} again, but now considering only the simulated \Euclid\ (+ancillary) data. We used exactly the same \texttt{LePhare} settings as for the UltraVISTA DR4 catalogue, described in Sect.\,\ref{sec:sed} (the flat absolute magnitude prior is now applied to the \HEuc\ band). Despite the low number of photometric bands, \texttt{LePhare} finds a redshift solution for $>99$\,\% of the sources from \Euclid\ photometry alone. We did not repeat our checks for stellar solutions or compatibility with short wavelengths, but the latter is discussed in Sect.\,\ref{sec:discussvalidity}. 

We aim to illustrate how the incorporation of ancillary data improves the performance of the photometric redshift recovery. Therefore, we compare the derived redshifts (to which we refer as simulated redshifts) with the fiducial redshifts obtained from our UltraVISTA and remaining COSMOS photometry (28 bands in total) in three scenarios: \Euclid, \textit{Euclid}+H20, and \textit{Euclid}+Rubin DDF. The results are shown in Fig.\,\ref{fig:zphot}. In each panel, we identify catastrophic outliers as sources satisfying the condition 
\begin{equation}\label{eq:outlier}
    \frac{|z_{\mathrm{sim}} - z_{\mathrm{fid}}|}{1+z_{\mathrm{fid}}} > 0.15, 
\end{equation}
where $z_{\mathrm{sim}}$ is the simulated redshift and $z_{\mathrm{fid}}$ the fiducial redshift. We calculated the catastrophic outlier fraction (OLF) in two redshift bins separately, namely $z=1$--6 (intermediate-$z$) and $z=6$--8 (high-$z$). We note that the OLF only quantifies the quality of the photometric redshifts in these fiducial redshift bins; contamination of the intermediate-$z$ bin to the high-$z$ bin is addressed in Sect.\,\ref{sec:contaminants}.

From Fig.\,\ref{fig:zphot} it is evident that the addition of ancillary data improves the redshift estimation for both intermediate- and high-$z$ galaxies. First, we discuss the photometric redshift quality when only the four \Euclid\ bands are utilised. At $z_{\mathrm{fid}}>6$, the redshifts are already quite accurate, with an OLF of 7.6\,\%. This is the result of the wavelength range covered by the \Euclid\ bands, as for galaxies at $z=6$--8, they sample the rest-frame UV and optical continuum, enabling the identification of $z>6$ galaxies with the Lyman-break drop-out technique \citep{steidel1996}. On the contrary, the redshift recovery at intermediate redshifts is considerably poorer. The majority ($>65$\,\%) of the intermediate-$z$ sample consists of $z_{\mathrm{fid}}=1$--2 galaxies (see Fig.\,\ref{fig:highzsample}), and so the \Euclid\ filters sample the rest-frame optical continuum redwards of the Lyman-$\alpha$ line. With no constraint on the Lyman break, galaxies with red UV slopes (either from dust attenuation or old age) are easily confused for higher-redshift objects. Simultaneously, young non-dusty sources at $z_{\mathrm{fid}}=1$--2 that have a mostly flat UV and optical continuum become highly degenerate with $z\sim5$ galaxies, as a flat SED without a strong $4000\,\AA$ break can be confused for a UV-bright high-redshift object. Interestingly, only the latter type of degeneracy is clearly present in Fig.\,\ref{fig:zphot}; $z_{\mathrm{fid}}=1$--2 galaxies are predominately scattered between $z_{\mathrm{sim}}=4$ and $z_{\mathrm{sim}}=6$, with only a few sources at $z_{\mathrm{sim}}>7$. 
   
We find that the inclusion of deep optical data, either from Rubin observations or the H20 survey, reduces the number of catastrophic outliers, especially at intermediate redshifts. At $z=6$--8, this can be explained since the short-wavelength bands sample the spectrum bluewards and redwards of the Lyman-$\alpha$ break, such that the photometric redshift estimation becomes more precise. In addition, the inclusion of optical data rules out a low-redshift nature for nearly all $z_{\mathrm{fid}}>6$ galaxies. At intermediate redshifts, we see how the inclusion of optical data strongly improves the OLF. Moreover, with data from the Rubin DDF, which constitutes the deepest optical ancillary data considered in this work, the degeneracy between $z_{\mathrm{fid}}=4$--6 and $z_{\mathrm{sim}}>6$ galaxies is almost completely lifted in our analysis. 


{\renewcommand{\arraystretch}{1.5}
\begin{table*}
\caption{Contaminants amongst $z>6$ galaxies.}             
\label{table:fraction_contaminants}      
\centering                         
\begin{tabular}{l c c c c }        
\hline\hline                 
Filters & True $z>6$ & Contaminants & Contamination Fraction & Completeness\\    
\hline                        
   \Euclid\ & 287 & 65 & $0.18^{+0.07}_{-0.06}$ & 91\,\%\\  
   \hline
   \Euclid\ + Rubin & 291 & 71 & $0.20^{+0.07}_{-0.09}$  & 92\,\%\\
   \Euclid\ + Rubin DDF  & 301 & 13 & $0.04^{+0.03}_{-0.02}$ & 96\,\%\\
   \Euclid\ + H20 & 286 & 65 & $0.19^{+0.07}_{-0.09}$ & 91\,\%\\
   \hline
   \Euclid\ + \textit{Spitzer} & 188 & 27 & $0.13^{+0.04}_{-0.05}$ & 93\,\%\\
   \hline
   \Euclid\ + Rubin + \textit{Spitzer} & 193 & 26 & $0.12^{+0.04}_{-0.04}$ & 95\,\%\\
   \Euclid\ + Rubin DDF + \textit{Spitzer}  & 191 & 2 & $0.01^{+0.0006}_{-0.0001}$ & 94\,\%\\
   \Euclid\ + H20 + \textit{Spitzer} & 186 & 23 & $0.11^{+0.04}_{-0.04}$ & 92\,\%\\
\hline                                   
\end{tabular}
\tablefoot{Number of true $z>6$ galaxies (that is, galaxies at fiducial $z_{\mathrm{fid}}=6$--8 that are recovered at $z_{\mathrm{sim}}>6$) and $z>6$ contaminants (galaxies at fiducial $z_{\mathrm{fid}}=1$--5.8 recovered at $z_{\mathrm{sim}}>6$), from various combinations of \Euclid\ and ancillary data. In addition, we report the fraction of contaminants amongst the total apparent $z>6$ population, for which uncertainties from ten random realisations of the contaminant and true $z>6$ source photometry. We also report the completeness, which represents the percentage of fiducial $z=6$--8 galaxies that are correctly identified as $z>6$ sources.} 
\end{table*}}
\subsection{Identification of $z>6$ contaminants} \label{sec:contaminants}

For each \Euclid\ (+ancillary) data scenario, we identify three populations from the redshift comparison plots shown in Fig.\,\ref{fig:zphot}: (i)  galaxies with fiducial redshifts $z_{\mathrm{fid}}=1$--6  that stay in that same redshift bin when the photometric redshift is obtained with \Euclid\ (+ancillary) photometry, to which we refer as the `stable'  intermediate-$z$ galaxy population; (ii)  galaxies with fiducial redshifts $z_{\mathrm{fid}}>6$  that stay at these high redshifts when the photometric redshift is obtained with \Euclid\ (+ancillary) photometry, which are the `true' $z>6$ galaxies; and (iii)  galaxies with fiducial redshifts $z_{\mathrm{fid}}=1$--5.8 and \Euclid\ (+ancillary) redshifts $z>6$, which we consider to be the intermediate-redshift contaminants to the high-$z$ galaxy sample. The latter population constitutes the main subject of study in this paper. We set an upper redshift cut at a fiducial redshift $z_{\mathrm{fid}}=5.8$ for the purpose of defining contaminants to avoid discussing sources that may end up populating the $z>6$ regime simply due to a random error scattering of the photometric redshifts. Therefore, galaxies with $z_{\mathrm{fid}}=5.8$--6 that are falsely recovered at $z>6$ are discarded from our analysis, as they would constitute only 6\,\% the true $z>6$ population (\Euclid\ alone) and the majority are recovered at  $z_{\mathrm{sim}}=6$--6.2. Lastly, we acknowledge a fourth population consisting of galaxies with fiducial redshifts $z_{\mathrm{fid}}>6$ that appear as intermediate-$z$ galaxies when observed with \Euclid. These sources constitute 9\,\% of the fiducial $z=6$--8 galaxy sample (\Euclid\ alone). The study of this population is outside the scope of this paper. 
 
For each data scenario, in Table \ref{table:fraction_contaminants} we present the number of true $z>6$ galaxies, the number of $z>6$ contaminants and the following fraction of contaminants amongst the apparent $z>6$ galaxy population. By apparent $z>6$ galaxies we mean all the galaxies assigned a photometric redshift $z>6$ based on the \Euclid\ (+ancillary) simulated photometry, independently of being truly at these redshifts or not. We also report the completeness in Table \ref{table:fraction_contaminants}, which represents the percentage of fiducial $z_{\mathrm{fid}}=6$--8 galaxies that are correctly identified as $z_{\mathrm{sim}}>6$ sources. The missing galaxies in our reported completeness are those with fiducial $z_{\mathrm{fid}}>6$ redshifts, but which are falsely recovered at $z_{\mathrm{sim}}<6$ with the \Euclid\ (+ancillary) photometry. We remind the reader that not all galaxies in our fiducial $z=1$--8 galaxy sample are IRAC-detected, explaining the lower numbers of $z>6$ contaminants and true $z>6$ galaxies in scenarios where \textit{Spitzer} data are considered.

We calculated the uncertainties in the contamination fraction by producing ten randomised flux catalogues of the $z>6$ contaminants and the true $z>6$ galaxies, for which we derived the photometric redshifts with \texttt{LePhare}. Subsequently, we assigned a probability of correct identification to each source by counting in how many realisations the source is re-identified as a contaminant, and identically for the true $z>6$ sample. Using the average probability of correct identification, we calculated upper and lower limits on both the number of contaminants and the number of true $z>6$ galaxies, which were propagated into an upper and lower limit on the contamination fraction. We adopt this approach as a compromise because producing ten realisations of the entire $z=1$--8 flux catalogue is too computationally expensive. 

Our main findings on the contamination fraction are as follows. First, the fraction of contaminants amongst the apparent $z>6$ population is already relatively low when only data from the four \Euclid\ bands are available: 18\,\% of all apparent $z>6$ galaxies are actually intermediate-$z$ contaminants. In addition, the $z>6$ completeness is very high in all data scenarios, even with \Euclid\ photometry alone. 

Second, for sources at the UltraVISTA depth, the inclusion of ancillary optical data produces a negligible effect in the fraction of contaminants and the $z>6$ completeness level. This is because the Rubin and H20 surveys are both shallower in the optical regime than the Euclid Deep Survey (see Table \ref{table:euclid_rubin_bands}), and as such their data are of little help in preventing intermediate-$z$ galaxies from being misidentified as $z>6$ galaxies. In fact, the contamination fractions from \textit{Euclid}+Rubin and \textit{Euclid}+H20 are slightly higher than that from \Euclid\ photometry alone, although the difference is not significant within the uncertainties. We expect the H20 data to perform better in the redshift recovery compared to Rubin, even though the latter includes the additional $y$ band coverage. This can be explained as the H20 data are slightly deeper, especially in the $z$ band.

\hspace{0.4cm} Only with the ultra-deep photometry from the Rubin DDF does the contamination fraction improve, as the Rubin DDF $r$ and $i$ data will be 0.3 and 0.1 mag deeper than the \IEuc\ photometry. Clearly, the Rubin DDF data provide such stringent constraints on the photometric depth that even the faintest intermediate-$z$ galaxies in our sample cannot be confused for high-redshift sources. However, we note that this is the most idealised scenario we consider in this paper, and only with these data can the contamination fraction be taken to very low levels. As a safety measure, we tested the scenario where we combine simulated Rubin DDF and \textit{Spitzer} data, without including \Euclid\ photometry. In this case, we find a contamination fraction of 0.08 and a completeness of 92\,\%. This demonstrates that whereas the Rubin DDF photometry is incredibly effective at reducing the degree of contamination, \Euclid\ photometry is essential to achieve virtually no contamination.

Third, \textit{Spitzer} data are moderately helpful in preventing the incidence of intermediate-redshift contaminants to the $z>6$ sample. The majority of contaminants are at $z_{\mathrm{fid}}=4$--6, and produce redder fiducial $(H-[3.6])$ contaminants than actual $z>6$ galaxies. Hence, IRAC detections enable one to distinguish between a red SED slope from intermediate-$z$ interlopers and a flat SED slope that one would observe for young galaxies at high redshifts. However, using \textit{Euclid}+\textit{Spitzer} data, the contamination fraction is still 0.13, so the improvement is marginal compared to the \textit{Euclid}+Rubin DDF scenario. We believe this is mostly due to the typical uncertainty of IRAC fluxes, given that \textit{Spitzer} sources are severely blended in crowded fields such as COSMOS. In addition, we investigate the signal-to-noise ratio (S/N) in the simulated IRAC bands and conclude that $67$\,\% of the contaminants identified from \textit{Euclid}+\textit{Spitzer} data have a $F_\nu/\sigma_{F_\nu} \geq 5$ detection in both IRAC bands. 

\hspace{0.4cm} It is possible that the particular de-blending treatment used to obtain IRAC photometry (see \citealt{deshmukh18} for a detailed overview) leads to a slight underestimation of the flux errors. Given that in this scenario redshifts are based on only six bands, any uncertainties in the IRAC photometry carry more weight in the SED fitting as compared to the fit on the fiducial, 28-band photometry. We note that the limited effectiveness of the IRAC photometry is independent of our specific method for measuring the IRAC fluxes; when we estimate the contamination fractions using the COSMOS2020 catalogue as presented in Appendix \ref{appendix}, for which the IRAC photometry was derived in a completely independent manner, we find it has essentially no effect on the contamination fraction. 

Finally, combining \Euclid\ photometry with both optical and infrared data yields the best contamination fractions; evidently, with more photometric bands available for the SED fitting, the redshift recovery steadily improves. In the \textit{Euclid}+Rubin DDF+\textit{Spitzer} scenario, the deep, 11-band photometry is highly successful at correctly identifying $z>6$ galaxies, and so contamination from intermediate-$z$ interlopers is virtually non-existent and the $z>6$ completeness is very high at 94\,\%. 

Apart from the eight combinations of \Euclid\ and ancillary bands considered throughout this work, we evaluate a few other scenarios to gain more insight into preventing intermediate-$z$ interlopers.

First, we tested the importance of the Rubin $y$ band for the selection of high-$z$ galaxies, given that \Euclid\ itself will create very deep imaging in the \YEuc\ band (see Fig.\,\ref{fig:throughput} for the respective filter transmission curves). Presumably, $y$-band observations are of key importance to the $z>7$ galaxy selection, since at $z=7$--8 the Lyman-$\alpha$ break at $1216\,\AA$ is sampled by this band. In this paper we have assumed that the Rubin DDF will be 0.1 mag shallower in the $y$ band as compared to the Euclid Deep Fields. We derive the contamination fraction from \textit{Euclid}+Rubin DDF photometry whilst excluding the $y$ band, and find that it is 0.07 as compared to 0.04 whilst including the $y$ band (see Table \ref{table:fraction_contaminants}). Simultaneously, we find that a 0.5 mag increase in the Rubin DDF $y$-band depth does not improve the contamination fraction any further. Therefore, we conclude that even though the central wavelengths and filter widths of the Rubin $y$ and \Euclid\ \YEuc\ band differ, ultra-deep Rubin $y$-band photometry is only marginally effective when \YEuc-band imaging is readily available. 

Second, we tested how robust our results on the contamination fraction are when we vary the full final depth of the ancillary data. The magnitude limits adopted throughout this paper are, with the exception of the \textit{Spitzer} data, not definite as the observations have not commenced or are not completed yet. Therefore, an assessment of how the degree of contamination is dependent on the final survey depths is important. We determine that if the H20 survey was 0.5 mag deeper across all five bands, the fraction of contaminants amongst the apparent $z>6$ sample would decrease from 0.19 to 0.10. Simultaneously, by making the Rubin DDF 0.5 mag shallower across all six bands, the contamination fraction worsens from 0.04 to 0.08. Clearly, the final depth of the optical data has strong implications for the contamination fraction. 

Finally, we {\color{black} investigated} which optical band contributes the most to accurate $z>6$ galaxy selection, using \textit{Euclid}+H20 data. We {\color{black} increased} the $5\sigma$ depth by 0.5 mag for each H20 band individually whilst the photometry in the other bands remains unchanged, creating five different flux catalogues. Subsequently, we derive the contamination fraction for all five realisations, and find the Subaru HSC $i$ band is most important for excluding intermediate-$z$ interlopers, reducing the contamination fraction to 0.14 (as compared to 0.19 from \textit{Euclid}+H20 with no depth variations). This is unsurprising as the majority of our fiducial $z>6$ sample is at $z=6$--7 and so the $i$ band provides a strong constraint on the Lyman-$\alpha$ break. The second most important band is the Subaru HSC $z$ band (contamination fraction of 0.15). Conversely, we find that increasing the survey depth in the CFHT $u$ band leaves the contamination fraction unimproved. We emphasise that these results on the importance of individual bands concern the contamination of $z>6$ sources. In fact, the CFHT $u$ band is most important for normal galaxies at lower-redshifts, that is, the OLF of stable intermediate-$z$ galaxies as defined in Fig.\,\ref{fig:zphot} moderately improves to 11.2\,\% from increasing the $u$ band depth. 

Depending on the research purpose, certain detection threshold requirements may be imposed on potential \Euclid\ high-redshift galaxies. Therefore, we explore how a $5\sigma$ detection threshold requirement in at least one of the NIR bands for \Euclid\ high-redshift galaxies may result in lower degree of contamination by intermediate-$z$ sources. Considering only \Euclid\ data, the contamination fraction is reduced to $0.12^{+0.04}_{-0.04}$ with this measure. Generally, we find a moderate improvement in the contamination fraction in each \Euclid\ (+ancillary) data scenario, but no significant differences within the error bars. We further explore the usefulness of S/N cuts in Sect.\,\ref{sec:colourdiagrams}. 


\subsection{Separation of contaminants from true $z>6$ galaxies} \label{sec:colourdiagrams}

Having quantified the incidence of intermediate-redshift contaminants in the $z>6$ sample, now we aim to develop a method to cleanly separate the contaminants from the true $z>6$ galaxies, based on the photometry available in the Euclid Deep Fields. To achieve this, we investigate which photometric and SED properties can separate the two populations. Specifically, we investigate the usefulness of colour diagrams. For instance, \citet{bisigello2020} already showed how \textit{Euclid} colour-colour selection techniques can effectively separate star-forming and quiescent galaxies at $z=0$--3.

\begin{figure}
 \centering
 \includegraphics[width=\hsize]{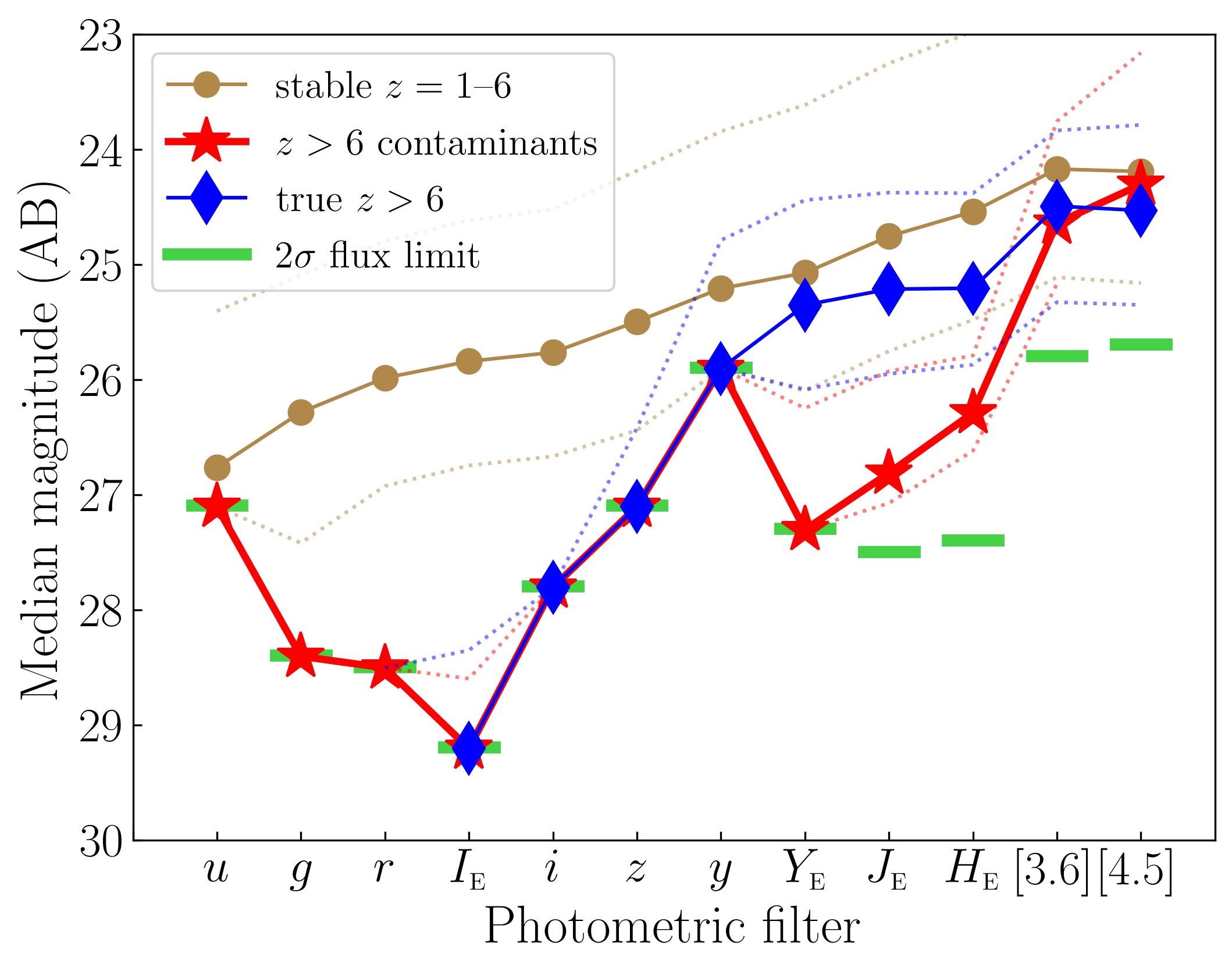}
 \caption{Median magnitudes of our simulated sources in the \Euclid, Rubin, and \textit{Spitzer} filters. Light brown circles represent stable intermediate-$z$ galaxies, red stars $z>6$ contaminants, and blue diamonds true $z>6$ galaxies. For each population, the dotted lines indicate the 16\textsuperscript{th} and 84\textsuperscript{th} percentiles. For each band, the $2\sigma$ flux limit is shown with a green bar. The selection of intermediate-$z$, contaminants, and $z>6$ galaxies is based on \textit{Euclid}+Rubin+\textit{Spitzer} data. The median magnitudes for $z>6$ galaxies in the Rubin $u, g, r$ bands are equal to $-99$ (no intrinsic flux) and therefore omitted from this figure.}
 \label{fig:median_magnitude}
\end{figure}

\begin{figure*}
\begin{subfigure}{17cm}
\centering 
{\renewcommand{\arraystretch}{1.5}
\setlength{\tabcolsep}{2.7pt} 
\hfill
\begin{tabularx}{15.9cm}{|R|R|}
\hline\hline                 
Colour cut & Condition \\
\hline\hline
\cellcolor{darkblue!75} {\color{white}$\boldsymbol{(\mIEuc-\mYEuc)>2.0\ \&\ (\mYEuc-\mJEuc)<1.4}$} & all true $z>6$ galaxies with a \IEuc\ and/or \YEuc\ detection included \\ 
\cellcolor{cyan!75}$\boldsymbol{(\mIEuc-\mYEuc)>2.6\ \&\ (\mYEuc-\mJEuc)<1.4}$ & 95\,\% of true $z>6$ galaxies with a \IEuc\ and/or \YEuc\ detection included \\
\cellcolor{yellow!75}$\boldsymbol{(\mIEuc-\mYEuc)>2.8\ \&\ (\mYEuc-\mJEuc)<1.4}$ & 90\,\% of true $z>6$ galaxies with a \IEuc\ and/or \YEuc\ detection included \\ 
\cellcolor{darkred!75} {\color{white} $\boldsymbol{(\mIEuc-\mYEuc)>3.4\ \&\ (\mYEuc-\mJEuc)<0.9}$} & all contaminants with a \IEuc\ and/or \YEuc\ detection excluded \\
\hline
\end{tabularx}}
\end{subfigure}
\newline
\begin{subfigure}{8.5cm}
  \centering
  \includegraphics[trim={0cm 1.3cm 0.2cm 0.1cm },clip,width=8.5cm,height=6.0cm]{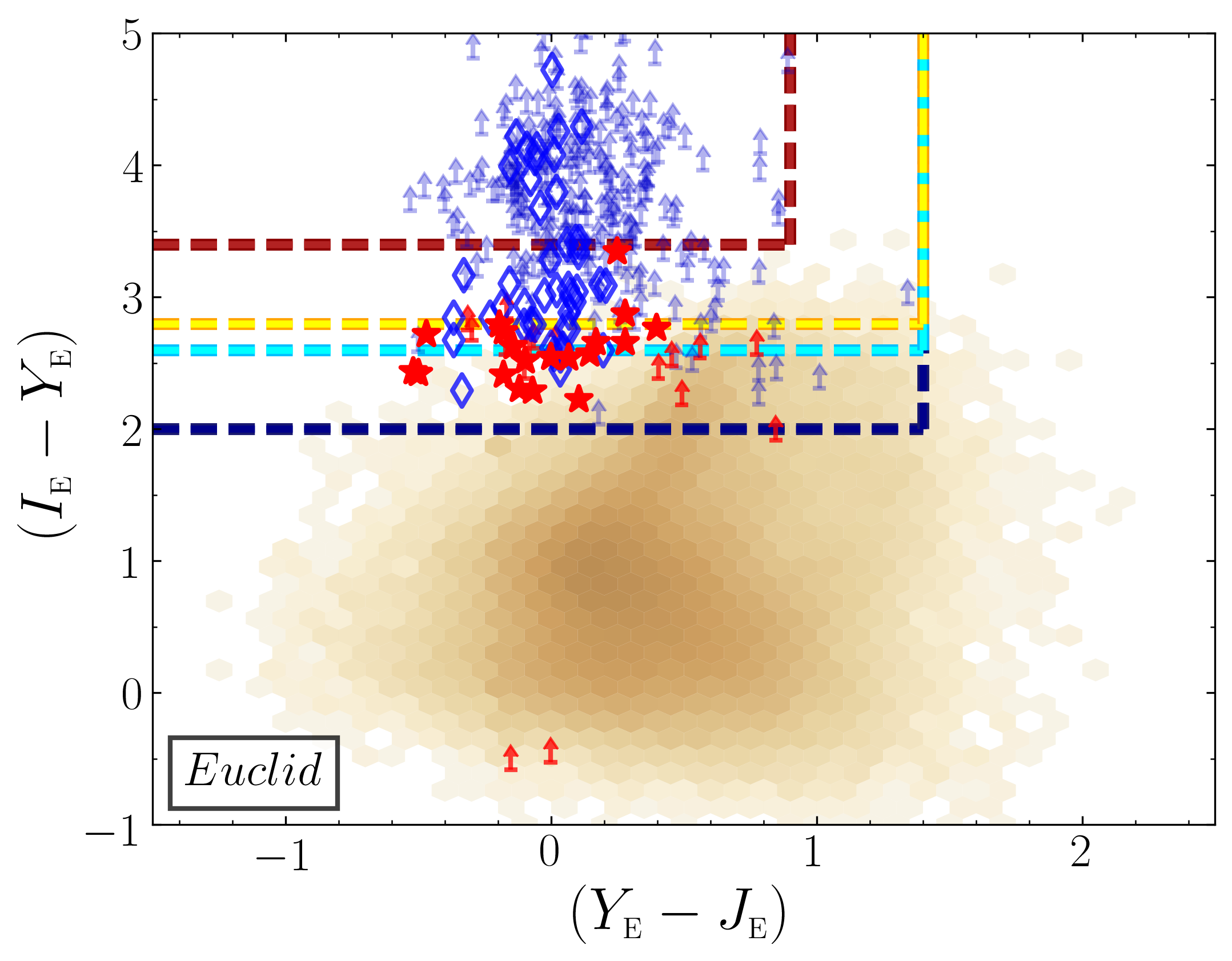}   
\end{subfigure}
\begin{subfigure}{8.5cm}
  \centering
  \includegraphics[trim={1.3cm 1.3cm 0.2cm 0.1cm },clip,width=8.5cm,height=6.0cm]{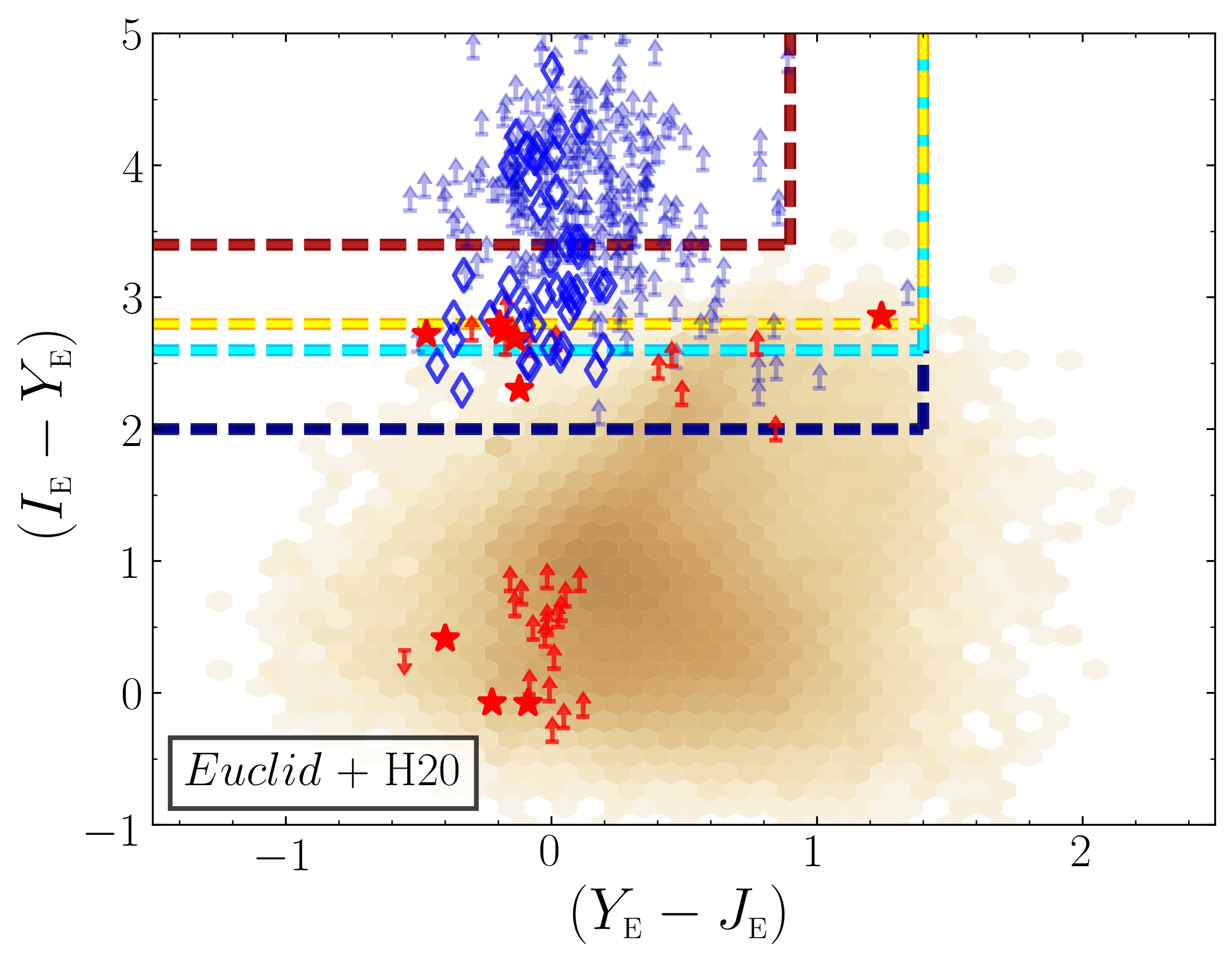}
\end{subfigure}
\newline
\begin{subfigure}{8.5cm}
  \centering
  \includegraphics[trim={0cm 0cm 0.2cm 0.3cm },clip,width=8.5cm,height=6.65cm]{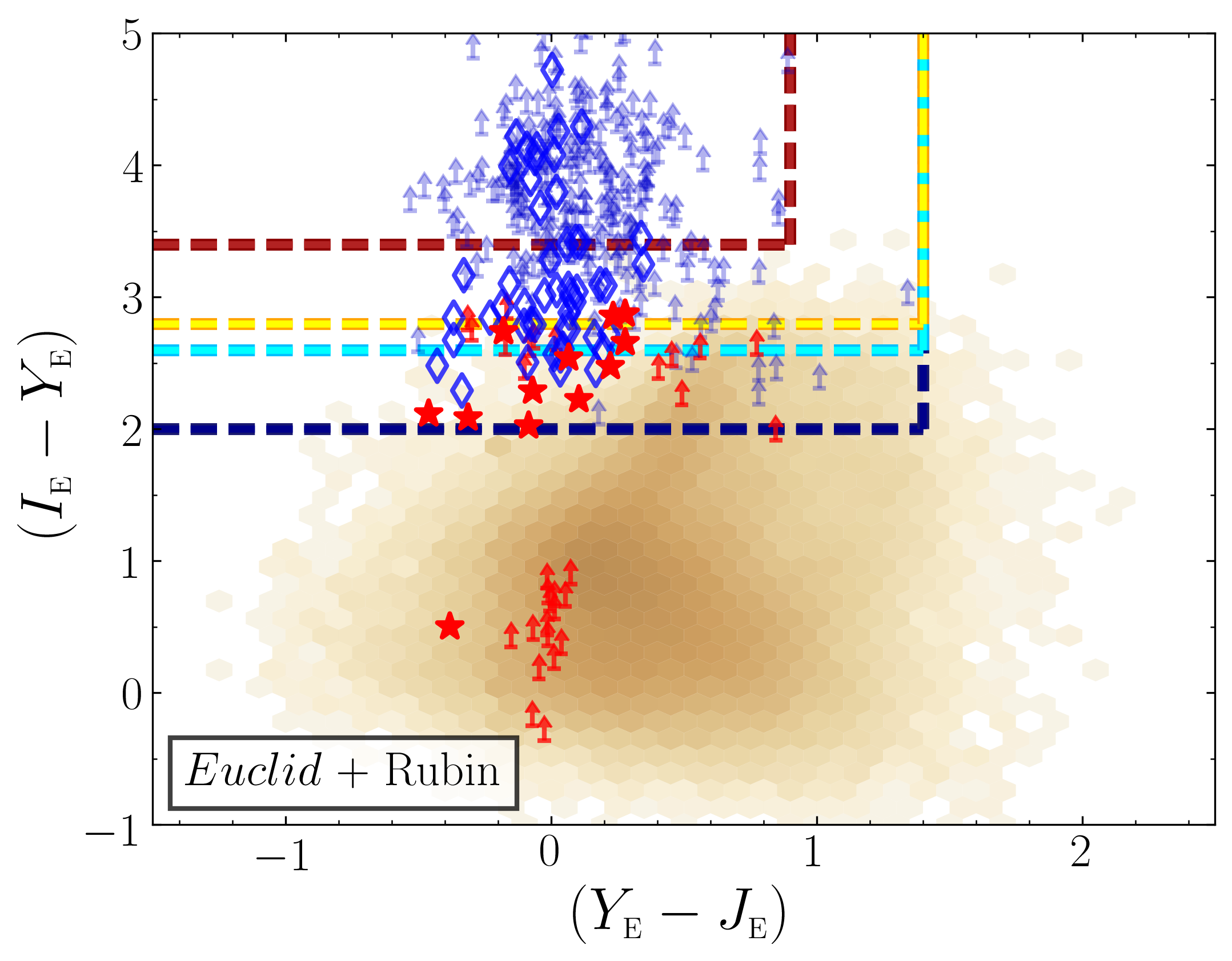}
\end{subfigure}
\begin{subfigure}{8.5cm}
  \centering
  \includegraphics[trim={1.3cm 0cm 0.2cm 0.3cm },clip,width=8.5cm,height=6.65cm]{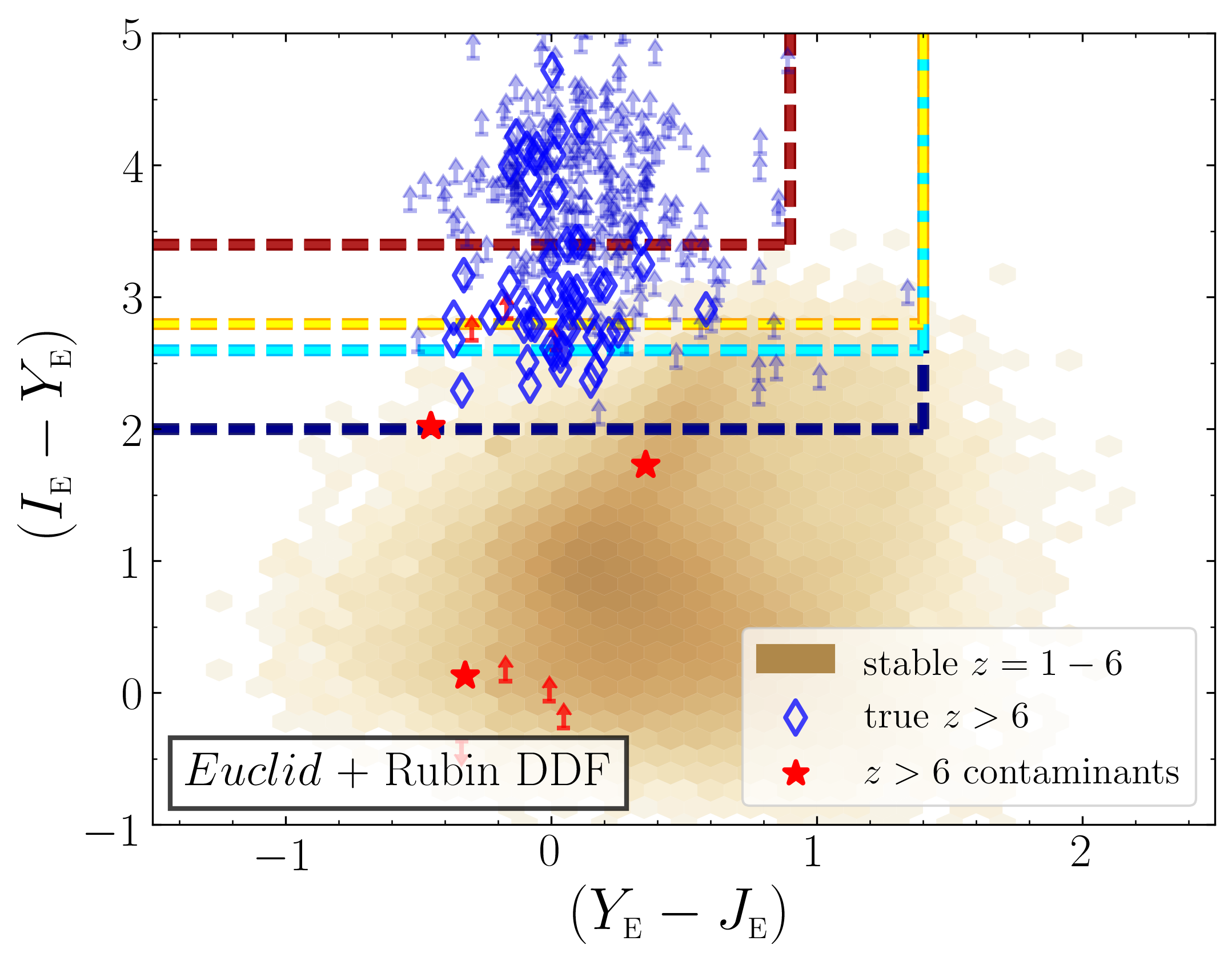}
\end{subfigure}
\caption{$(\mIEuc-\mYEuc)$ versus $(\mYEuc-\mJEuc)$ colour diagrams in eight cases of combinations of \Euclid\ and ancillary data, showing only sources with a flux measurement in the \IEuc\ and/or \YEuc\ band. In the scenario of \Euclid\ data only, this means that 1.4\,\% and 46\,\% of the true $z>6$ sources and contaminants are left out of the colour-colour diagram. The $z>6$ contaminants and true $z>6$ galaxies are shown with red stars and blue diamonds, respectively (lower limits are shown with arrows in corresponding colours). The stable intermediate-$z$ sources are shown in light brown hexagonal bins, where the colour intensity corresponds to the number of galaxies in each bin (darker colours correspond to more sources). The various colour selection criteria and the conditions they meet are listed at the top, and the criteria are indicated in the colour diagrams with dashed coloured lines.} 
\label{fig:colourdiagrams}
\end{figure*}

\begin{figure*} \ContinuedFloat
\begin{subfigure}{8.5cm}
  \centering
  \includegraphics[trim={0cm 1.3cm 0.2cm 0.3cm },clip,width=8.5cm,height=6.0cm]{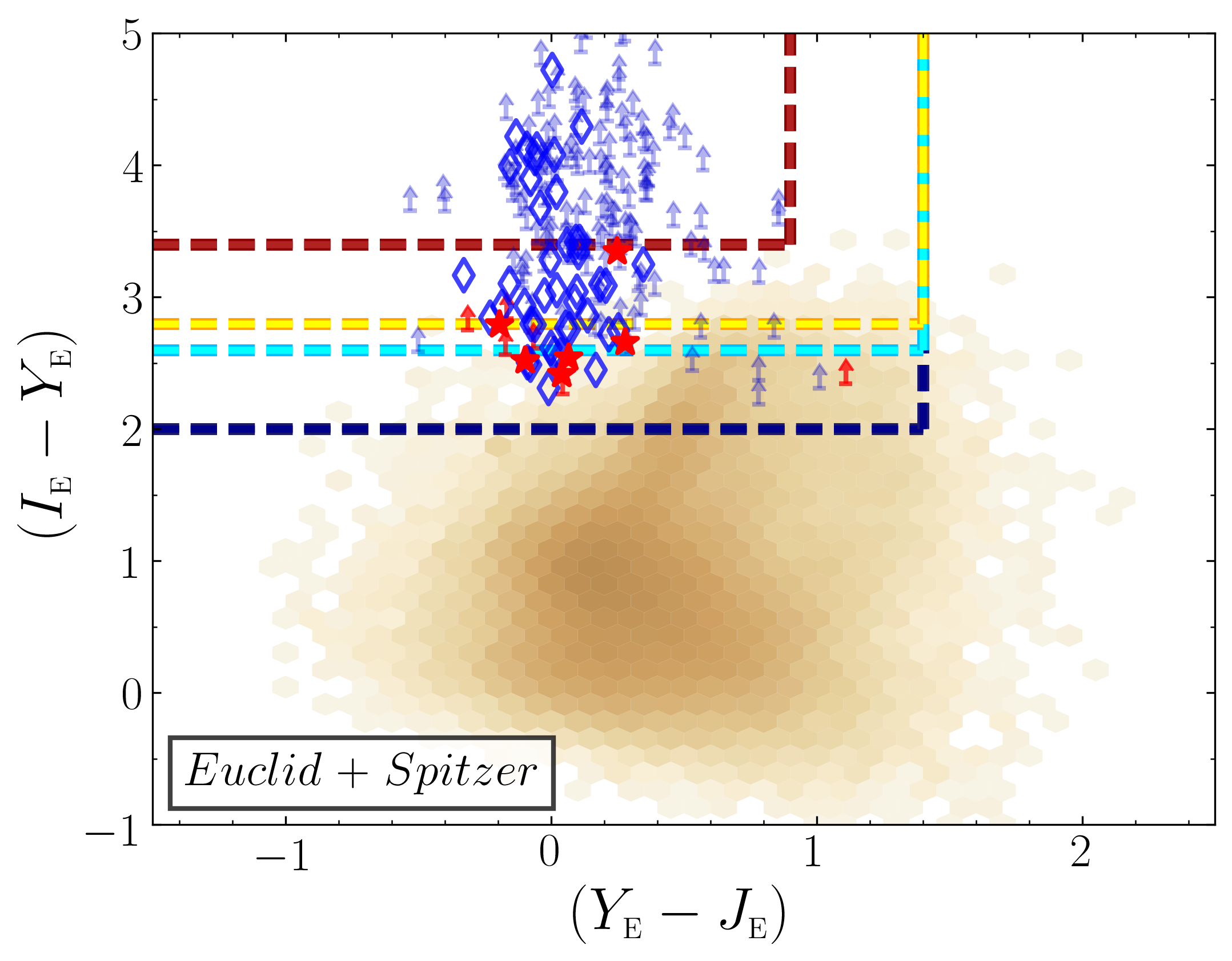}
\end{subfigure}
\begin{subfigure}{8.5cm}
  \centering
  \includegraphics[trim={1.3cm 1.3cm 0.2cm 0.3cm },clip,width=8.5cm,height=6.0cm]{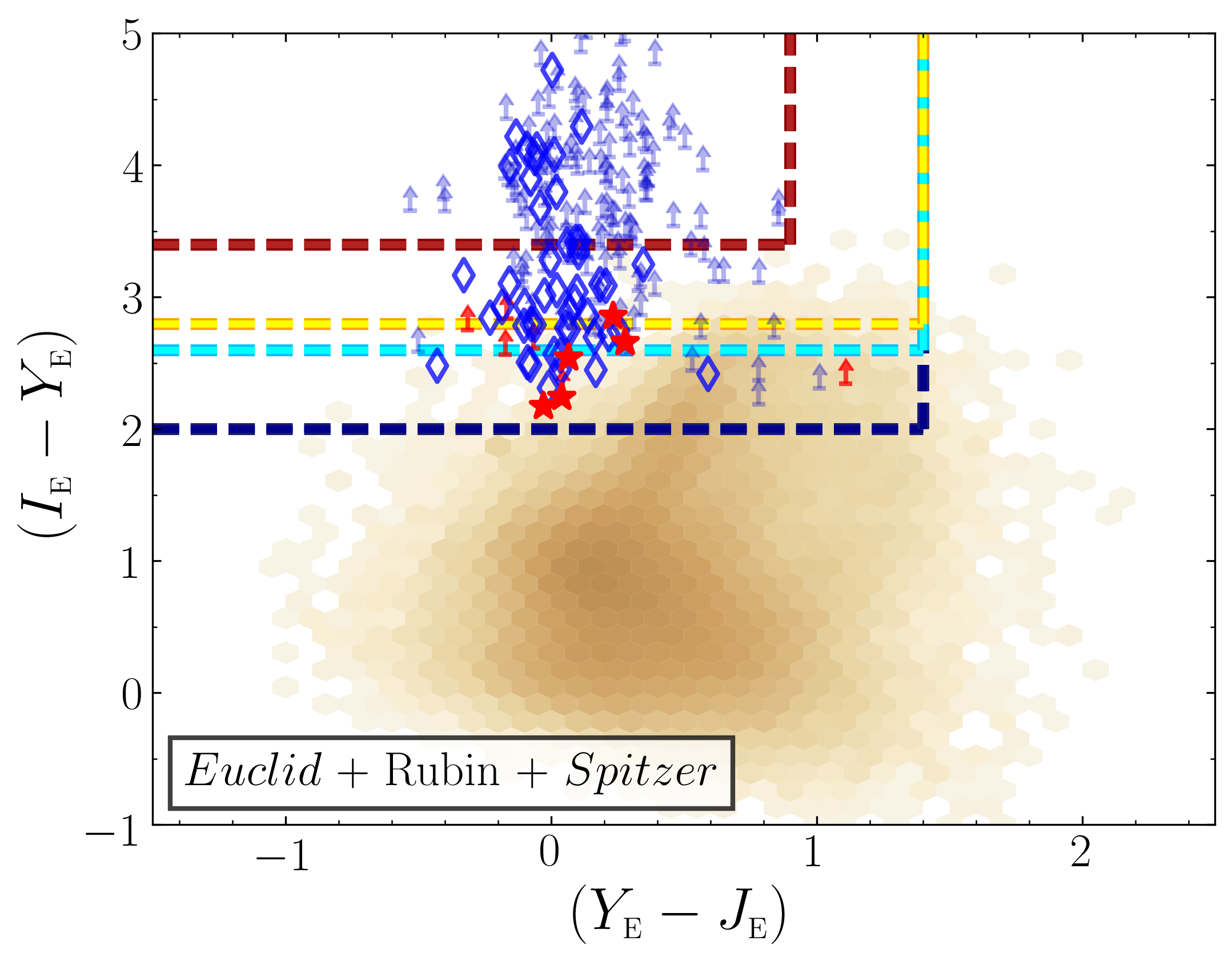}
\end{subfigure}
\newline
\begin{subfigure}{8.5cm}
  \centering
  \includegraphics[trim={0cm 0cm 0.2cm 0.3cm },clip,width=8.5cm,height=6.65cm]{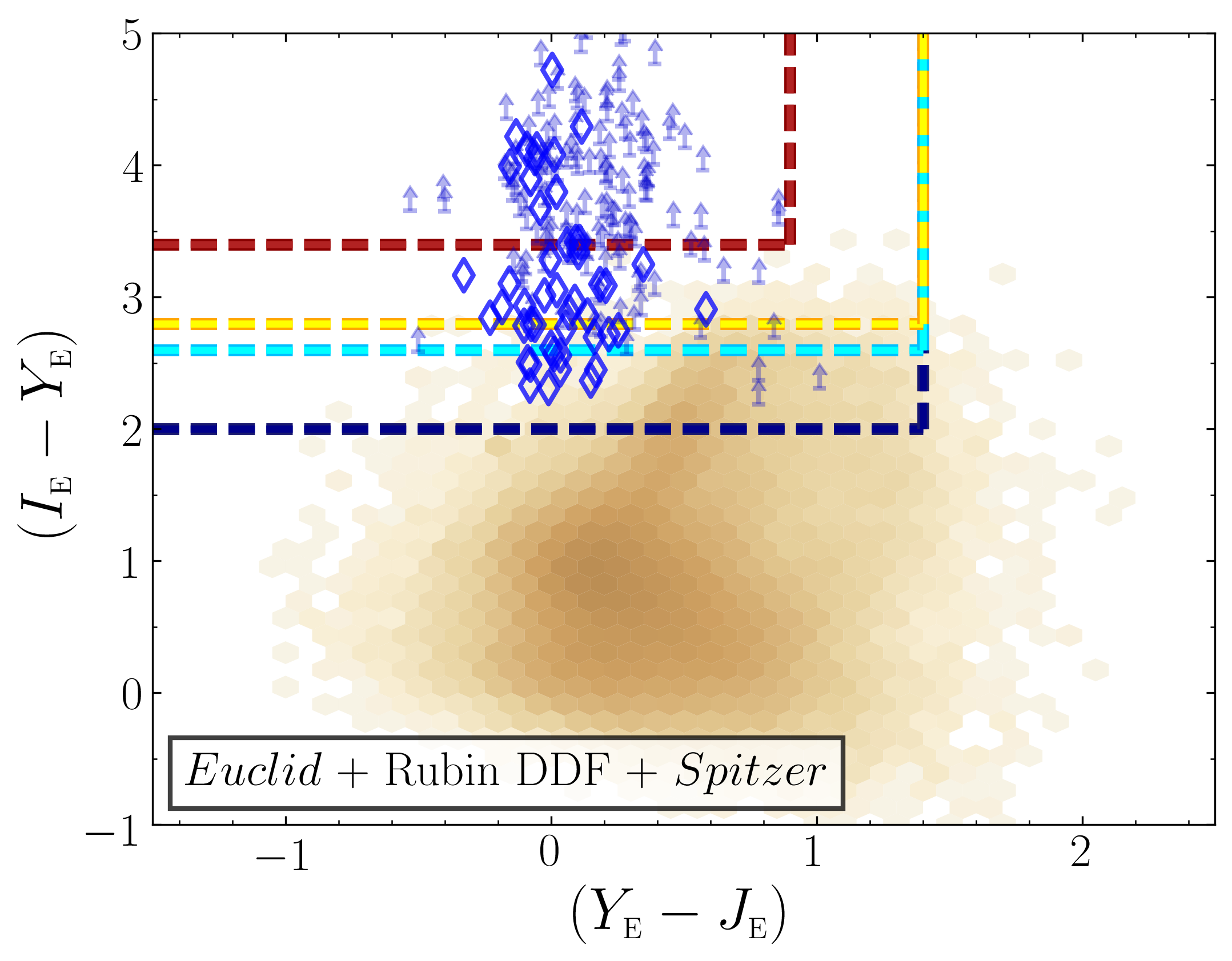}
\end{subfigure}
\begin{subfigure}{8.5cm}
  \centering
  \includegraphics[trim={1.3cm 0cm 0.2cm 0.3cm },clip,width=8.5cm,height=6.65cm]{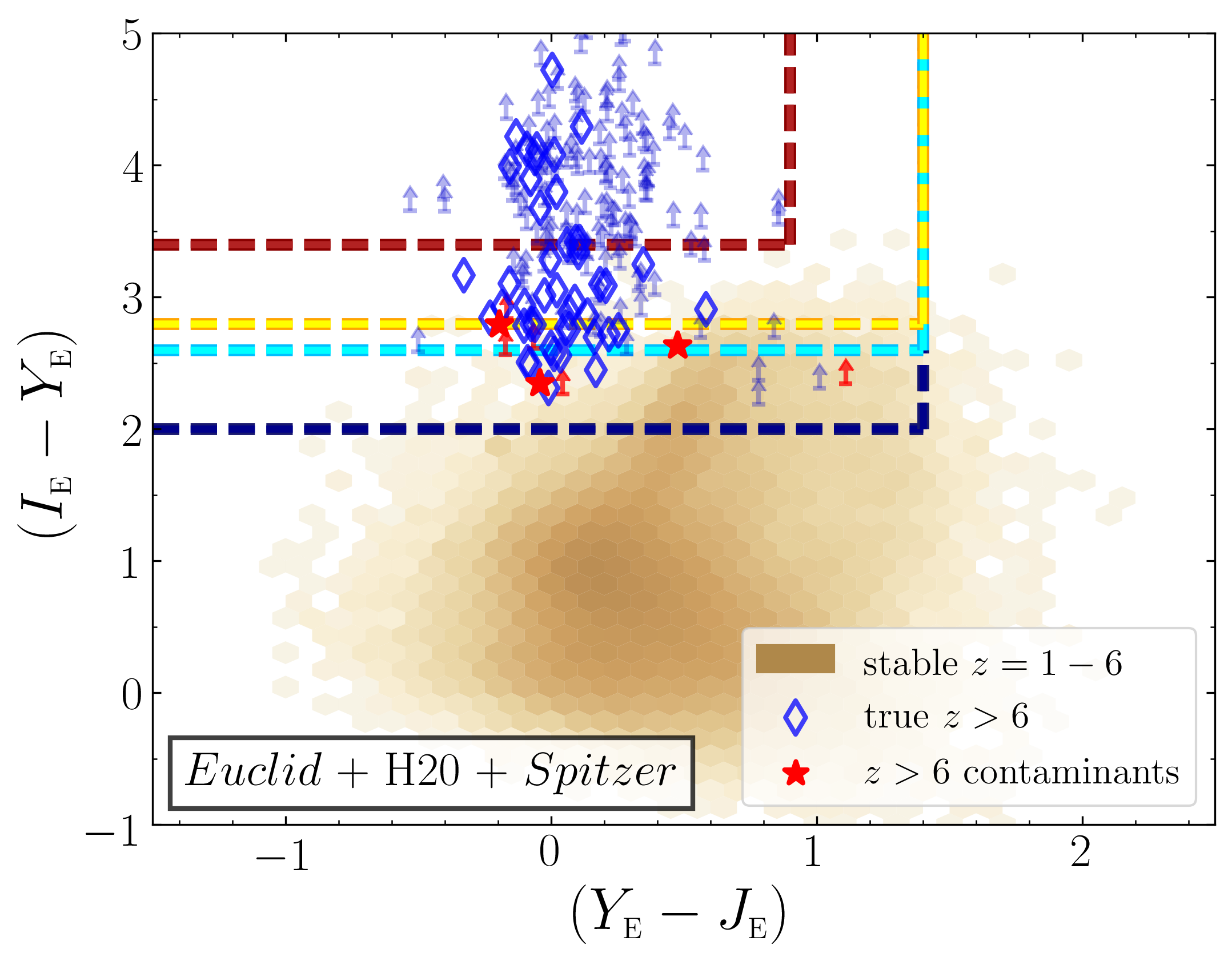}
\end{subfigure}
\caption{Continued.}
\label{fig:colourdiagrams2}
\end{figure*}

{\renewcommand{\arraystretch}{1.5}
\setlength{\tabcolsep}{2.7pt}
\begin{table*}
\caption{Contamination fraction and completeness for various $(\mIEuc-\mYEuc)$ \& $(\mYEuc-\mJEuc)$ colour cuts applied to all sources with an \IEuc\ and/or \YEuc\ flux measurement in the UltraVISTA-like bright sample.}             
\label{table:colourcuts}      
\centering    
\begin{tabular}{l|cc|cc|cc|cc|cc|cc|cc|cc|}        
\hline\hline                 
 Colour cut &\multicolumn{2}{c}{\Euclid}&\multicolumn{2}{c}{\Euclid}&\multicolumn{2}{c}{\Euclid}&\multicolumn{2}{c}{\Euclid}&\multicolumn{2}{c}{\Euclid}&\multicolumn{2}{c}{\Euclid}&\multicolumn{2}{c}{\Euclid}&\multicolumn{2}{c}{\Euclid} \\ [-1ex]
 
$(\mIEuc-\mYEuc)$&\multicolumn{2}{c}{}&\multicolumn{2}{c}{+H20}&\multicolumn{2}{c}{+Rubin}& \multicolumn{2}{c}{+Rubin}&\multicolumn{2}{c}{+\textit{Spitzer}}&\multicolumn{2}{c}{+Rubin}&\multicolumn{2}{c}{+Rubin}&\multicolumn{2}{c}{+H20}\\ [-1ex]

\& $(\mYEuc-\mJEuc)$ &\multicolumn{2}{c}{}&\multicolumn{2}{c}{}&\multicolumn{2}{c}{}&\multicolumn{2}{c}{DDF}&\multicolumn{2}{c}{}&\multicolumn{2}{c}{+\textit{Spitzer}}&\multicolumn{2}{c}{DDF}&\multicolumn{2}{c}{+\textit{Spitzer}} \\[-1ex]

&\multicolumn{2}{c}{}&\multicolumn{2}{c}{}&\multicolumn{2}{c}{}&\multicolumn{2}{c}{}&\multicolumn{2}{c}{}&\multicolumn{2}{c}{}&\multicolumn{2}{c}{+\textit{Spitzer}}&\multicolumn{2}{c}{} \\

\hline \hline
   \cellcolor{darkblue!75} {\color{white} $\boldsymbol{2.0\ \&\ 1.4}$} & 0.10 & 90\%& 0.05 & 90\%& 0.08 & 91\%& 0.01 & 94\%& 0.06 & 91\%& 0.05 & 94\%& 0.0 & 93\%& 0.04 & 90\%\\  
   \cellcolor{cyan!75} $\boldsymbol{2.6\ \&\ 1.4}$ & 0.05 & 86\%& 0.03 & 85\%& 0.03 & 86\%& 0.01 & 89\%& 0.03 & 85\%& 0.03 & 86\%& 0.00 & 86\%& 0.03 & 85\%\\
   \cellcolor{yellow!75} $\boldsymbol{2.8\ \&\ 1.4}$ & 0.01 & 81\%& 0.01 & 81\%& 0.01 & 81\%& 0.00 & 82\%& 0.01 & 78\%& 0.01 & 79\%& 0.00 & 78\%& 0.01 & 77\%\\  
   \cellcolor{darkred!75} {\color{white} $\boldsymbol{3.4\ \&\ 0.9}$} & 0.00 & 58\%& 0.00 & 59\%& 0.00 & 58\%& 0.00 & 59\%& 0.00 & 55\%& 0.00 & 55\%& 0.00 & 55\%& 0.00 & 55\%\\  
\hline                                   
\end{tabular}
\tablefoot{ Contamination fraction and completeness for four different $(\mIEuc-\mYEuc)$ \& $(\mYEuc-\mJEuc)$ colour cuts, in eight scenarios of \Euclid\ (+ancillary) photometry, once applied to all sources that have a \IEuc\ and/or \YEuc\ flux measurement. The colour cuts are listed in the first column, while the other columns correspond to the different data combinations. In these eight columns, we show the fraction of intermediate-$z$ interlopers amongst the apparent $z>6$ galaxy sample after applying the colour cut (on the left) and the percentage of fiducial $z>6$ galaxies correctly identified as high-$z$ sources that survive the cut (on the right). The colouring of the rows in this table is equal to how the colour selection criteria are plotted with coloured, dashed lines in Fig.\,\ref{fig:colourdiagrams}. }
\end{table*}}

Figure \ref{fig:median_magnitude} shows the median observed magnitude in each filter for intermediate-$z$ galaxies, $z>6$ contaminants and true $z>6$ galaxies identified from \textit{Euclid}+Rubin+\textit{Spitzer} photometry.  This figure demonstrates that contaminants of $z>6$ galaxies are the faintest amongst intermediate-redshift galaxies, that is, falsely identified $z>6$ galaxies tend to be much fainter than secure $z=1$--6 galaxies. On average, contaminants are $\sim1.9$ mag fainter than stable intermediate-$z$ galaxies in the \Euclid\ and Rubin filters. Only in the IRAC bands are the contaminants similarly bright to the stable intermediate-$z$ sources. The photometry of contaminants is dominated by $2\sigma$ flux upper limits. This is in agreement with \citet{vulcani2017}, who have shown that low-$z$ contaminants of drop-out selected $z>5$ galaxies are located near the detection limit of galaxy surveys. Because of the numerous upper limits, the SED fitting of the contaminants is poorly constrained, that is, their redshift parameter space becomes highly degenerate. 
 
Moreover, Fig.\,\ref{fig:median_magnitude} clearly shows how contaminants differ from true $z>6$ galaxies based on their \IEuc, \YEuc, \JEuc, and \HEuc\ photometry. True $z>6$ galaxies have very red $(\mIEuc-\mYEuc)$ colours in addition to very flat $(\mYEuc-\mJEuc)$ and $(\mJEuc-\mHEuc)$ colours. Conversely, $z>6$ contaminants show on average a gradual brightening over the same wavelength range, with bluer $(\mIEuc-\mYEuc)$ colours than true $z>6$ galaxies. The physical interpretation of this difference is straightforward. At $z_{\mathrm{fid}}=6$--8, the \IEuc\ and \YEuc\ bands sample the rest-frame spectrum blue- and redwards of Lyman-$\alpha$ line at $\lambda = 1216\,\AA$, resulting in a strong, red $(\mIEuc-\mYEuc)$ colour. On the contrary, at $z_{\mathrm{fid}}=1$--5.8, the \IEuc\ and \YEuc\ bands sample the UV continuum mostly redwards from the Lyman-$\alpha$ line. At this wavelength range, the fiducial SEDs of the $z>6$ contaminants are particularly faint: they are below the assumed Euclid Survey depth and therefore have significantly different $(\mIEuc-\mYEuc)$ colours than true $z>6$ galaxies.

For true $z>6$ galaxies, we obtain median $(\mYEuc-\mJEuc)=0.14$ and $(\mJEuc-\mHEuc)=0.01$ colours as the \YEuc, \JEuc, and \HEuc\ filters sample the rest-frame UV and blue optical continuum. This is the result of our input fiducial $z=6$--8 galaxies, which have similarly flat UltraVISTA $(Y-J)$ and $(J-H)$ colours and are typically UV bright (median $M_{\mathrm{1500\AA}}= -21.7$ mag). Additionally, the majority of true $z>6$ galaxies have fiducial redshift $z_{\mathrm{fid}}=6$--7 (77\,\%), so the median $(\mYEuc-\mJEuc)$ colour is marginally influenced by the red $(\mYEuc-\mJEuc)$ colour caused by the Lyman-$\alpha$ break of $z_{\mathrm{sim}} > 7$ sources. The flat colour signature is typical for high-redshift Lyman-break galaxies \citep{salmon2020}, as they have high specific star-formation rates \citep{barros2014} and virtually no dust attenuation \citep{bouwens2012}, resulting in flat UV spectra because of the dominance of young stellar populations. 
 
Although Fig.\,\ref{fig:median_magnitude} is based on \textit{Euclid}+Rubin+\textit{Spitzer} data, the observed general trends are present in all eight combinations of photometry. In all scenarios, we find that $z>6$ contaminants comprise the faintest intermediate-redshift galaxies, and that contaminants have significantly different $(\mIEuc-\mYEuc)$ and $(\mYEuc-\mJEuc)$ colours from the true $z>6$ galaxies. 

These results clearly indicate the importance of the \YEuc\ band in separating contaminants from the real $z>6$ galaxies. Therefore, we construct colour-colour diagrams considering this band, namely $(\mIEuc-\mYEuc)$ versus $(\mYEuc-\mJEuc)$, which are shown in Fig.\,\ref{fig:colourdiagrams}. We emphasise that these colour diagrams can only be constructed for galaxies up to $z_{\mathrm{fid}}=8$, as sources beyond this redshift are \YEuc\  dropouts and as such do not have a meaningful $(\mIEuc - \mYEuc)$ colour. Therefore, the below proposed colour criteria cannot be used to identify \Euclid\ $z>8$ galaxies. 

Similarly, because the contaminants comprise the faintest galaxies in our sample, many have flux upper limits in both the \IEuc\ and \YEuc\ band, so that their $(\mIEuc-\mYEuc)$ colour is meaningless. Therefore, this analysis considers only true $z>6$ galaxies and $z>6$ contaminants that have a detection in the \IEuc\ and/or \YEuc\ band. In the scenario of \Euclid\ data alone, this means that 1.4\,\% and 46\,\% of the true $z>6$ galaxies and $z>6$ contaminants are excluded, respectively. These numbers are representative for the other combinations of \Euclid\ and ancillary photometry. Furthermore, the vast majority of true $z>6$ galaxies in all scenarios have a detection in the \YEuc\ band but a flux upper limit in the \IEuc\ band; as such, their $(\mIEuc-\mYEuc)$ colour is actually a lower limit and may be even redder in reality.  

To separate the contaminants from the true $z>6$ galaxies, we present an array of $(\mIEuc-\mYEuc)$ \& $(\mYEuc-\mJEuc)$ colour criteria that produce different degrees of contamination and $z>6$ completeness. The four colour cuts were derived using only \Euclid\ data and are based on sources with a solid flux measurement in at least the \IEuc\ or \YEuc\ band. The colour criteria and the conditions that they meet are listed in Fig.\,\ref{fig:colourdiagrams}. We subsequently applied these colour cuts to the other \Euclid\ (+ancillary) data scenarios, and derived the completeness and contamination fraction of the surviving galaxy sample. The results are presented in Table \ref{table:colourcuts}. The completeness is defined as the fraction of recovered $z_{\mathrm{fid}}=6$--8 galaxies compared to the entire fiducial $z_{\mathrm{fid}}=6$--8 sample, consisting of 315 galaxies.

First, we emphasise that the achieved contamination fractions from applying the colour criteria are lower limits, given that generally half of the $z>6$ contaminants are not included in this analysis. For the true $z>6$ galaxies and contaminants that have a \IEuc\ and/or \YEuc\ detection, the colour criteria are highly successful at preventing the intermediate-$z$ interlopers from entering the high-redshift galaxy sample. Specifically, the colour cut $(\mIEuc-\mYEuc)>2.8$ \& $(\mYEuc-\mJEuc)<1.4$ (in yellow) reduces the contamination fraction to 0.01 while simultaneously preserving 90\,\% of the true $z>6$ galaxies (\Euclid\ alone); the resulting completeness of fiducial $z=6$--8 galaxies is 81\,\%. When additional optical photometry, Spitzer data, or a combination of both are considered, this cut reduces the contamination fraction to $\leq0.01$, whilst still preserving $\sim80$\,\% of the high-redshift galaxies.

Here we comment on the usability of these \Euclid\ colour diagrams for the selection of $z=6$--8 galaxies. In the case of \Euclid\ data alone, 30 out of the 65 contaminants cannot be included in these diagrams because they do not have a well-constrained $(\mIEuc-\mYEuc)$ colour. Considering the colour criteria $(\mIEuc-\mYEuc)>2.8$ \& $(\mYEuc-\mJEuc)<1.4$, this means that in the worst-case scenario, 30 additional contaminants could actually survive this selection, and so the contamination fraction would be 0.11 instead of 0.01. Therefore, the purity of the apparent $z>6$ sample would still improve from applying colour selection criteria, although possibly not as drastically as presented in Table \ref{table:colourcuts}. On the contrary, given that the $(\mIEuc-\mYEuc)$ colour of most true $z>6$ galaxies is a lower limit, the recovered completeness with the colour criteria may be higher in reality. In conclusion, the presented colour criteria in this work are useful for selecting a relatively pure sample of $z>6$ galaxies whilst maintaining acceptable completeness, although it is not possible to exactly state to which degree. 

\begin{figure}
\centering
\includegraphics[width=\hsize]{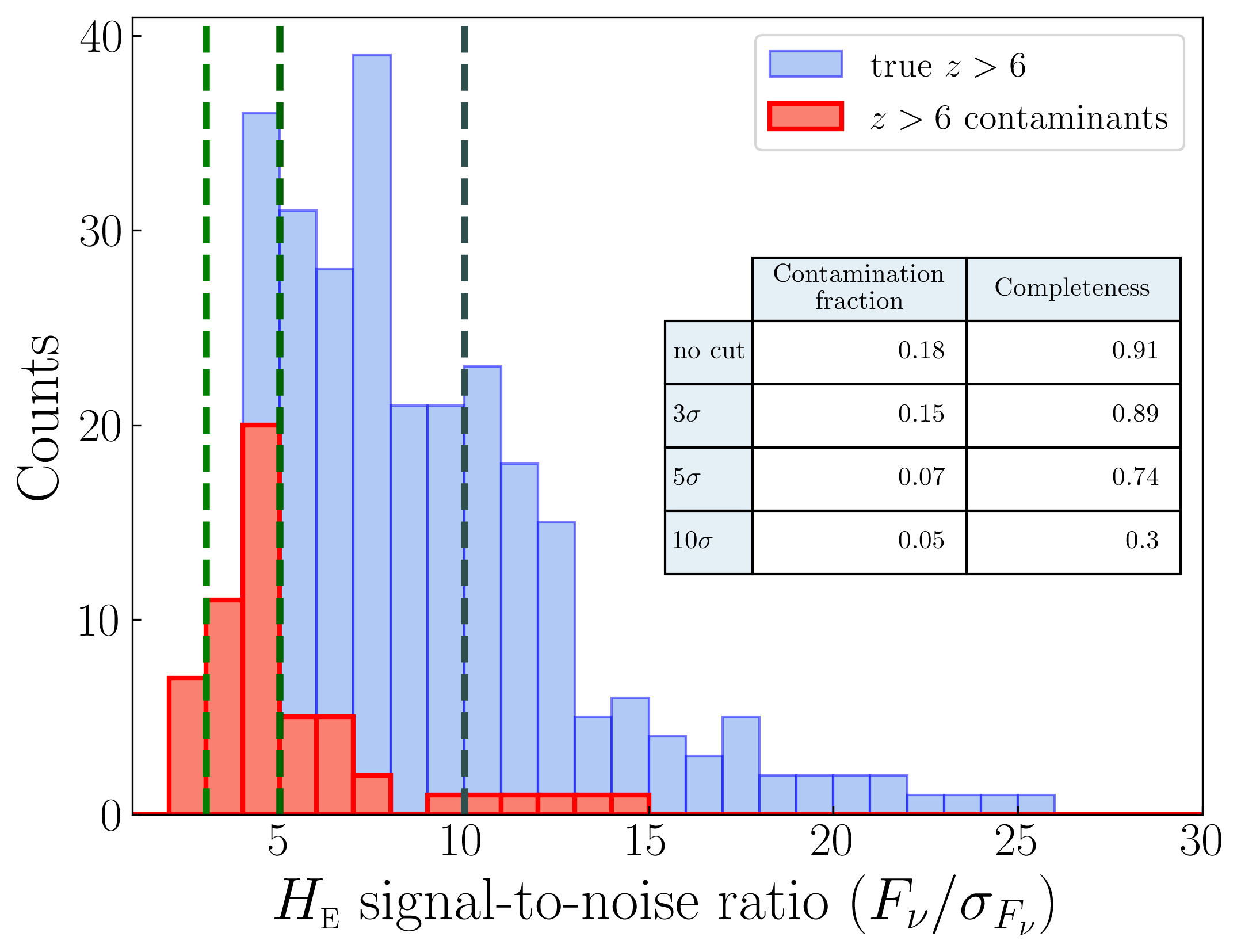}
\caption{Histogram of the \HEuc-band S/N.\ Contaminants and true $z>6$ galaxies as identified from \Euclid\ data alone are displayed in red and blue graphs, respectively. The vertical dashed lines indicate the 3$\sigma$, 5$\sigma$, and 10$\sigma$ detection cutoffs. The contamination fraction and completeness computed for the three cuts are indicated in the table.}
\label{fig:snr_hist}
\end{figure}

As mentioned in Sect.\,\ref{sec:contaminants}, an alternative strategy that is often used to ensure intermediate-$z$ galaxies do not enter the high-redshift galaxy sample is to require a detection with a certain S/N for high-$z$ candidates. Here we explore in depth how imposing a 3-, 5-, and 10$\sigma$ detection threshold requirement on the apparent \Euclid\ $z>6$ sample could reduce the contamination fraction. Figure \ref{fig:snr_hist} shows the distribution of the \HEuc-band S/N for contaminants and true $z>6$ galaxies, where the three S/N cuts are indicated with dashed lines. We {\color{black}imposed} each S/N cut on the apparent $z>6$ sample and subsequently {\color{black}recomputed} the contamination fraction and $z>6$ completeness. Ultimately, we find that even a $10\sigma$ S/N requirement merely reduces contamination to 5\,\% whilst preserving only 30\,\% of the actual high-$z$ galaxies. For reference, the most stringent colour cut presented in Table \ref{table:colourcuts} is able to maintain 58\,\% completeness. We find similar results for the \YEuc\ and \HEuc\ bands. We conclude that for the UltraVISTA-like bright sample, the colour cuts are more effective for identifying intermediate-$z$ interlopers than imposing a S/N requirement. 

Finally, we explore how effective a combination of a S/N requirement with colour selection criteria would be. First, we {\color{black}applied} a 5$\sigma$ \HEuc-band S/N detection threshold requirement to the apparent $z>6$ sample recovered with \Euclid\ data alone. Subsequently, we {\color{black}applied} the same colour criteria presented in this paragraph to the restricted sample, and highlight the results from the colour criteria $(\mIEuc-\mYEuc)>2.8$ \& $(\mYEuc-\mJEuc)<1.4$. Imposing these criteria, the contamination fraction is reduced to 0.01 whilst preserving 70\,\% of the high-redshift sources. However, even with the detection threshold requirement, 5 out of 18 contaminants do not have a constrained  $(\mIEuc-\mYEuc)$ colour and, therefore, cannot be included in the colour diagrams. Therefore, if all of these sources survived these colour criteria and continued to populate the recovered high-redshift sample, the contamination fraction would be 0.03. In summary, the combination of a S/N requirement and colour selection criteria is able to recover a high-redshift sample with very high purity, but at the cost of the $z>6$ completeness; as such, whether or not this combination should be used will depend on the research purpose.   

\begin{figure*}
\begin{subfigure}{8.5cm}
 \centering
\includegraphics[width=\hsize]{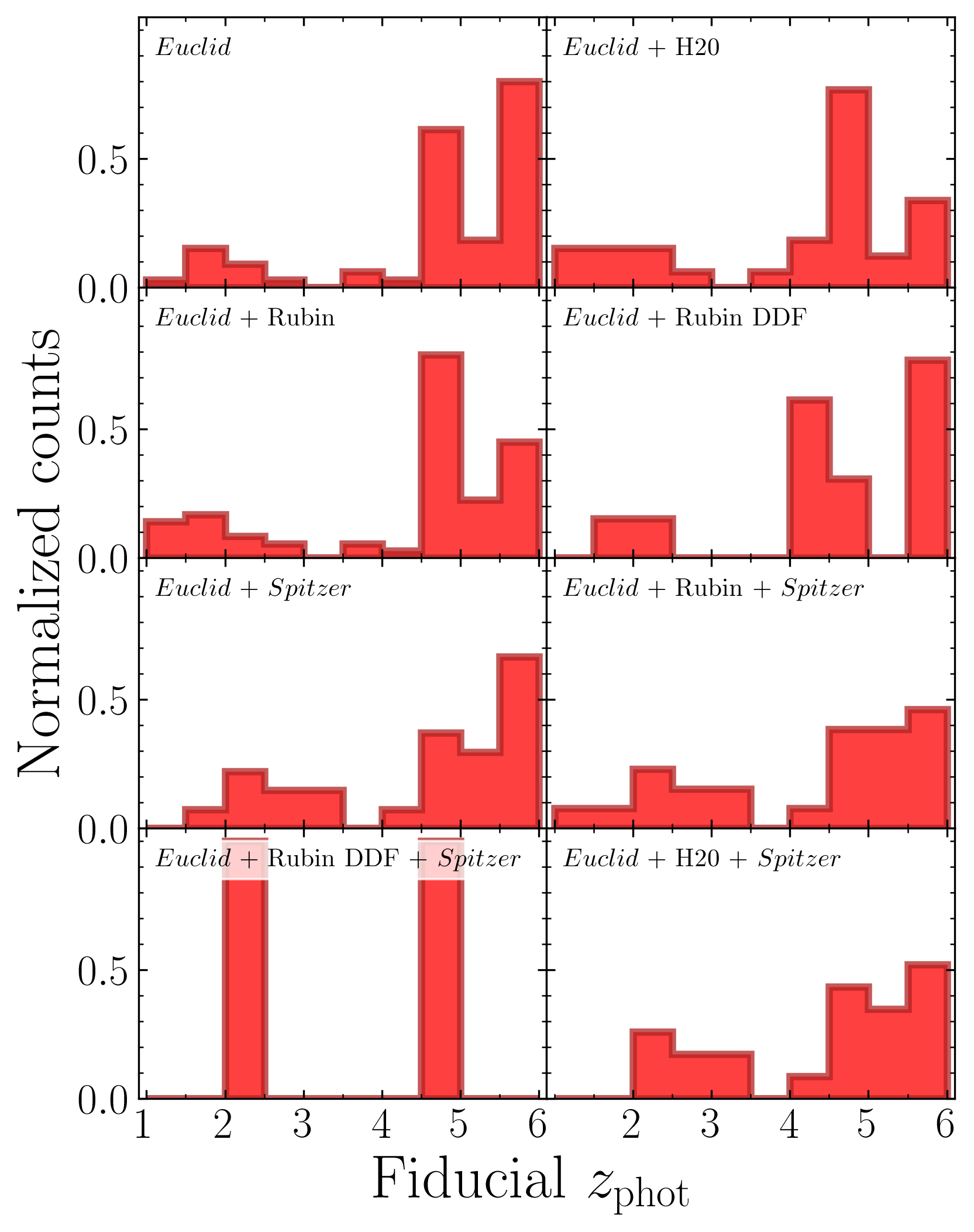}
\end{subfigure}
\begin{subfigure}{8.5cm}
 \centering
 \includegraphics[width=\hsize]{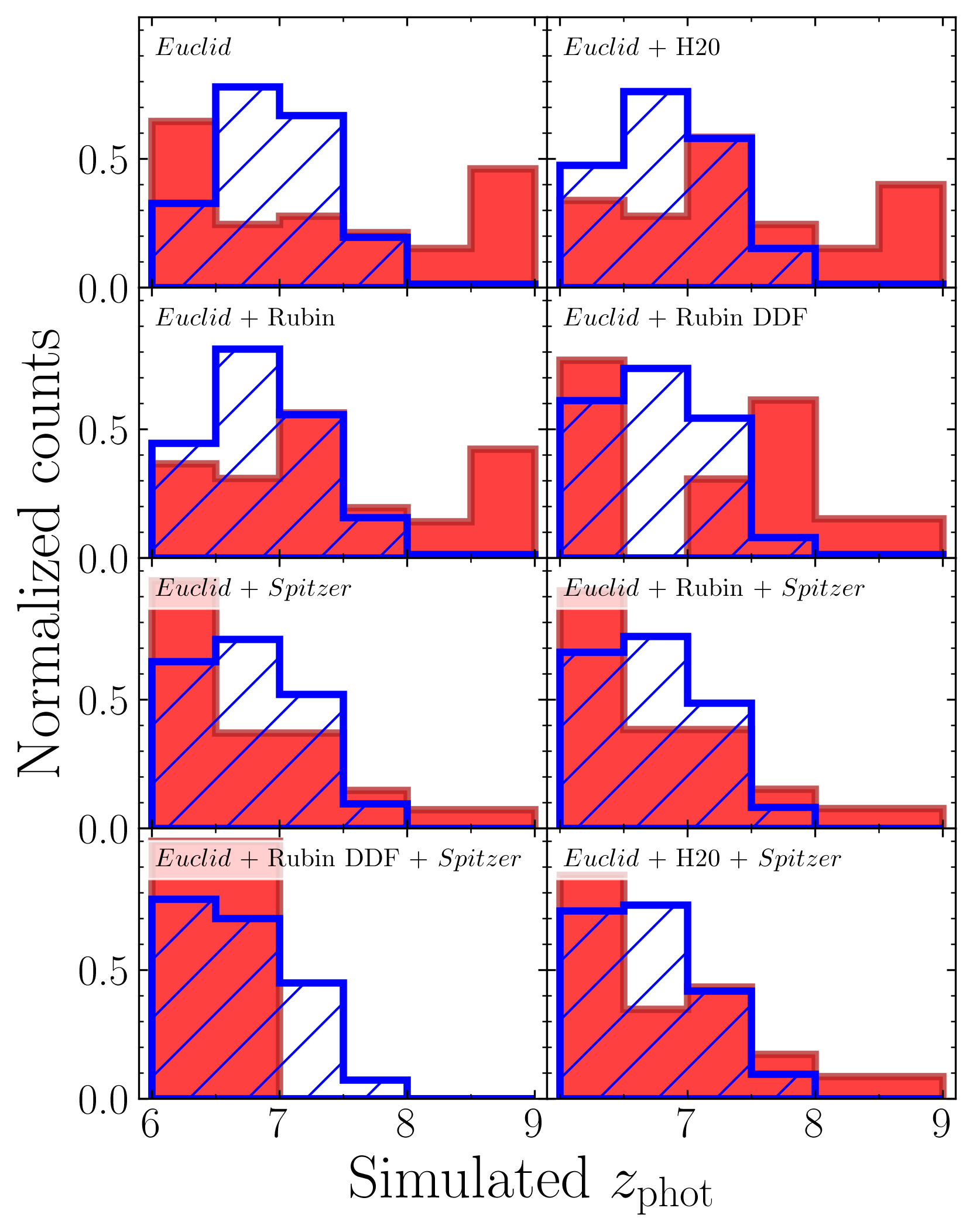}
\end{subfigure}
 \caption{{\color{black} Photometric redshifts distributions of the intermediate-$z$ interlopers.} (\textit{Left panel}) Normalised fiducial redshift distribution of the $z>6$ contaminants in each scenario of \Euclid\ (+ancillary) photometry. (\textit{Right panel}) Normalised \Euclid\ (+ancillary) photometry-derived redshift distributions of the true $z>6$ galaxies (in hatched blue) and $z>6$ contaminants (in red). }
 \label{fig:zphot_histogram}
\end{figure*}

\subsection{The nature of the $z>6$ contaminants}
\noindent Now that we have quantified the expected degree of contamination in the Euclid Deep $z>6$ galaxy selection and showed how this can be reduced, we aim to understand the nature of the contaminant sources. We do this by inspecting the fiducial physical parameters of these galaxies, as obtained from \texttt{LePhare} in the original run based on COSMOS 28-band photometry. Subsequently, we compare the \Euclid-derived SED properties of true $z>6$ galaxies and contaminants, in order to investigate if the two populations can be further segregated based on their recovered physical parameters.

\subsubsection{Photometric redshift distributions}

\noindent Figure \ref{fig:zphot_histogram} shows the fiducial and simulated normalised redshift distribution of $z>6$ contaminants and true $z>6$ galaxies, in each \Euclid\ (+ancillary) data scenario. Between the different photometric scenarios, the shape of the fiducial redshift distribution of the contaminants is roughly similar: we consistently identify broad peaks at $z_{\mathrm{fid}}\sim 1$--3 and $z_{\mathrm{fid}}\sim 4.0$--6, and typically very few sources at $z_{\mathrm{fid}}\sim 3$--4. In addition, for almost all scenarios, the majority of contaminants have $z_{\mathrm{fid}}\sim 4.0$--6. Therefore, we conclude that the underlying galaxy populations that are likely to be misidentified as high-$z$ galaxies are similar, regardless of the external data available to complement \Euclid\ data in the SED fitting analysis. 

Figure \ref{fig:zphot_histogram} also shows the redshift distribution of $z>6$ contaminants once constrained with simulated \Euclid\ (+ancillary) photometry. Here we can see how the contaminants affect different redshift bins at $z>6$, which varies according to the considered data combination. Generally we find that without constraints from \textit{Spitzer} photometry, the contaminants are systematically placed at higher redshifts than in scenarios where \textit{Spitzer} data are available. 
From Fig.\,\ref{fig:zphot_histogram} it is evident that the recovered redshift distributions of true $z>6$ galaxies and $z>6$ contaminants are different but largely overlapping, and thus the $z>6$ contaminants cannot be identified solely based on their recovered redshifts.

\begin{figure}[ht!]
 \centering
 \includegraphics[width=\hsize]{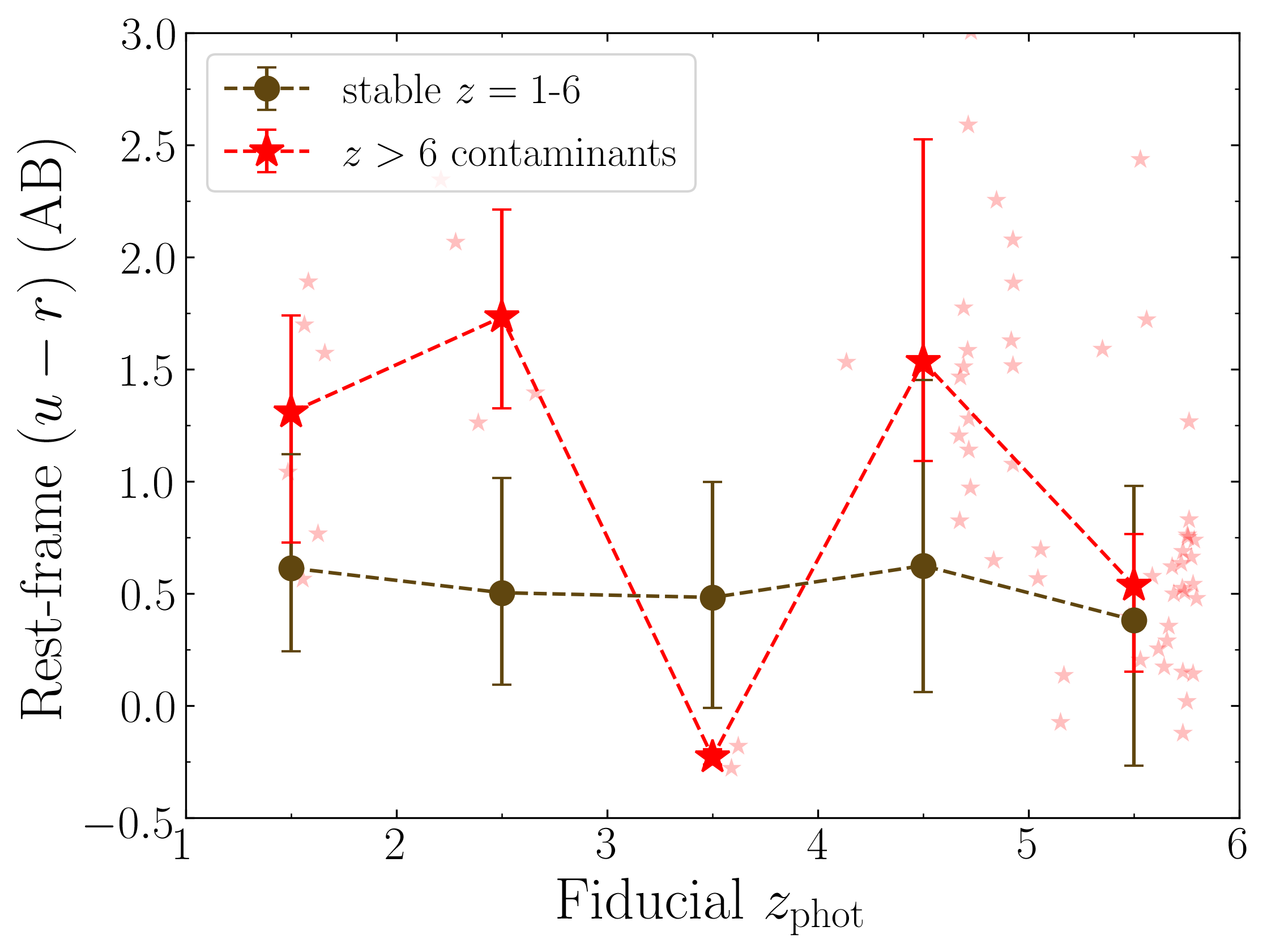}
 \caption{Rest-frame $(u-r)$ colour of the $z>6$ contaminants and stable intermediate-$z$ galaxies against the fiducial redshift, based on real COSMOS photometry. The $z>6$ contaminants and stable intermediate-$z$ galaxies are identified through \Euclid\ photometry alone. For both populations, we plot the median $(u-r)$ colour in $\Delta z=$1.0 fiducial redshift bins, where the error bars represent the 16\textsuperscript{th} and 84\textsuperscript{th} percentiles. The individual data points of $z>6$ contaminants are shown with small red stars. }
 \label{fig:red_ur}
\end{figure}

\begin{figure*}
\begin{subfigure}{17cm}
 \centering
 \includegraphics[width=\hsize]{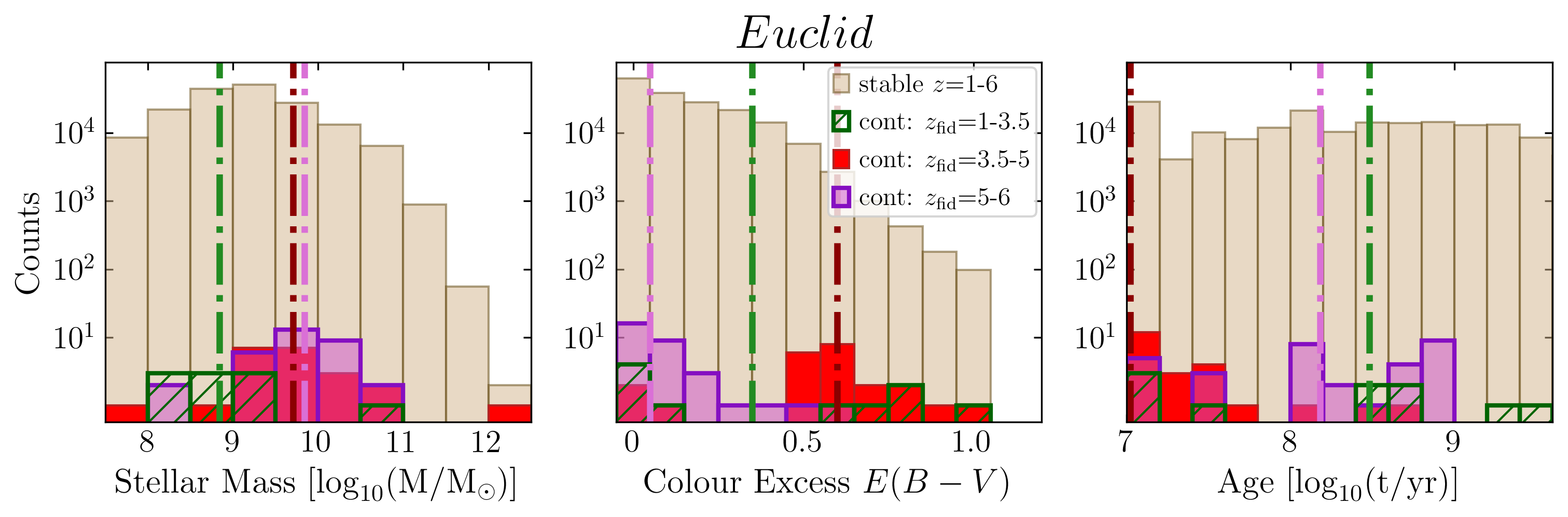}
\end{subfigure}
\newline 
\begin{subfigure}{17cm}
 \centering
 \includegraphics[width=\hsize]{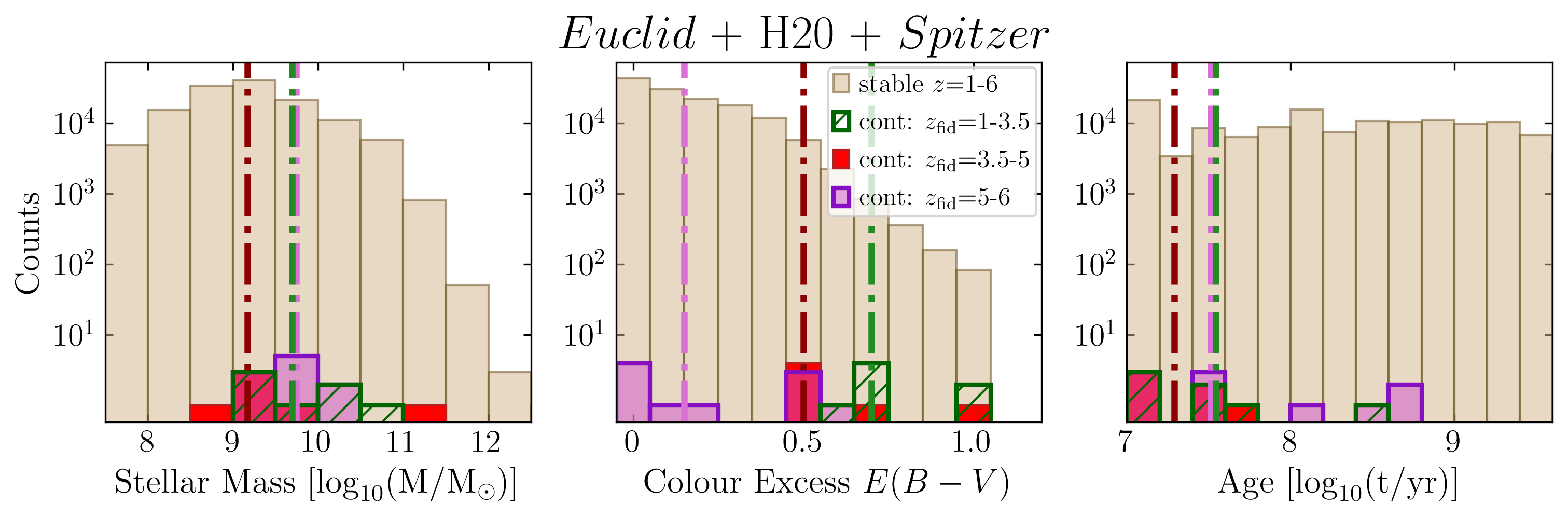}
\end{subfigure}
 \caption{Fiducial COSMOS photometry-derived stellar mass, colour excess, and age distributions of the stable intermediate-$z$ galaxies (in light brown) and the $z>6$ contaminants. The latter are divided into three samples: contaminants with fiducial $z=1$--3.5 (in hatched green), contaminants with fiducial $z=3.5$--5 (in red) and contaminants with fiducial $z=5$--6 (in purple). The median of each distribution is indicated with a vertical line in a corresponding colour. The upper panels show contaminants identified from \Euclid\ data alone, and the lower panels show contaminants identified from \textit{Euclid}+H20+\textit{Spitzer} data.  }
 \label{fig:sedparamuvista}
\end{figure*}

\begin{figure}[ht!]
 \centering
 \includegraphics[width=\hsize]{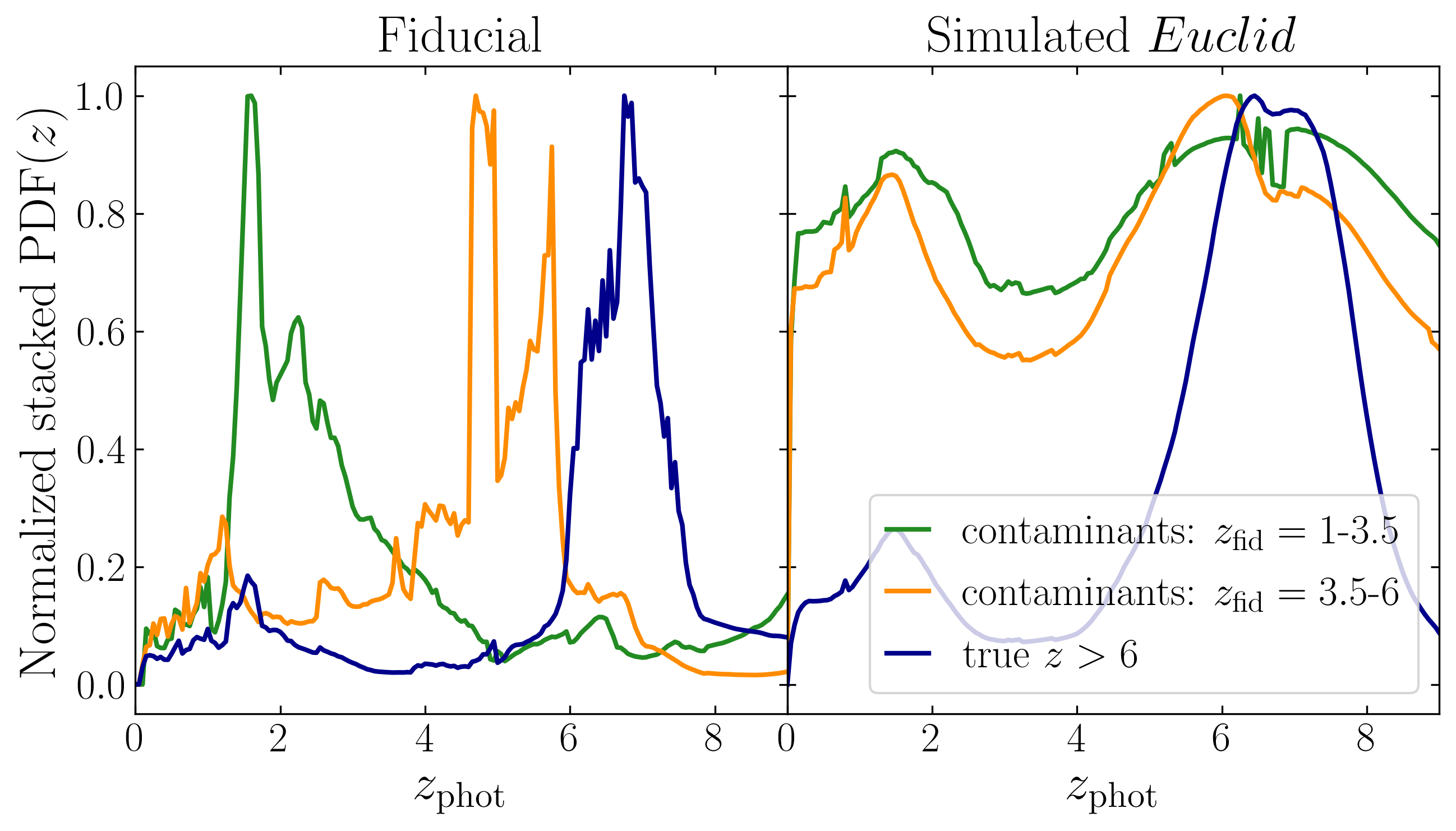}
 \caption{Normalised, stacked probability distribution function of the redshift, PDF($z$), of the $z>6$ contaminants and true $z>6$ galaxies identified from \Euclid\ data. The fiducial PDF($z$) derived from COSMOS 28-band photometry is shown in the left panel; the simulated PDF($z$) derived from \Euclid\ photometry is shown on the right. }
 \label{fig:pdf}
\end{figure}

\begin{figure*}
 \centering
 \includegraphics[width=\hsize]{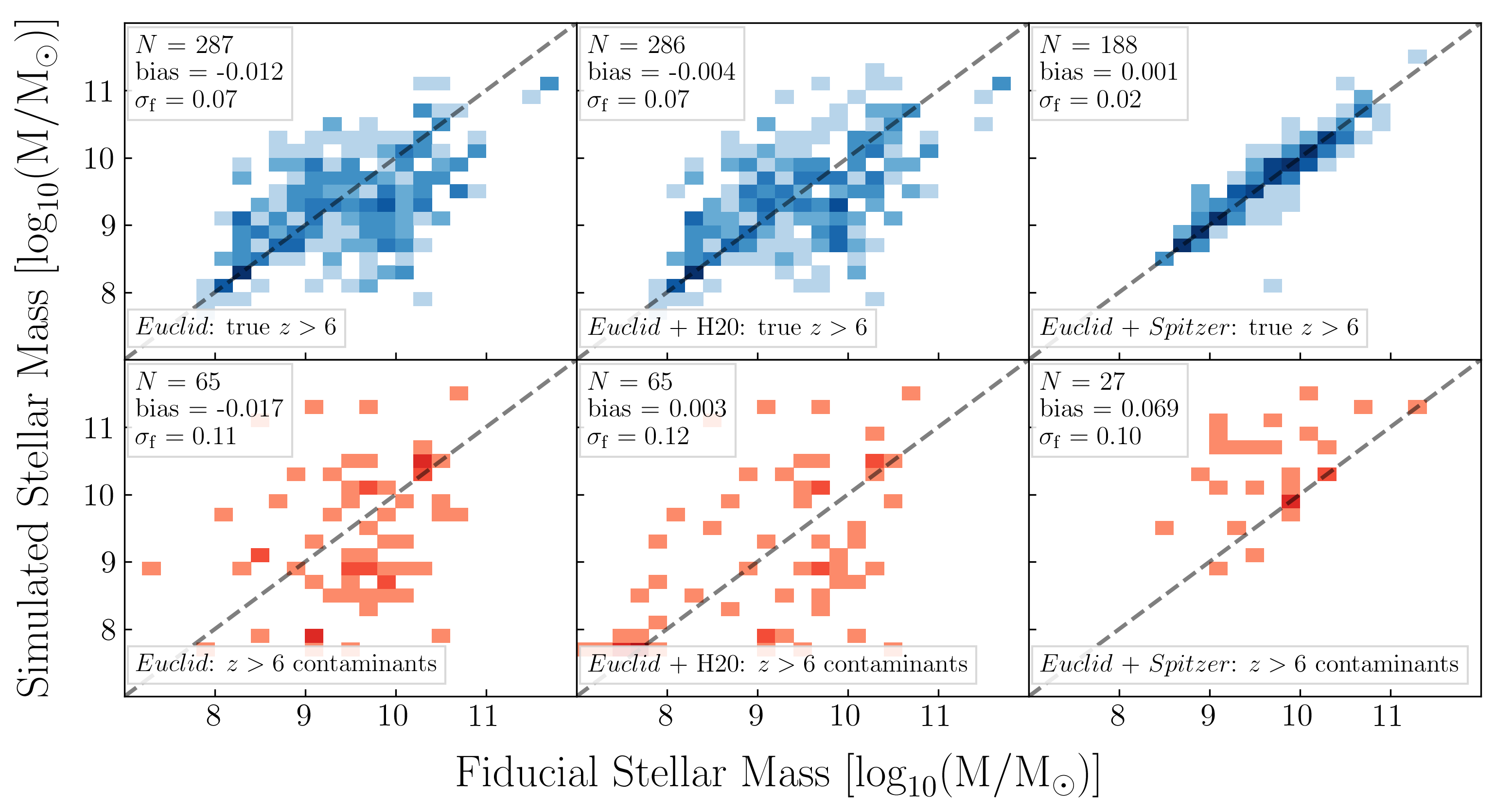}
 \caption{Fiducial stellar mass versus \Euclid\ (+ancillary) photometry-derived stellar mass for true $z>6$ galaxies (upper panels) and $z>6$ contaminants (lower panels), for three scenarios of \Euclid\ and ancillary photometry. The number of galaxies shown is indicated in the upper-left corner of each panel, as is the bias and scatter $\sigma_{\rm f}$. The bias is defined as $mean[(\mathrm{M_{sim}-M_{fid})/M_{fid}}]$ and the scatter as $rms[(\mathrm{M_{sim}-M_{fid})/M_{fid}}]$. The colour intensity corresponds to the number of galaxies in each bin (darker colours correspond to more sources).  }
 \label{fig:sedparameuclid}
\end{figure*}

\subsubsection{Physical parameters}

So far, we have established that, regardless of which ancillary data are added to the \Euclid\ photometry, $z>6$ contaminants are amongst the faintest sources in our intermediate-$z$ galaxy sample and primarily correspond to galaxies with fiducial redshifts around $z_{\mathrm{fid}}\sim2.5$ or $z_{\mathrm{fid}}\sim5$. Here we analyse the physical properties of these galaxies, to better understand why specifically these populations are sensitive to the $z>6$ degeneracy. 
 
In Fig.\,\ref{fig:red_ur} we show the rest-frame $(u-r)$ colour against the fiducial redshift for intermediate-$z$ galaxies and $z>6$ contaminants, as identified from \Euclid\ photometry. The rest $(u-r)$ colour is derived from the fiducial COSMOS photometry and corresponding best-fit SED, and samples the rest-frame optical continuum at $\lambda\sim 3800\,\AA$ ($u$) and $\lambda \sim 6300\,\AA$ ($r$). First, we note that the blue $(u-r)$ colours of contaminants at $z_{\mathrm{fid}}=3.0$--4.0 are a coincidence rather than a real degeneracy: there are only two galaxies in this bin that are both faint and have poorly constrained fiducial SEDs. Overall, we find that $z>6$ contaminants typically have redder $(u-r)$ colours than stable intermediate-$z$ galaxies. Only at $z_{\mathrm{fid}}=5.5$--6.0 are the rest-frame optical colours of $z>6$ contaminants similar to those of stable $z_{\mathrm{fid}}=5.5$--6.0 galaxies; the former are contaminants simply because they are much fainter compared to the latter. Lastly, for other combinations of \Euclid\ and ancillary photometry, contaminants at $z_{\mathrm{fid}}=3.0$--4.0 display similarly red colours as those in other fiducial redshift bins. In conclusion, including the result from Fig.\,\ref{fig:median_magnitude}, the $z>6$ contaminants are generally characterised as faint, systematically reddened, intermediate-$z$ galaxies, in good agreement with previous studies (e.g. \citealt{oesch2012}; \citealt{bowler2014}). 
 
In order to uncover what produces the redder optical colours of the contaminants, we inspect their fiducial stellar masses, dust extinctions and ages as derived from the original (COSMOS) photometry. We also investigate the \Euclid\ (+ancillary) photometry-derived physical parameters of the contaminants, to see if these sources that appear as $z>6$ galaxies occupy a different parameter space than the true $z>6$ galaxies. 
 
In Fig.\,\ref{fig:sedparamuvista} we show the fiducial stellar mass, colour excess, and age distributions of the $z>6$ contaminants and the stable intermediate-$z$ galaxies, as derived from the original SED fitting with COSMOS photometry. We show the comparison for two scenarios: one where interlopers are identified from \Euclid\ photometry alone, and one where they are identified from \textit{Euclid}+H20+\textit{Spitzer} data. Furthermore, the contaminants are split into three samples based on their fiducial redshift, that is, $z_{\mathrm{fid}}=1$--3.5, $z_{\mathrm{fid}}=3.5$--5 and  $z_{\mathrm{fid}}=5$--6, following the typical redshift distinction we observed in Fig.\,\ref{fig:zphot_histogram}, and considering that contaminants at $z_{\mathrm{fid}}=3.5$--5 and $z_{\mathrm{fid}}=5$--6 display different $(u-r)$ colours. Apart from the three physical properties investigated in Fig.\,\ref{fig:sedparamuvista}, we also inspect the characteristics of the best-fit SEDs of these sources, that is, their metallicity and SFH. 

First and foremost, it is clear from Fig.\,\ref{fig:sedparamuvista} that $z>6$ contaminants and stable intermediate-$z$ galaxies occupy the same parameter space for any property investigated in this paper. Therefore, we conclude from this that the galaxies driving the \Euclid\ $z>6$ contamination are not part of some specific population, but are primarily interlopers because of their faintness. 

When we investigate the three samples of $z>6$ contaminants as identified from \Euclid\ photometry, we find that contaminants at $z_{\mathrm{fid}}=1$--3.5 are typically moderately massive galaxies that have red optical colours either because they are young with considerable dust attenuation, or because they are old with well established stellar populations. Towards higher redshifts, we observe that contaminants at $z_{\mathrm{fid}}=3.5$--5 constitute almost solely of young, massive galaxies that are strongly affected by dust. Finally, contaminants at $z_{\mathrm{fid}}=5$--6 are comparably massive, typically not dusty and of average age. Summarising, the sources that produce similar \Euclid\ colours to actual $z>6$ galaxies are either intermediate-$z$ galaxies that become degenerate through typical confusion of the Lyman-$\alpha$ and $4000\,\AA$ breaks (either due to strong dust attenuation or old age), or faint galaxies with flat SEDs bordering $z\sim6$ that become degenerate because of the limited measurements available to properly constrain them.

Given that we have demonstrated how the inclusion of ancillary photometry reduces the number of intermediate-$z$ interlopers, we also inspect the physical parameters of contaminants identified from \textit{Euclid}+H20+\textit{Spitzer} data. These surviving galaxies constitute the core of the $z>6$ contamination, since they produce similar colours as high-redshift galaxies in not just the four \Euclid\ bands, but in seven ancillary optical and infrared filters as well. As shown in the lower panels of Fig.\,\ref{fig:sedparamuvista}, the remainder of $z>6$ contaminants show a similar albeit narrower mass distribution compared to the scenario with only \Euclid\ photometry. Conversely, contaminants at $z_{\mathrm{fid}}=1$--3.5 and $z_{\mathrm{fid}}=3.5$--5.0 are now solely young and dusty galaxies. Apparently, the degeneracy between older, intermediate-$z$ galaxies with well-developed $4000\,\AA$ breaks and actual $z>6$ galaxies is at least partially broken through the addition of ancillary photometry, whereas the degeneracy between dust-reddened galaxies and $z>6$ galaxies mostly remains. 

We show the normalised, stacked probability distribution function of the redshift, PDF($z$), of the $z>6$ contaminants and true $z>6$ galaxies in Fig.\,\ref{fig:pdf}. For clarity, we divide the contaminants into only two redshift bins, namely $z_{\mathrm{fid}}=1$--3.5 and $z_{\mathrm{fid}}=3.5$--6. Interestingly, although their fiducial PDF($z$) is generally broad considering it was derived from 28-band photometry, the $z>6$ contaminants do not show significant probability for secondary redshift solutions at $z>6$. When observed with \Euclid\ alone, the PDF($z$) of any $z>6$ contaminant is highly degenerate. This means that even though the contaminants are falsely identified as high-redshift galaxies with \Euclid, one cannot possibly exclude a low-redshift nature based on the PDF($z$) that is recovered. Upon further inspection, we find that even with additional H20 and \textit{Spitzer} photometry, the majority of contaminants produce highly degenerate results for the PDF($z$); only with the ultra-deep photometry from the Rubin DDF do we retrieve $z>6$ contaminants that have a PDF($z$) solely defined at $z>6$. Therefore,  one would never know from the PDF($z$) that these sources are actually misidentified intermediate-$z$ galaxies. 

Now that we have characterised the galaxy population that drives the $z>6$ contamination, we inspect the physical properties as derived with \Euclid\ (+ancillary) photometry of the true $z>6$ galaxies and $z>6$ contaminants. Since the stellar mass is the most important physical parameter second to the photometric redshift, we first inspect the stellar-mass recovery in Fig.\,\ref{fig:sedparameuclid}. It is clear that in the scenarios of \Euclid\ and \textit{Euclid}+H20 data, where the true $z>6$ galaxies are unconstrained in the NIR, the stellar-mass recovery is poor. Conversely, when IRAC photometry is available, the recovery of the stellar mass from only the six \Euclid\ and \textit{Spitzer} bands is very efficient. In addition, when we inspect the age and SFH recovery of the true $z>6$ galaxies, we see (not shown in this paper) a similar trend. \textit{Spitzer} photometry is therefore far more effective than optical data for the recovery of these parameters. Without IRAC data, the rest-frame stellar continuum beyond the $4000\,\AA$ break of an apparent $z>6$ source is completely unconstrained, and so physical parameters directly related to the older stellar population (the stellar mass and age of the galaxy) are unfounded. 

Figure \ref{fig:sedparameuclid} also shows the stellar-mass recovery of the $z>6$ contaminants. Naturally, we do not expect a tight correlation, as the contaminants are by definition misidentified as vastly different galaxies from \Euclid\ (+ancillary) photometry. However, the abundant scatter in the distribution is noteworthy; we have established that distinct populations of galaxies cause the $z>6$ contamination, but we see no signs of bimodal behaviour in the stellar-mass recovery. By including \textit{Spitzer} photometry, the light from the older stellar populations is actually constrained: for 74\,\% of the contaminants, their recovered stellar mass is higher than their fiducial stellar mass. This is expected, as even though we know the contaminants are relatively faint compared to their stable intermediate-$z$ counterparts, their fluxes can only be attributed to massive $z>6$ galaxies since the stellar mass is directly derived from the SED normalisation. Unfortunately, the simulated stellar mass distributions of true $z>6$ galaxies and intermediate-$z$ interlopers overlap considerably, and so we cannot further separate these populations based on their stellar masses, nor on their ages or SFH models.

\begin{figure}
 \centering
 \includegraphics[width=\hsize]{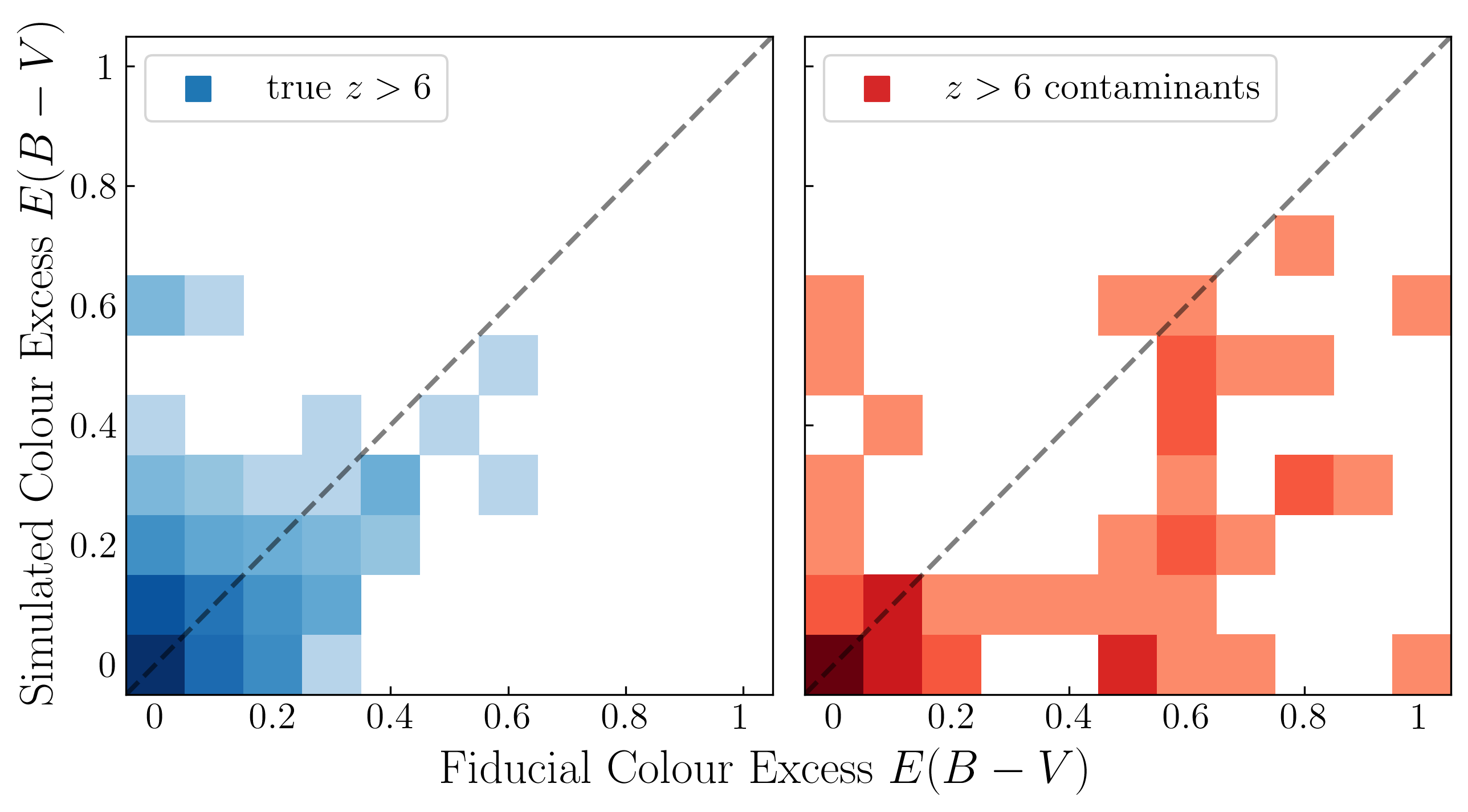}
 \caption{Fiducial colour excess versus \Euclid\ photometry-derived colour excess for true $z>6$ galaxies (left panel) and $z>6$ contaminants (right panel). The colour intensity corresponds to the number of galaxies in each bin (darker colours correspond to more sources).  }
 \label{fig:colourexcessrecovery}
\end{figure}

Figure \ref{fig:colourexcessrecovery} shows the colour excess recovery for $z>6$ contaminants and intermediate-$z$ galaxies derived from \Euclid\ data alone. We do not show any other combination of \Euclid\ and ancillary photometry, as the results are universal. The vast majority of true $z>6$ galaxies (93\,\%) have very low fiducial dust extinction, $E(B-V) \leq 0.2$. These values are well recovered once observed with \Euclid. This is different for the $z>6$ contaminants; as expected, dusty intermediate-$z$ galaxies are mistaken for high-redshift sources with considerably less dust attenuation. Unfortunately, the colour excess cannot be employed to separate interlopers from actual $z>6$ galaxies; 72\,\% of the $z>6$ contaminants have recovered dust extinction $E(B-V) \leq 0.2$.  

We employ the two-parameter Kolmogorov-Smirnov test \citep{smirnov1939} to compare the parameter distributions of $z>6$ contaminants and true $z>6$ galaxies. In virtually all scenarios of data availability, the test shows that the stellar mass and colour excess values of contaminants and actual $z>6$ galaxies are unlikely to come from the same parent distribution, with $p$-values generally below 0.01 (in the scenarios with Rubin DDF photometry we cannot properly compare the distributions due to the low number of contaminants). The age distributions of interlopers and actual $z>6$ galaxies are more similar. Still, the parameter spaces of each physical property significantly overlap for the two populations. Therefore, we conclude that there exists no obvious separation between intermediate-$z$ contaminants and true $z>6$ galaxies based on their recovered stellar mass, dust extinction and age parameters.

\section{Complementary tests based on a faint mock galaxy sample} \label{sec:mock}

In this work we used real data from the COSMOS field to investigate the expected contamination of $z>6$ galaxies in the Euclid Deep Fields. However, especially in the optical regime, the Euclid Deep Fields will be considerably deeper than our fiducial UltraVISTA catalogue. In fact, the $5\sigma$ limiting magnitude of the UltraVISTA $H$-band photometry included in our fiducial UltraVISTA catalogue is 25.2, whereas the expected $5\sigma$ depth in \Euclid\ \HEuc\ band is 26.4 magnitude. Therefore, our results on the contamination fraction in Table\,\ref{table:fraction_contaminants} are biased towards the brightest and most massive galaxies at $z_{\mathrm{fid}}=6$--8, and therefore possibly too optimistic. 

To uncover how successful \Euclid\ data will be at identifying faint high-redshift galaxies, we {\color{black}created} a sample of faint mock galaxies from our fiducial UltraVISTA catalogue and {\color{black}repeated} our analysis of the contamination fraction with this sample. To create the mocks, we {\color{black}shifted} the entire fiducial best-fit SEDs of our UltraVISTA galaxies at $z_{\mathrm{fid}}=1$--8 by 1.2 magnitude, which is the difference in depth between the UltraVISTA $H$- and \Euclid\ \HEuc-band images. From these modified SEDs, we {\color{black}selected} all galaxies with $25.3 \leq H < 27.0$ magnitude, since sources with $H<25.3$ are already discussed in the bright UltraVISTA-like sample and $H=27.0$ corresponds to the $H$-band $3\sigma$ flux limit in the Euclid Deep Survey.

To ensure the faint mock sample follows a realistic photometric redshift distribution, especially at $z_{\mathrm{fid}}=6$--8, we {\color{black}made} use of the CANDELS photometric redshift catalogue in the COSMOS field produced by \citet{nayyeri2016}, which consists of 38\,671 sources identified with the HST and contains photometric data in 42 bands. This catalogue is approximately 2.4 magnitudes deeper in the $H$ band than our UltraVISTA catalogue. From this CANDELS catalogue, we {\color{black}selected} all galaxies with $z=1$--8 and $25.3\leq H<27.0$, the latter based on the \textit{HST}/WFC3 F160W band at 1.6 $\micron$. Subsequently, in $\Delta z=0.2$ bins, we {\color{black}selected} galaxies from our constructed faint mock galaxy sample at random to replicate the re-normalised CANDELS redshift distribution in that bin. Within each redshift bin, we also altered the scaling of the modified SEDs of individual sources such that their $H$ magnitude distribution (in $\Delta \rm{m}=0.1$ mag bins) is identical to the re-normalised CANDELS $H$ magnitude distribution in the same redshift bin. This method ensures that our faint mock sample follows quite closely the CANDELS luminosity functions at $z=6$--8. 

\begin{figure}
 \centering
 \includegraphics[width=\hsize]{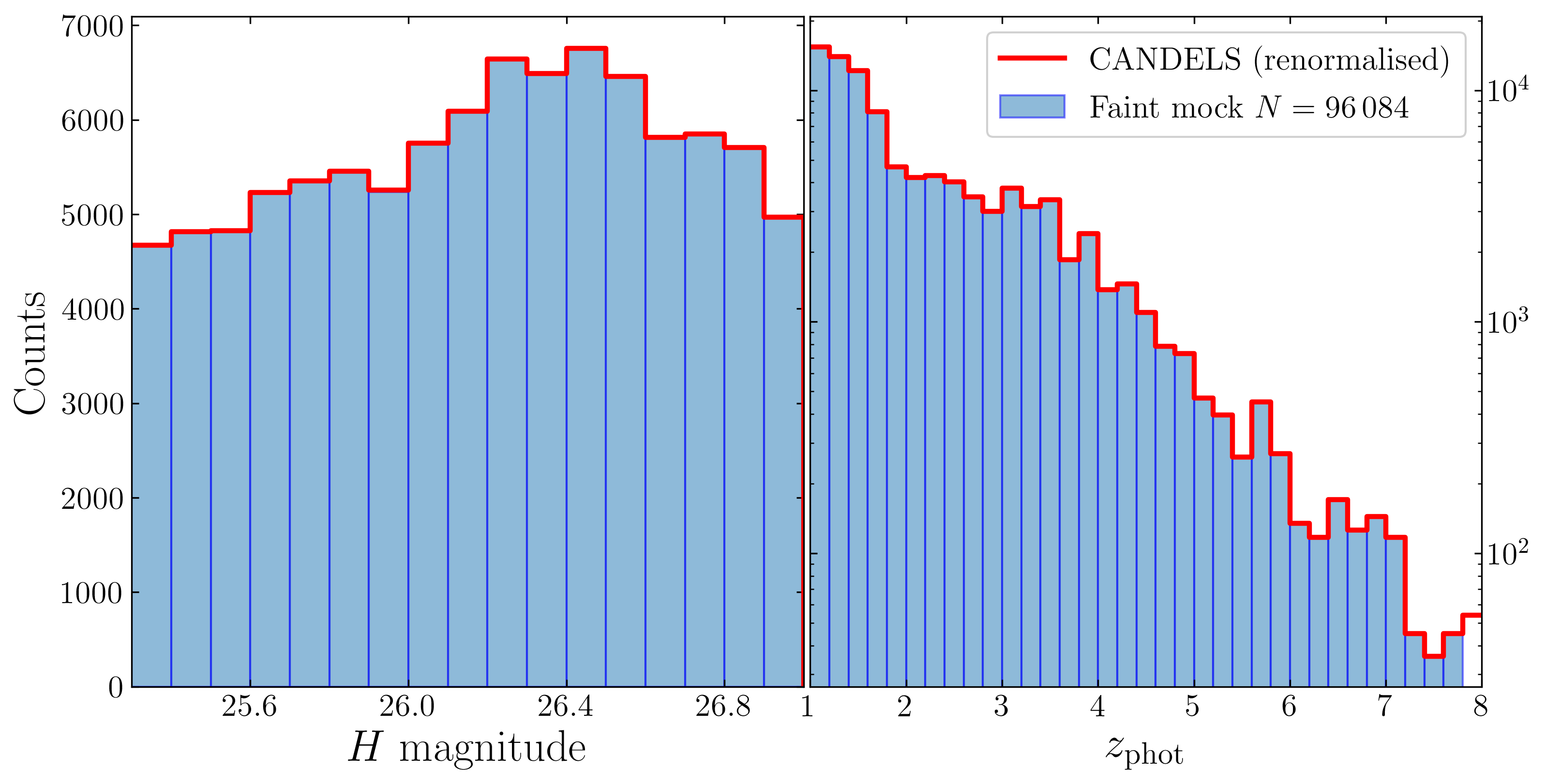}
 \caption{Template UltraVISTA $H$ magnitude distribution in $\Delta \rm{m} = 0.1$ mag bins (left panel) and fiducial photometric redshift distribution in $\Delta z = 0.2$ bins (right panel) of the faint mock sample. In both panels, the corresponding re-normalised CANDELS distributions are shown in red outlines.}
 \label{fig:hmag-zphot-hist}
\end{figure}

We note that the CANDELS $z=7.8$--8.0 redshift distribution cannot be reproduced, as our UltraVISTA catalogue does not contain sources in that redshift bin. However, for simplicity, we globally refer to the faint mock high-redshift sample as mock galaxies with $z_{\mathrm{fid}}=6$--8 redshifts. Therefore, the final faint mock sample contains 96\,084 sources, of which 936 are at $z_{\mathrm{fid}}=6$--8. Out of this sample, 41\,775 sources are detected in at least IRAC channel 1 or channel 2, and 408 of these sources are at fiducial $z_{\mathrm{fid}}=6$--8. The fiducial photometric redshift and $H$ magnitude distribution of the faint mock sample are shown in Fig.\,\ref{fig:hmag-zphot-hist}, together with the corresponding re-normalised CANDELS distributions. 

We {\color{black}sampled} the \Euclid\ (+ancillary) fluxes directly from the modified SEDs of the faint mock galaxies, and {\color{black}obtained} the final, randomised \Euclid\ (+ancillary) photometry following the methods described in Sect.\,\ref{sec:sim_data}, except for the simulated \textit{Spitzer} flux errors. Instead, these {\color{black}were} not taken directly from the observed UltraVISTA/SMUVS photometry, but scaled along with the fiducial SEDs to preserve the S/N and subsequently adopted as the simulated \textit{Spitzer} flux errors. 

We note that some of the UltraVISTA sources, especially at $z_{\mathrm{fid}}=6$--8, are included in the faint mock sample multiple times, as there are relatively more high-redshift sources in the CANDELS catalogue than in the scaled-down UltraVISTA catalogue. However, fiducial duplicates can be considered as individual sources for the purpose of this analysis, as the simulated Euclid (+ ancillary) photometry is independently randomised for each instance. 

{\renewcommand{\arraystretch}{1.5}
\begin{table*}
\caption{Contaminants amongst $z>6$ faint mock galaxies.}             
\label{table:fraction_contaminants_faint}      
\centering                         
\begin{tabular}{l c c c c }        
\hline\hline                 
Filters & True $z>6$ & Contaminants & Contamination Fraction & Completeness\\    
\hline                        
   \Euclid\ & 824 & 523 & $0.39^{+0.10}_{-0.11}$ & 88\,\%\\  
   \hline
   \Euclid\ + Rubin & 824 & 540 & $0.40^{+0.10}_{-0.12}$ & 88\,\%\\
   \Euclid\ + Rubin DDF & 863 & 63 & $0.07^{+0.03}_{-0.03}$ & 92\,\%\\
   \Euclid\ + H20 & 831 & 489 & $0.37^{+0.10}_{-0.11}$ & 89\,\%\\
   \hline
   \Euclid\ + \textit{Spitzer} & 355 & 201 & $0.36^{+0.09}_{-0.09}$ & 87\,\%\\
   \hline
   \Euclid\ + Rubin + \textit{Spitzer} & 358 & 191 & $0.35^{+0.08}_{-0.10}$ & 88\,\%\\
   \Euclid\ + Rubin DDF + \textit{Spitzer}  & 378 & 24 & $0.06^{+0.03}_{-0.02}$ & 93\,\%\\
   \Euclid\ + H20 + \textit{Spitzer} & 357 & 181 & $0.34^{+0.08}_{-0.09}$ & 88\,\%\\
\hline                                   
\end{tabular}
\tablefoot{Number of true $z>6$ galaxies and $z>6$ contaminants, from various combinations of \Euclid\ and ancillary data, based on the faint mock sample. As in Table \ref{table:fraction_contaminants}, we report the contamination fraction and completeness for each data availability scenario.} 
\end{table*}}

\begin{figure}
 \centering
 \includegraphics[width=\hsize]{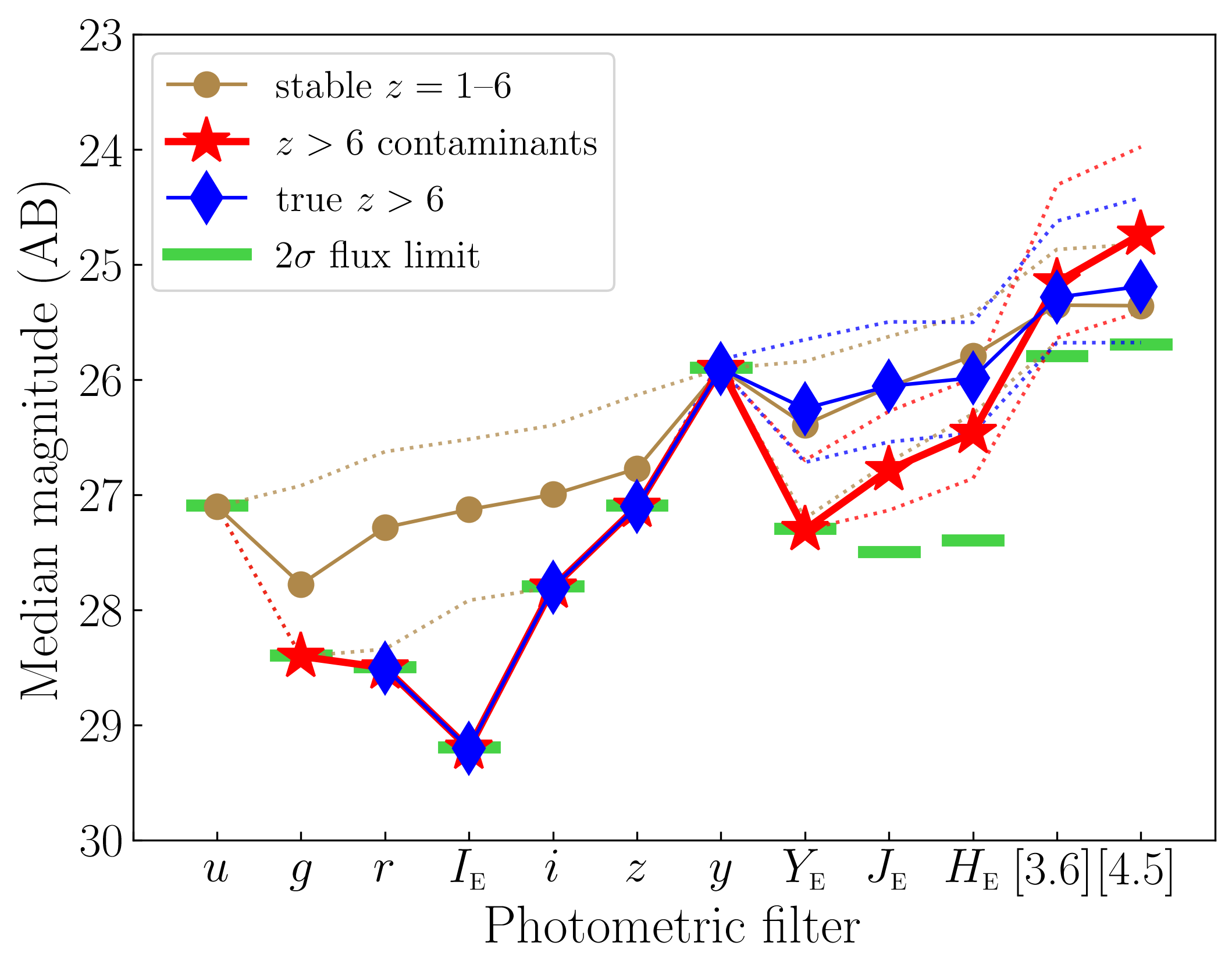}
 \caption{Median simulated magnitudes of the faint mock sample in the \textit{Euclid}, Rubin, and \textit{Spitzer} filters. Light brown circles represent stable intermediate-$z$ galaxies, red stars $z>6$ contaminants, and blue diamonds true $z>6$ galaxies. For each population, the dotted lines indicate the 16\textsuperscript{th} and 84\textsuperscript{th} percentiles. The $2\sigma$ flux limits are shown with green bars. The selection of intermediate-$z$, contaminants, and $z>6$ galaxies from the faint mock sample is based on \textit{Euclid}+Rubin+\textit{Spitzer} data. The median magnitudes for $z>6$ galaxies in the Rubin $u$ and $g$ bands and for contaminants in the Rubin $u$ band are equal to $-99$ (no intrinsic flux) and therefore omitted from this figure. }
 \label{fig:medianmagmockfaint}
\end{figure}

We derive the contamination fraction and $z>6$ completeness of the mock sample in all eight \Euclid\ (+ancillary) data scenarios, which are shown in Table\,\ref{table:fraction_contaminants_faint}. The uncertainties on the contamination fraction were derived in the same manner for the UltraVISTA-like bright sample (see Sect.\,\ref{sec:contaminants}). First and foremost, the $z>6$ completeness ranges from 87\,\% to 93\,\% (in the most optimistic scenario \textit{Euclid}+Rubin DDF+\textit{Spitzer}), which is lower than the 91-96\,\% completeness obtained with the bright UltraVISTA-like sample. This is unsurprising, since we have demonstrated how the $z_{\mathrm{fid}}=6$--8 galaxies are characterised by having very red $(\mIEuc-\mYEuc)$ colours and 2$\sigma$ flux upper limits in the \IEuc\ band. By shifting the fiducial SED downwards, the simulated \YEuc\ fluxes of $z_{\mathrm{fid}}=6$--8 are fainter and the $(\mIEuc-\mYEuc)$ colours not as red, which makes identification of these sources more difficult. This is illustrated in Fig.\,\ref{fig:medianmagmockfaint}, where we show the median magnitude in the \textit{Euclid}, Rubin and \textit{Spitzer} bands of the faint mock galaxies. Nonetheless, a completeness of $>80$\,\% for faint high-redshift sources identified with \Euclid\ will be excellent for most astronomy science purposes.  

Simultaneously, we find that the contamination fraction of the mock sample is $0.39$ with \Euclid\ photometry alone, compared to $0.18$ with the bright UltraVISTA-like sample. Again, this stark increase is not surprising, as we have demonstrated how intermediate-$z$ interlopers are primarily characterised by their faintness. By shifting the UltraVISTA photometry downwards, intermediate-$z$ galaxies that previously remained at the same redshifts when observed with \Euclid\ are now assigned \Euclid\ (+ancillary) fluxes much closer to their respective $2\sigma$ flux limits, and therefore are more likely to be mistaken for $z>6$ galaxies.

With the faint mock sample, we find similar trends regarding the usefulness of ancillary optical data; only the ultra-deep Rubin DDF photometry truly improves the contamination fraction, reducing it to $0.07$. In addition, we find that including the simulated \textit{Spitzer} photometry of the mock galaxies barely improves the contamination fraction, contrary to the moderate improvement obtained with the regular UltraVISTA sample. We explain this from the average $(\mYEuc-[3.6])$ colour of $z>6$ contaminants and true $z>6$ galaxies. As can be seen in Fig.\,\ref{fig:medianmagmockfaint}, contaminants identified in the faint mock sample do not produce the extremely red $(\mYEuc-[3.6])$ colours of contaminants in the bright UltraVISTA-like sample. Because of this, the $(\mYEuc-[3.6])$ colours of $z>6$ contaminants and true $z>6$ galaxies are more similar, meaning \textit{Spitzer} photometry barely helps in distinguishing between these sources.

As was done for the UltraVISTA-like bright sample, we explore what happens to the contamination fraction once we impose a NIR detection threshold requirement for the source selection. This should be especially relevant for the faint mock sample, given that many of these sources have flux measurements close to the $2\sigma$ flux limits in each band. If we require that faint mock sources have a $5\sigma$ detection in at least one of the \Euclid\ NIR bands, 70\,831 sources (74\,\% of the sample) survive, of which 669 galaxies have fiducial $z_{\mathrm{fid}}=6$--8. We recalculate the contamination fraction and $z>6$ completeness on this restricted sample for all eight data combinations, and show the results in Table\,\ref{table:fraction_contaminants_mockfaint_5sigma}. We emphasise that the completeness is calculated as the fraction of all fiducial $z_{\mathrm{fid}}=6$--8 galaxies that satisfy the detection threshold requirement; otherwise, variations in the completeness could be simply attributed to the restriction imposed on the sample. 

From Table\,\ref{table:fraction_contaminants_mockfaint_5sigma}, it is clear that a NIR detection threshold requirement for faint, apparent $z>6$ galaxies considerably reduces contamination from intermediate-$z$ sources. In any data scenario, the $z>6$ completeness is comparable to that of the UltraVISTA-like bright sample. This is expected, as the faintest and therefore most poorly constrained high-redshift sources do not survive the detection threshold requirement. Between the different combinations of \Euclid\ and ancillary photometry, we find that including \textit{Spitzer} photometry does not improve the redshift recovery, nor is it able to distinguish contaminants from true $z>6$ sources. Again, the only truly valuable addition to \Euclid\ data is ultra-deep photometry from Rubin DDF.  

To conclude, the high level of $z>6$ galaxy recovery together with the low contamination fraction are only representative of bright galaxies equivalent to those in the UltraVISTA catalogue. As we showed here, the $z>6$ completeness level becomes lower and the contamination fraction significantly higher for fainter sources, which will be part of the Euclid Deep Survey as well. Although we have shown that a NIR detection threshold requirement only moderate reduces the degree of contamination in the UltraVISTA-like bright sample, it is very useful for the faint mock sample.  

{\renewcommand{\arraystretch}{1.5}
\begin{table*}
\caption{Contaminants amongst $z>6$ galaxies in the restricted faint mock sample.}             
\label{table:fraction_contaminants_mockfaint_5sigma}      
\centering                         
\begin{tabular}{l c c c c }        
\hline\hline                 
Filters & True $z>6$ & Contaminants & Contamination Fraction & Completeness\\    
\hline                        
   \Euclid\ & 609 & 204 & $0.25^{+0.07}_{-0.07}$ & 91\,\%\\  
   \hline
   \Euclid\ + Rubin & 609 & 202 & $0.25^{+0.07}_{-0.07}$ & 91\,\%\\
   \Euclid\ + Rubin DDF & 636 & 28 & $0.04^{+0.02}_{-0.01}$ & 95\,\%\\
   \Euclid\ + H20 & 617 & 172 & $0.22^{+0.06}_{-0.06}$ & 92\,\%\\
   \hline
   \Euclid\ + \textit{Spitzer} & 312 & 107 & $0.26^{+0.06}_{-0.07}$ & 91\,\%\\
   \hline
   \Euclid\ + Rubin + \textit{Spitzer} & 315 & 97 & $0.24^{+0.06}_{-0.07}$ & 92\,\%\\
   \Euclid\ + Rubin DDF + \textit{Spitzer}  & 329 & 18 & $0.05^{+0.02}_{-0.02}$ & 96\,\%\\
   \Euclid\ + H20 + \textit{Spitzer} & 309 & 94 & $0.23^{+0.06}_{-0.06}$ & 90\,\%\\
\hline                                   
\end{tabular}
\tablefoot{Number of true $z>6$ galaxies and $z>6$ contaminants, from various combinations of \Euclid\ and ancillary data, based on the restricted faint mock sample. The restricted faint mock sample consists of sources that have a 5$\sigma$ flux measurement in at least one the \textit{Euclid} NIR bands. We report the contamination fraction and $z>6$ completeness, where the latter is calculated as the fraction of all fiducial $z_{\mathrm{fid}}=6$--8 sources that satisfy the detection threshold requirement. } 
\end{table*}}

\section{Discussion}\label{sec:discussion}

\subsection{Validity of the simulated high-$z$ solutions} \label{sec:discussvalidity}
\noindent We present numbers of $z>6$ contaminants for each scenario of simulated photometry in Table \ref{table:fraction_contaminants}. These contaminants are identified from their simulated redshift, with no further checks on the compatibility of the photometry and the best-fit photometric redshifts. Therefore, we explore if additional validation of their redshift could reduce the fraction of contaminants amongst $z>6$ galaxies. 

We checked if the photometric redshifts are compatible with detections at short wavelengths. At $z_{\mathrm{sim}}>6$, the Lyman limit at $\lambda = 912\,\AA$ is redwards of the Rubin $u$ and $g$ filters, and CFHT $u$ and HSC $g$ filters. Therefore, any apparent $z>6$ galaxy with a $>2\sigma$ detection limit in these bands can be ruled out. Similarly, at $z_{\mathrm{sim}}>6.5$ the Lyman limit is redwards of the Rubin $r$ and HSC $r$ bands, so a detection in said bands should not exceed its $2\sigma$ limit. Additionally, due to Lyman series absorption of neutral hydrogen in the intergalactic medium \citep{inoue+2014}, the Lyman break of a galaxy spectrum shifts to the  Lyman-$\alpha$ line at $\lambda = 1216\,\AA$; as such, sources at $z_{\mathrm{sim}}>7$ with Rubin $r$ and $i$, HSC $r$ and $i$, and \IEuc\ detections exceeding their $2\sigma$ limit are ruled out, as are $z_{\mathrm{sim}}>8.1$ galaxies with a $>2\sigma$ detection in the \YEuc\ band. 

Based on the bright UltraVISTA-like sample, imposing these conditions on the $z>6$ contaminant sample has strongly varying consequences depending on the photometric availability: 2.0\,\% of the \Euclid-observed contaminants are discarded; 20\% of the \textit{Euclid}+Rubin-observed contaminants are discarded; 38\,\% of the \textit{Euclid}+Rubin DDF-observed contaminants are discarded; 28\,\% of the \textit{Euclid}+H20-observed contaminants are discarded; and 0\,\% of the \textit{Euclid}+\textit{Spitzer}-observed contaminants are discarded. We find considerably lower values for the faint mock sample; between 0.2 and 4\,\% of contaminants can be discarded from these checks, depending on the combination of \Euclid\ and ancillary data. Clearly, it is useful to compare the best-fit photometric redshifts with their detections in ancillary optical bands, as it will improve the contamination fraction. Nonetheless, we conclude that $z>6$ contaminants generally are assigned valid redshifts by the SED fitting based on their simulated photometry. Therefore, the majority of contaminants are not the result of poor SED fitting in which short-wavelength photometry is ignored by the algorithm. 

\subsection{Independent check of contamination fraction with COSMOS2020}
We verified our estimates of the contamination amongst \Euclid\ $z>6$ galaxies by repeating our analysis of the bright UltraVISTA-like sources using a different photometric catalogue and other SED fitting codes. We used the COSMOS2020 catalogue \citep{weaver2021} to simulate \Euclid\ (+ancillary photometry) following the same method presented in Sect.\,\ref{sec:sim_data}. The COSMOS2020 catalogue contains the latest updated data in the COSMOS field \citep{scoville2006}, with photometry for over 1.7 million sources in 44 optical and infrared bands. In this work we have made use of the so-called COSMOS2020 Classic catalogue,  which contains aperture photometry performed on PSF-homogenised images in all bands, except for the IRAC photometry, which was derived from PSF-fitting with the \texttt{IRACLEAN} software \citep{hsieh2012}. We provide a detailed description on how we selected a sample of $z=1$--8 galaxies from this catalogue in Appendix \ref{appendix}.

Using the COSMOS2020 Classic photometry, we performed three independent tests to derive the contamination fraction in various \Euclid\ (+ancillary) data scenarios. Here we briefly list our tests for which fitting codes were used (detailed descriptions of the tests are provided in Appendix \ref{appendix}):
(i) The new \texttt{C++} version of \texttt{LePhare} (\texttt{LePhare++}), following the two-step SED fitting strategy used in \citet[][see our Appendix \ref{appendix:lepharepp}]{weaver2021}. (ii) The conventional \texttt{LePhare} algorithm, adopting the photometric redshifts from \citet{weaver2021} but deriving their corresponding best-fit templates using the SED fitting parameters from this paper (see Appendix \ref{sec:appendixlephare}). (iii) The new \texttt{Python} version of the \texttt{EAZY} template fitting algorithm (version \texttt{0.5.2.dev7}; \citealt{brammer2008}; see our Appendix \ref{sec:appendixeazy}).

Using these photometric redshift codes, we simulated \Euclid\ (+ancillary) photometry from the COSMOS2020 catalogue. For each test, we recomputed the photometric redshifts based on this photometry and tabulated the resulting estimates of the contamination fraction and $z>6$ completeness in Appendix \ref{appendix}.

Here we reflect on the reproducibility and therefrom resultant validity of our \Euclid\ $z>6$ galaxy sample contamination estimates. We find generally good agreement between the contamination fractions derived from our own UltraVISTA catalogue and from the COSMOS2020 catalogue processed by either version of the \texttt{LePhare} code. Therefore, we conclude that regardless of the fiducial photometry, between 15 and 20\,\% of apparent $z>6$ galaxies identified from \Euclid\ photometry with \texttt{LePhare} are actually intermediate-$z$ interlopers. As we have shown in Sect.\,\ref{sec:mock}, this number is likely too optimistic, since \Euclid\ will probably detect fainter $z>6$ galaxies than what is currently possible with COSMOS. The contamination is neither surprising nor unanticipated, as huge efforts are being made to collect ancillary photometry in all three Euclid Deep Fields. Overall, this result emphasises how we should be careful with \Euclid\ high-redshift candidates for which ancillary photometry is not available. 

Between the \texttt{LePhare} strategies, the contamination fractions with \Euclid\ data alone agree well, although those derived from the COSMOS2020 photometry are slightly smaller. Interestingly, we find conflicting results regarding the importance of ancillary data for preventing $z>6$ contamination. Overall, \textit{Spitzer} photometry simulated from COSMOS2020 data barely contributes to reducing the contamination fraction, contrary to the moderate improvement achieved with our own UltraVISTA photometry, whereas the COSMOS2020 catalogue actually contains the deepest possible IRAC photometry over the COSMOS field. In addition, ancillary optical photometry from the H20 survey improves the contamination fraction in the test with \texttt{LePhare++}. We believe this is possible given that the COSMOS2020 catalogue contains very deep optical imaging from the Subaru HSC. However, we do not find the same improvement from the H20 survey data in the test with \texttt{LePhare}. Nonetheless, we observe the same general trend that optical imaging from the Rubin DDF is most effective at distinguishing intermediate-$z$ interlopers from actual $z>6$ galaxies, which confirms that the relative depth of the ancillary optical data is of key importance.

We used \texttt{EAZY} to reproduce our results in three data scenarios: \Euclid, \textit{Euclid}+H20, and \textit{Euclid}+Rubin DDF. These scenarios were chosen as they summarise the different levels of expected depths of the ancillary data. Interestingly we find that the contamination fractions as derived with \texttt{EAZY} are extremely low in all three scenarios. In addition, the redshift recovery of actual $z>6$ galaxies is poorer than when using \texttt{LePhare}, with a completeness of $81$\,\% with \Euclid\ alone. Contrary to what is expected, the level of completeness drops considerably from adding ancillary data: in the \textit{Euclid}+H20 scenario, the $z>6$ completeness is only 45\,\%. In all three scenarios, the missing $z_{\rm fid}=6$--8 galaxies are misidentified as $z_{\rm sim}=1$--2 and $z_{\rm sim}=5.5$--6 galaxies, which is in accordance with the typical redshift degeneracies we identified for the intermediate-$z$ contaminants. These results demonstrate how our \texttt{EAZY} routine appears reluctant to assign high-redshift solutions at all, which is remarkable, as we did not employ an apparent magnitude prior for the SED fitting of the \Euclid\ (+ancillary) photometry. The lower $z>6$ completeness may be partially due to the different flux upper-limit strategies between \texttt{LePhare} and \texttt{EAZY}. With the former code, $2\sigma$ upper limits in the optical bands impose very strong constraints on the photometric redshift of a true $z>6$ galaxy; \texttt{LePhare} will outright reject most intermediate-$z$ solutions as they produce higher fluxes than the upper limits in optical bands. With \texttt{EAZY}, the upper limits are less constraining as they are treated as actual detections with large flux errors that reflect the $2\sigma$ detection level. Therefore, false intermediate-$z$ solutions for actual $z>6$ galaxies may be unlikely, but not immediately rejected. 

In conclusion, our tests using different SED fitting codes on the same dataset demonstrate that the choice of code could have strong implications for both the degree of contamination and the redshift recovery of $z>6$ galaxies. 

\subsection{{\color{black}Identification of $z>6$ galaxies from colour-colour diagrams alone}}

\noindent We {\color{black}showed} how, once the photometric redshifts are determined, a simple $(\mIEuc-\mYEuc)$ \& $(\mYEuc-\mJEuc)$ colour cut can weed out $z>6$ contaminants from true $z>6$ galaxies in all scenarios of photometric availability, but only for sources that have a flux measurement in at least the \IEuc\ or \YEuc\ band. Regardless, it makes sense to wonder to what degree it is possible to use directly this colour-colour plot for a $z>6$ galaxy selection in \Euclid, without the need for photometric redshifts in the first place. In fact, based on the bright UltraVISTA-like sample, we find that not any $z_{\rm fid}=1$--6 galaxy (with a detection in \IEuc\ and/or \YEuc), contaminant or not, populates the $(\mIEuc-\mYEuc) > 3.4$ and $(\mYEuc-\mJEuc) < 0.9$ colour box (\Euclid\ data alone). Compared to 184 true $z>6$ galaxies that do, this means one can potentially obtain a pure sample whilst preserving 58\,\% of the fiducial $z>6$ galaxies. This suggests that direct selection of $z>6$ galaxies may possible from \Euclid\ colours alone, circumventing the need for photometric redshifts. 

However, once we take into account the faint mock sample, it becomes clear that photometric redshifts are absolutely necessary. In the scenario of \Euclid data alone, 66\,\% of contaminants in the faint mock sample have an unconstrained $(\mIEuc-\mYEuc)$ colour. If we would apply the colour criteria $(\mIEuc-\mYEuc) > 3.4$ and $(\mYEuc-\mJEuc) < 0.9$ to the faint mock sample, virtually none of the contaminants that have a $(\mIEuc-\mYEuc)$ colour survive, but neither do the actual high-redshift galaxies: the completeness drops to 13\,\%. In addition, considering that more than half of the contaminants cannot be included in this analysis at all, we can conclude that using colour cuts to identify faint high-redshift sources with \Euclid\ is unfeasible. 

For the bright UltraVISTA-like sample, we explore if combining the $(\mIEuc-\mYEuc)$ and $(\mYEuc-\mJEuc)$ colours with additional criteria can achieve pure $z>6$ galaxy selection with acceptable completeness. Based on the \Euclid\ photometry, we inspect the $(\mYEuc-\mJEuc)$ versus $(\mJEuc-\mHEuc)$ colour diagrams for all sources that survive the $(\mIEuc-\mYEuc) > 2.8$ \& $(\mYEuc-\mJEuc) < 1.4$ colour cut, including all galaxies with an unconstrained $(\mIEuc-\mYEuc)$ colour. We find that the true $z>6$ galaxies are well-separated from stable intermediate-$z$ galaxies, but not from contaminants. Additionally, many contaminants that have no $(\mIEuc-\mYEuc)$ colour actually have $(\mYEuc-\mJEuc)$ and $(\mJEuc-\mHEuc)$ colours similar to true $z>6$ galaxies, so applying a succeeding colour cut in this colour space is mostly unhelpful {\color{black}for weeding out contaminants}.  

Finally, before we can draw conclusions on the feasibility of \Euclid\ $z>6$ galaxy selection from colours alone, we have to address a caveat in our analysis: we have not inspected the colours of fiducial $z_{\mathrm{fid}}=0$--1 galaxies, given that by definition our intermediate-$z$ sample spans $z_{\mathrm{fid}}=1$--6. Upon inspection of the \Euclid\ colours of UltraVISTA galaxies at $z=0$--1, virtually none of these sources survive the $(\mIEuc-\mYEuc)$ \& $(\mYEuc-\mJEuc)$ colour criterion, and therefore contamination of $z>6$ galaxies by $z=0$--1 galaxies is negligible.

To summarise, our results demonstrate that one should not select a $z>6$ galaxy sample from \Euclid\ colours alone, as the colours $z>6$ contaminants are poorly constrained, especially in the faint regime. Accurate knowledge of photometric redshifts cannot be circumvented, as even the most stringent colour cut presented in this work is not guaranteed to eliminate intermediate-$z$ contamination, while it strongly sacrifices the $z>6$ completeness.

In general, we note that photometric redshift algorithms are by nature more efficient than colours alone for accurate selection of galaxy samples. As an alternative to the proposed colour selection criteria, one could make efforts to fine-tune \texttt{LePhare} (or any other SED fitting tool) to assign more weight to the \YEuc\ band or the $(\mYEuc-\mJEuc)$ colour, thereby optimising the algorithm to further lift the degeneracy between intermediate- and high-redshift galaxies. More work is warranted to investigate how to optimise SED fitting codes to improve the discrimination between true $z>6$ and intermediate-$z$ interlopers. 

In addition, machine learning methods as alternatives to traditional SED fitting are becoming increasingly popular, especially with the advent of large sky surveys such as the Euclid Survey. Recent works using mock \Euclid\ photometry have demonstrated how machine learning approaches typically outperform template-fitting algorithms to retrieve the photometric redshift and galaxy classification at $z<1$ (\citealt{desprez2020}; Humphrey et al. 2021 in prep). At higher redshifts, machine learning methods become increasingly unreliable because extensive spectroscopic training samples are lacking. However, \Euclid\ will obtain spectroscopy and high-quality photometry in several reference fields that already contain deep ancillary data. Therefore, machine learning methods for selecting $z>6$ galaxies are likely to become more viable in the future and will possibly reduce contamination of intermediate-$z$ galaxies further.

\subsection{Possible caveats in this work}

As a few final words of caution, a possible caveat in this work is how the \Euclid\ (+ancillary) photometry is simulated from best-fit SED templates, as compared to extracting fluxes from real \Euclid\ (+ancillary) images or even simulated images that emulate the real data more directly. By sampling the photometry directly from models, we do not take into account factors that could degrade the quality of the photometric measurements, such as telescope defects and source confusion. We do note, however, that \Euclid\ will have an excellent ({HST}-like) angular resolution, and, thus, source blending is unlikely to be a major problem, considering the depths of the Euclid Deep Fields. Nonetheless, the simulated data in this work are idealised to a certain degree, and as such photometric redshift degeneracies could be more prevalent for the real data. Moreover, we used the same template set to simulate the \Euclid\ (+ancillary) photometry and to recompute the photometric redshifts based on these simulated data. The template fluxes were randomised as we described in Sect.\,\ref{sec:sim_data}, but nonetheless, our results on the contamination fractions could be affected by this.

Second, throughout this work we have not distinguished between normal galaxies and galaxies that host an active galactic nucleus (AGN), even though we know that these types of galaxies are present in the COSMOS field \citep{brusa2010, delvecchio2017,chang2017}. We cross-correlated our UltraVISTA DR4 catalogue with the C-COSMOS X-ray catalogue \citep{civano2016} and identified 996 X-ray AGN sources with host galaxy redshifts between $z=1$--5.3. These sources are rare, as they constitute only 0.6\,\% of the total galaxy population between $z=1$--5.3. Moreover, we find that the simulated redshifts of these sources lie below $z=6$ in any scenario of \Euclid\ (+ancillary) data. Therefore, we conclude that our X-ray selected AGNs do not contaminate the \Euclid\ high-redshift galaxy sample. However, alternative methods to identify AGNs exist, so more work regarding this AGN contamination of $z>6$ galaxies is desired.

Lastly, throughout this work we have assumed the expected full depths for the \Euclid\ and Rubin photometry, finalised once the nominal missions have been completed. With this assumption, we have demonstrated how ancillary photometry is only marginally helpful to prevent intermediate-$z$ contaminants from entering the high-redshift sample, and how $z>6$ galaxies are well-recovered from \Euclid\ data alone for both bright and faint galaxies. This does not imply that the ancillary data have an insignificant role in the study of high-$z$ galaxies. The efforts to gather ancillary imaging in the Euclid Deep Fields are very important. For a long time the ancillary data in the optical bands will be deeper than that provided by \Euclid\ \IEuc. In fact, the final \IEuc\ depth assumed here will not be achieved until late stages of the \Euclid\ operations. Given that the \Euclid\ and Rubin missions will run over many years, interim datasets at intermediate depths will be released. Therefore, ancillary imaging will be very important for all the studies of high-$z$ galaxies performed in the first years of \Euclid\ data analysis. In general, we conclude that more work is required to analyse the contamination fraction estimates at intermediate \Euclid\ and Rubin depths.

\section{Summary and conclusions}\label{sec:conclusion}

We have investigated the contamination from intermediate-$z$ interlopers ($z=1$--5.8) of the $z>6$ galaxy population as expected for the Euclid Deep Fields. Our tests are based on $\sim$176\,000 real galaxies at $z=1$--8 in a $\sim$0.7 deg$^2$ area selected from the UltraVISTA ultra-deep survey in the COSMOS field and an additional sample of $\sim$96\,000 mock galaxies with $25.3 \leq H<27.0$ that follow a CANDELS-like photometric redshift and $H$ magnitude distribution. For both datasets, we simulated \Euclid\ and ancillary photometry from fiducial 28-band photometry and subsequently re-derived photometric redshifts for eight scenarios of data availability: (i) \Euclid\ data only; (ii) \Euclid\ and Rubin $ugrizy$ data; (iii) \Euclid\ and ultra-deep Rubin $ugrizy$ data from the Rubin DDF; (iv) \Euclid\ and CFHT $u$ and Subaru HSC $griz$ data from the H20 survey; (v) \Euclid\ and \textit{Spitzer} IRAC 3.6 and 4.5 \micron\ data; (vi) \Euclid, Rubin, and \textit{Spitzer} data; (vii) \Euclid, Rubin DDF, and \textit{Spitzer} data; and (viii) \Euclid, H20, and \textit{Spitzer} data. We emphasise that the findings presented below are only representative of galaxies up to $z=8$ due to the limitations of our fiducial sample and that we cannot assess the photometric redshift recovery of \Euclid\ $z>8$ galaxies.  

\begin{enumerate}
    \item We determined the fraction of intermediate-$z$ contaminants (fiducial $z=1$--5.8) amongst the apparent $z>6$ population as identified from \Euclid\ (+ancillary) data. Based on the bright UltraVISTA-like sample, we estimate the contamination fraction of $z>6$ galaxies observed with \Euclid\ to be $0.18^{+0.07}_{-0.06}$. Contrarily, when we consider the faint mock sample, the contamination fraction of $z>6$ galaxies with \Euclid\ is significantly higher at $0.39^{+0.10}_{-0.11}$.
    
    \item Based on the bright UltraVISTA-like sample, we find that the contamination fraction is reduced to $0.13^{+0.04}_{-0.05}$ when including \textit{Spitzer} IRAC photometry, and to $0.04^{+0.03}_{-0.02}$ when including Rubin DDF photometry. In our most optimistic scenario, where we combined \Euclid, Rubin DDF, and \textit{Spitzer} photometry, virtually any possible contamination from intermediate-$z$ interlopers is ruled out; the contamination fraction is $0.01^{+0.0006}_{-0.0001}$. Conversely, when we consider the faint mock sample, the contamination fraction is only significantly reduced once we include the ultra-deep Rubin DDF photometry, to $0.07^{+0.03}_{-0.03}$. 
    
    \item We replicated our analysis of the bright UltraVISTA-like sample on the COSMOS2020 galaxy catalogue, which contains independent photometry over the same field, in combination with various SED fitting routines. We find generally good agreement between the contamination fraction estimated from our own photometry and those derived using \texttt{LePhare}/\texttt{LePhare++} with the COSMOS2020 catalogue. However, we obtain different results using the \texttt{EAZY} code, which returns a lower fraction of contaminants as well as a lower completeness in the recovery of $z>6$ galaxies. 
    
    \item The contaminants of $z>6$ galaxies have distinctly different $(\mIEuc-\mYEuc)$ colours from true $z>6$ galaxies, so colour selection criteria can be used to separate these populations a posteriori of obtaining the \textit{Euclid}-like photometric redshifts. However, many contaminants have unconstrained $(\mIEuc-\mYEuc)$ colours and as such cannot be included in this analysis. Regardless, we have presented a grid of $(\mIEuc-\mYEuc)$ \& $(\mYEuc-\mJEuc)$ colour cuts that balance the contamination fraction and $z>6$ galaxy completeness. For the bright UltraVISTA-like sample, the colour cut $(\mIEuc-\mYEuc)>2.8$ \& $(\mYEuc-\mJEuc)<1.4$ when applied to the apparent $z>6$ galaxies (that have a flux measurement in the \IEuc\ and/or \YEuc\ bands; identified from \Euclid\ data) reduces the contamination fraction to 0.01 while preserving 81\,\% of the fiducial $z>6$ galaxies (i.e. galaxies at fiducial $z=6$--8 that are recovered at $z>6$ from \Euclid\ photometry). Considering the faint mock sample, these proposed colour selection criteria are not useful, given that the majority of contaminants are undetected in both the \IEuc\ and \YEuc\ bands. 
    
        \item Alternatively, we find that imposing a $5\sigma$ detection threshold requirement in at least one of the \Euclid\ NIR bands is useful for obtaining a purer apparent $z>6$ sample, specifically for the faint mock sample. By employing this detection threshold requirement, the contamination fraction of the faint mock $z>6$ galaxies is $0.25^{+0.07}_{-0.07}$ (\Euclid\ data alone), and we maintain a $z>6$ completeness of 91\,\%.\ As such, the identification of faint high-redshift sources in the Euclid Deep Fields will be very efficient, with a moderate contamination of intermediate-$z$ galaxies.
        
        \item We investigated if one can select a pure sample of $z>6$ galaxies from \Euclid\ colours alone. Based on the bright UltraVISTA-like sample, we have shown how a strict colour cut at $(\mIEuc-\mYEuc)>3.4$ \& $(\mYEuc-\mJEuc)<0.9$ could possibly select a pure sample of high-redshift galaxies, whilst maintaining a $z>6$ completeness of 58\,\%. However, this cannot be achieved for the faint mock sample at all, and therefore we conclude that deriving photometric redshifts cannot be circumvented. 
    
    \item We compared the fiducial physical properties (based on 28 bands in COSMOS) of the intermediate-redshift galaxies that are $z>6$ contaminants and stable intermediate-$z$ galaxies (i.e. $z=1$--6 galaxies that stay in the same redshift bin when observed with \Euclid\ (+ancillary) data). The analysis of the physical parameters is solely based on the bright UltraVISTA-like sample. We find that the $z>6$ contaminants reside primarily at fiducial redshifts $z_{\mathrm{fid}} \sim 1$--3 and $z_{\mathrm{fid}}\sim4.5$--6 and are primarily faint, regardless of which ancillary data are added to the \Euclid\ photometry. We identify three distinct populations: (i) moderately reddened galaxies at $z_{\mathrm{fid}}=$1--3.5 that are misidentified through the typical confusion of the Lyman-$\alpha$ break at $\lambda = 1216\,\AA$ and the $4000\,\AA$ break; (ii) a group of contaminants at $z_{\mathrm{fid}}=3.5$--5 that are strongly dust-reddened;  and (iii) a group at $z_{\mathrm{fid}}=5$--6 that have typically ill-defined, flat, fiducial SEDs and are mistaken for $z\sim6$ galaxies.
     
    \item Moreover, we compared the physical properties as derived with \Euclid\ (+ancillary) photometry of $z>6$ contaminants and true $z>6$ galaxies. We have demonstrated how \textit{Spitzer} photometry is essential for recovering the stellar masses of $z>6$ galaxies. The $z>6$ contaminants are most separated from true $z>6$ galaxies based on their colour excess, as the latter have virtually no dust attenuation. Although the parameter distributions of true $z>6$ galaxies and $z>6$ are generally statistically different, we find no selection criteria that effectively separate them based on their physical parameters. 
\end{enumerate}

{\color{black} \noindent Overall, we conclude that the \Euclid\ high-redshift recovery will be excellent for bright $z=6$--8 galaxies, and successful for faint galaxies as well. In addition, ultra-deep ancillary photometry is highly effective at reducing contamination from intermediate-$z$ interlopers to the $z>6$ galaxy sample.} 

\begin{acknowledgements}

Based on data products from observations conducted with ESO Telescopes at the Paranal Observatory under ESO program ID 179.A-2005 and on data products produced by TERAPIX and the Cambridge Astronomy Survey Unit on behalf of the UltraVISTA consortium. Also based in part on observations carried out with the \textit{Spitzer} Space Telescope, which is operated by the Jet Propulsion Laboratory, California Institute of Technology under a contract with NASA.  Also based on observations carried out by NASA/ESA Hubble Space Telescope, obtained and archived at the Space Telescope Science Institute; and the Subaru Telescope, which is operated by the National Astronomical Observatory of Japan. This research has made use of the NASA/IPAC Infrared Science Archive, which is operated by the Jet Propulsion Laboratory, California Institute of Technology, under contract with NASA.\\
SvM and KC acknowledge funding from the European Research Council through the award of the Consolidator Grant ID 681627-BUILDUP. PD acknowledges support from the European Research Council's starting grant ERC StG-717001 (DELPHI), from the NWO grant 016.VIDI.189.162 (ODIN) and the European Commission's and University of Groningen's CO-FUND Rosalind Franklin program. \AckEC We thank Smaran Deshmukh for useful discussions on the SMUVS catalogue photometry. We thank Marc Sauvage for carefully reading the manuscript and providing constructive comments for the Euclid Consortium internal review. 

\end{acknowledgements}

\bibliographystyle{aa} 
\bibliography{euclid_cont_z6_vanmierlo2022_fv.bib} 

\begin{appendix}
\section{COSMOS2020}\label{appendix}

In order to verify the validity of our estimates on the fraction of intermediate-$z$ interlopers amongst the \textit{Euclid}-observed $z>6$ galaxy population, we have repeated our analysis of the bright UltraVISTA-like sources on a different photometric catalogue and different SED fitting codes. 

We have used photometry from the COSMOS2020 catalogue (\citealt{weaver2021}; version of December 2020; hereafter W21), which comprises the latest data release of COSMOS (\citealt{scoville2006}). This catalogue includes new ultra-deep optical data from DR2 of the HSC Subaru Strategic Program \citep{aihara2019}, infrared data from the fourth release of the UltraVISTA survey \citep{mccracken2012}, and all \textit{Spitzer} data in four IRAC channels ever obtained in the COSMOS field \citep{moneti2021}. Two independent versions of the COSMOS2020 catalogue using different techniques of measuring photometry were constructed, and in this work we made use of the so-called Classic catalogue. For this version, source detection was performed with \texttt{SourceExtractor} on the combined $izYJHK_{\rm s}$ detection image, after which photometry was measured in $2''$ and $3''$ diameter apertures on the PSF-homogenised images in 40 optical and NIR bands. The IRAC fluxes were extracted separately using the \texttt{IRACLEAN} software \citep{hsieh2012}. The final catalogue consists of roughly $1\,700\,000$ sources. We refer the reader to W21 for a complete description of the COSMOS2020 Classic photometric catalogue. Throughout this paper we globally refer to the COSMOS2020 Classic photometry as the COSMOS2020 aperture photometry. 

In addition, W21 presented two independent sets of photometric redshifts for all sources in the COSMOS2020 Classic catalogue, derived from two different template-fitting routines, namely \texttt{LePhare} and \texttt{EAZY} (again, we refer the reader to the original publication for the full details on their methodology). Here we make use of both sets of photometric redshifts. 

For our tests, we only considered sources classified as galaxies (type=0) from this catalogue. In addition, we considered only the very deepest parts of COSMOS2020 catalogue and therefore cut the catalogue to ensure its area overlaps exactly with our own UltraVISTA catalogue (which comprises three out of the four ultra-deep stripes). These measures reduced the sample to 316\,698 galaxies. We used the aperture and Galactic extinction corrections from W21 (the latter based on dust maps from \citealt{schlafly2011}, consistent with our own analysis) to obtain total, dust-corrected photometry. 

Our goal is to derive \textit{Euclid}-like photometry and photometric redshifts based on the COSMOS2020 catalogue, in three independent tests where we use different SED fitting algorithms. 

\subsection{COSMOS2020 with the \texttt{LePhare++} routine} \label{appendix:lepharepp}

For our first test, we simply replicated the SED fitting process with \texttt{LePhare} as described in W21 to the best of our ability. We used an updated version of \texttt{LePhare} called \texttt{LePhare++}, which is based on the \texttt{Fortran} version of the code but has been migrated to \texttt{C++}. The set of galaxy templates included 19 empirical elliptical and spiral templates from \citet{poletta2007}, 12 star-forming galaxy models from the \citet[hereafter BC03]{bruzual2003} library and two BC03 templates with exponentially declining star-formation rates. The considered attenuation curves are from \citet{calzetti2000}, \citet{prevot1984}, and two realisations of the Calzetti law including the $2175\,\AA$ bump \citep{prevot1984} at different strengths. Dust extinction was varied between $E(B-V)=0.0$ and 0.5 in 0.05 steps and the considered redshift range was $z=0$--10 with $\Delta z=0.01$ increments. Emission lines were included in the fit following the original recipe from LePhare \citep{ilbert2009}, but using a new functionality in \texttt{LePhare++}, we allowed the modelled O{\sc iii} $\lambda\lambda4931,5007$ flux to vary between 50\,\% and 150\,\% (in 25\,\% steps) of the theoretical line flux. 

This run was based on 34 bands from the COSMOS2020 photometric catalogue: GALEX NUV; CFHT MegaCam u and u$^*$; Subaru HSC $g$, $r$, $i$, $z$, and $y$; Subaru Suprime-Cam $B$, $V$, $r^+$, $i^+$, $z^{++}$, and 14 intermediate- and narrow bands; VISTA $Y$, $J$, $H$, $K_{\rm s}$, and $NB118$; and IRAC channels 1 and 2. We applied photometric offsets as presented in Table 2 in W21 to the fluxes and added 0.02 mag errors in quadrature to the photometric errors in the optical bands, 0.05 mag errors to the VISTA $J$, $H$,  and $K_{\rm s}$ bands and IRAC channel 1 and the three narrow bands, and 0.1 mag to IRAC channel 2. 

Once we derived the photometric redshifts for the COSMOS2020 galaxies in this way, we fixed their redshifts to these values and used \texttt{LePhare++} with an alternate template library to derive their physical properties and best-fit SEDs from which we measure \textit{Euclid}-like photometry. The two-step strategy follows W21 and is necessary as the empirical galaxy templates cannot be used to derive physical parameters. In this second run, we used a set of 11 BC03 templates with nine exponentially declining and two delayed exponentially declining SFHs, assuming a \citet{chabrier2003} IMF. We adopted two metallicities: solar ($Z=0.02$) and half-solar ($Z=0.008$). The dust extinction was allowed to vary between $E(B-V)=0.0$ and 0.7 in 0.1 steps. The extinction laws used were the \citet{calzetti2000} law and a curve that has a slope in between the \citet{calzetti2000} law and the \citet{prevot1984} curve ($\lambda^{0.9}$; Appendix A of \citealt{arnouts2013}). Again, we add emission lines using the recipe from \citet{ilbert2009}, but employ no dispersion of the O{\sc iii} doublet. We emphasise again that this two-step method is directly taken from W21. 

Once we obtained the best-fit SEDs, we derived the \Euclid\ (+ancillary) photometry in precisely the same manner as presented in Sect.\,\ref{sec:sim_data}. An important caveat is that the source detection of the COSMOS2020 catalogue was performed using the UltraVISTA DR4 mosaics, which means the catalogue is deeper than our own DR3-derived UltraVISTA catalogue. Given that we want to exclude sources that lie beyond the detection abilities of \Euclid, we restrained the COSMOS2020 sample to sources that have at least a $3\sigma$ detection in both the simulated \JEuc\ and \HEuc\ bands.

In total, the COSMOS2020 photometry processed with the above described SED fitting routine, after applying the \JEuc- and \HEuc-band requirements, yields 164\,800 galaxies with fiducial redshifts $z_{\mathrm{fid}}=1$--8, of which 598 are at $z_{\mathrm{fid}}=6$--8. Out of this sample, 127\,985 galaxies are detected in at least IRAC channel 1 or channel 2, and 342 of these sources are at $z_{\mathrm{fid}}=6$--8.

For each of the eight \Euclid\ and ancillary data availability scenarios, we re-derived the photometric redshifts of the COSMOS2020 galaxies from their simulated \Euclid\ (+ancillary) photometry alone, using \texttt{LePhare++} with the same set of empirical and BC03 templates as described above in the first step of the two-step SED fitting strategy. Following the method outlined in Sect.\,\ref{sec:contaminants}, we calculated the high-redshift completeness and fraction of intermediate-$z$ interlopers amongst $z>6$ galaxies. The results are shown in Table \ref{table:fraction_contaminants_c2020_weaverroutine}. 

\subsection{COSMOS2020 with the \texttt{LePhare} routine used in this paper} \label{sec:appendixlephare}
For our second test, we used the COSMOS2020 aperture photometry in combination with \texttt{LePhare} (the \texttt{Fortran} version), using the SED fitting settings we adopted for our own UltraVISTA catalogue (Sect.\,\ref{sec:data}). As a reminder, our SED fitting routine considers among others a smaller redshift range ($z=0$--9), templates from the BC03 library only and a broader dust extinction range, that is, $E(B-V)=0$--1 with the \citet{calzetti2000} reddening law. We based this test on the same 34 optical and infrared band as listed for the previous test. We did not derive the fiducial photometric redshifts from scratch, instead fixing the redshift to those derived by W21 with \texttt{LePhare}. However, given that for this test, our \texttt{LePhare} parameters are not identical to those of W21, we minimised any discrepancies between our resulting best-fit SED and the fixed photometric redshift by deriving photometric offsets. This was done iteratively, until the difference between the input photometry and template fluxes converged. Subsequently, we sample the \Euclid\ (+ancillary) photometry from the final best-fit SED. 

With this second SED fitting method, the COSMOS2020 photometry yields 165\,190 galaxies at fiducial redshifts $z_{\mathrm{fid}}=1$--8 that exceed the $3\sigma$ flux in the \JEuc\ and \HEuc\ bands, of which 466 are at $z_{\mathrm{fid}}=6$--8. Out of this sample, 124\,520 galaxies are detected in at least IRAC channel 1 or channel 2, and 253 of these are at $z_{\mathrm{fid}}=6$--8.

Subsequently, photometric redshifts based on the \Euclid\ (+ancillary) photometry were recomputed using our own \texttt{LePhare} fitting routine. We derived the contamination fraction and $z>6$ completeness for each combination of \Euclid\ and ancillary photometry based on this second test with the COSMOS2020 catalogue, and show the results in Table \ref{table:fraction_contaminants_c2020}.

\subsection{COSMOS2020 with the \texttt{EAZY} code} \label{sec:appendixeazy}
For our third test, we used the COSMOS2020 aperture photometry but now with another SED fitting code, that is, the newly updated version of the \texttt{EAZY} code \citep{brammer2008}, which has been migrated to \texttt{Python} (version \texttt{0.5.2.dev7}). Similar to \texttt{LePhare}, \texttt{EAZY} is a template-fitting algorithm that assigns photometric redshifts to multi-wavelength photometry. This code was used by W21 to provide a second set of photometric redshifts based on the COSMOS2020 data. Therefore, we reproduce their process as closely as possible in order to recover the \texttt{EAZY} best-fit templates for the COSMOS2020 Classic catalogue. 

Following W21, \texttt{EAZY} uses a set of 17 templates derived from the Flexible Stellar Population Synthesis models \citep{conroy2009, conroy2010} with a range of dust attenuations and SFHs, and fits linear combinations of these templates to the observed photometry. The redshift is allowed to range from $z=0$--12 and a $K_{\rm s}$ apparent magnitude prior is applied in the process. We include the photometric offsets computed by W21 in the fit, derived iteratively from a subsample of
sources with spectroscopic redshifts. Multiplicative factors as presented in W21 were applied to the photometric errors, and the same magnitude errors as used for LePhare were added in quadrature. We selected a sample of $z=1$--8 galaxies from the EAZY COSMOS2020 catalogue and derived our own fiducial photometric redshifts with this method. \texttt{EAZY} was used to retrieve the observed frame fluxes in the \Euclid\ (+ancillary) filters.

Our strategy to ensure realistic depths for the simulated photometry is slightly different. With \texttt{LePhare}, one is able to set flux upper limits such that any template that produces fluxes higher than the upper limits in bands with non-detections is immediately discarded. \texttt{EAZY} performs the template fitting in linear flux density units, so it naturally accounts for the $\sigma$ flux limits we derive from the expected \Euclid\ (+ancillary) depths when we set the flux uncertainty in each band to half the $2\sigma$ flux. Subsequently, we re-derived the photometric redshifts based on the simulated photometry using the templates and settings as used for the COSMOS2020 photometry, although we did not employ the apparent magnitude prior. 

The COSMOS2020 photometry with \texttt{EAZY} yields 160\,260 galaxies at fiducial $z_{\mathrm{fid}}=1$--8 that satisfy the \JEuc- and \HEuc-band criteria we imposed, of which 369 are at $z_{\mathrm{fid}}=6$--8. We calculated the contamination fraction and high-redshift completeness for three scenarios of \Euclid\ and ancillary photometry: \Euclid, \textit{Euclid}+H20, and \textit{Euclid}+Rubin DDF. The results are shown in Table \ref{table:fraction_contaminants_c2020_eazy}. For our test with \texttt{EAZY}, we chose to investigate only these three scenarios as they reflect well the different levels of depth expected from the ancillary data. 

Lastly, we note that uncertainties on the contamination fraction were derived for all three tests, using the exact same method adopted for our photometric catalogue (see Sect.\,\ref{sec:contaminants}). 

{\renewcommand{\arraystretch}{1.5}
\begin{table*}
\caption{Contaminants amongst $z>6$ galaxies using COSMOS2020 with \texttt{LePhare++}.}             
\label{table:fraction_contaminants_c2020_weaverroutine}      
\centering                         
\begin{tabular}{l c c c c }        
\hline\hline                 
Filters & True $z>6$ & Contaminants & Contaminant Fraction & Completeness\\    
\hline                        
   \Euclid\ & 527 & 90 & $0.15^{+0.03}_{-0.04}$  &  88\,\% \\  
   \hline
   \Euclid\ + Rubin & 531 & 98 & $0.16^{+0.05}_{-0.06}$  &  89\,\% \\  
   \Euclid\ + Rubin DDF & 561 & 30 & $0.05^{+0.02}_{-0.02}$  &  94\,\% \\  
   \Euclid\ + H20 & 541 & 86 & $0.14^{+0.04}_{-0.04}$  &  90\,\% \\  
   \hline 
   \Euclid\ + \textit{Spitzer} & 292 & 55 & $0.16^{+0.05}_{-0.05}$ & 85\,\%  \\
   \hline 
   \Euclid\ + Rubin + \textit{Spitzer} & 301 & 57 & $0.16^{+0.04}_{-0.04}$ & 88\,\%  \\
   \Euclid\ + Rubin DDF + \textit{Spitzer} & 317 & 18 & $0.05^{+0.02}_{-0.01}$ & 93\,\%  \\
   \Euclid\ + H20 + \textit{Spitzer} & 300 & 54 & $0.15^{+0.04}_{-0.04}$ & 88\,\%  \\
\hline                                   
\end{tabular}
\tablefoot{Number of true $z>6$ galaxies and $z>6$ contaminants, using the COSMOS2020 Classic catalogue from \citet{weaver2021}. The fiducial photometric redshifts and \Euclid\ (+ancillary) photometry were derived using the methods described in \citet{weaver2021}. Only sources with a $3\sigma$ detection in both the \JEuc\ and \HEuc\ bands are considered here. } 
\end{table*}}

{\renewcommand{\arraystretch}{1.5}
\begin{table*}
\caption{Contaminants amongst $z>6$ galaxies using COSMOS2020 with \texttt{LePhare}.}             
\label{table:fraction_contaminants_c2020}      
\centering                         
\begin{tabular}{l c c c c }        
\hline\hline                 
Filters & True $z>6$ & Contaminants & Contaminant Fraction & Completeness\\    
\hline                        
   \Euclid\ & 418 & 82 & $0.16^{+0.05}_{-0.04}$  &  90\,\% \\  
   \hline
   \Euclid\ + Rubin & 414 & 81 & $0.16^{+0.06}_{-0.05}$  &  89\,\% \\  
   \Euclid\ + Rubin DDF & 424 & 18 &  $0.04^{+0.02}_{-0.02}$  &  91\,\% \\  
   \Euclid\ + H20  & 418 & 58 & $0.12^{+0.04}_{-0.04}$  &  90\,\% \\  
   \hline 
   \Euclid\ + \textit{Spitzer} & 227 & 43 & $0.16^{+0.06}_{-0.06}$ & 90\,\%  \\
   \hline 
   \Euclid\ + Rubin + \textit{Spitzer}  & 227 &  47 & $0.17^{+0.06}_{-0.05}$ & 88\,\%  \\
   \Euclid\ + Rubin DDF + \textit{Spitzer}  & 228 & 8 & $0.03^{+0.01}_{-0.01}$ & 90\,\%  \\
   \Euclid\ + H20 + \textit{Spitzer} & 228 & 28 & $0.11^{+0.05}_{-0.04}$ & 90\,\%  \\
\hline                                   
\end{tabular}
\tablefoot{Number of true $z>6$ galaxies and $z>6$ contaminants, using the COSMOS2020 Classic catalogue from \citet{weaver2021}. The fiducial photometric redshifts and \Euclid\ (+ancillary) photometry were derived using the same method adopted for the UltraVISTA/SMUVS photometry, described in Sect.\,\ref{sec:sed}. Only sources with a $3\sigma$ detection in both the \JEuc\ and \HEuc\ bands are considered here. } 
\end{table*}}

{\renewcommand{\arraystretch}{1.5}
\begin{table*}
\caption{Contaminants amongst $z>6$ galaxies using COSMOS2020 with \texttt{EAZY}.}             
\label{table:fraction_contaminants_c2020_eazy}      
\centering                         
\begin{tabular}{l c c c c }        
\hline\hline                 
Filters & True $z>6$ & Contaminants & Contaminant Fraction & Completeness\\    
\hline                        
   \Euclid\ & 299 & 30 & $0.09^{+0.05}_{-0.05}$  &  81\,\% \\  
   \Euclid\ + Rubin DDF & 263 & 2 &  $0.01^{+0.002}_{-0.006}$  &  71\,\% \\  
   \Euclid\ + H20  & 165 & 0 & $0.0^{+0.00}_{-0.00}$  &  45\,\% \\  

\hline                                   
\end{tabular}
\tablefoot{Number of true $z>6$ galaxies and $z>6$ contaminants, using the COSMOS2020 Classic catalogue from \citet{weaver2021}. The fiducial photometric redshifts and \Euclid\ (+ ancillary) photometry were derived with \texttt{EAZY}, following the methods outlined in \cite{weaver2021}. Only sources with a $3\sigma$ detection in both the \JEuc\ and \HEuc\ bands are considered here. } 
\end{table*}}

\end{appendix}

\end{document}